\documentclass{book}

\usepackage{wrapfig}
\usepackage{amsmath}
\usepackage{amssymb}
\usepackage{epsfig}
\usepackage{graphicx}

\newcommand{\booktitle}{A.K. Hartmann: Practical Guide to Computer Simulations}

\DeclareMathOperator{\E}{E}
\DeclareMathOperator{\Var}{Var}
\DeclareMathOperator{\Med}{Med}
\DeclareMathOperator{\MSE}{MSE}

\newenvironment{definition}
{{\bf Definition}}
{}

\newenvironment{proof}
{{\bf Proof}}
{\hfill $\Box$}

\newenvironment{theorem}
{{\bf Theorem}}
{}

\newenvironment{example}
{{\bf Example}}
{}

\newenvironment{exercises}
{
  \noindent{\large \bf Exercises}\\
 (\small solutions: see CD enclosed with book)\\

  \begin{enumerate}
}
{
   \end{enumerate}
}

\newcommand{\myvec}[1]{{\ensuremath\underline{#1}}}

\newcommand{\fd}[2]{\frac{d #1}{d #2}}
\newcommand{\pd}[2]{\frac{\partial #1}{\partial  #2}}
\newcommand{\Prob}{P}

\newlength{\sourcewidth}
\newcommand{\sources}[2]
{
 \begin{wrapfigure}[4]{r}{\sourcewidth} 
\vspace*{-4mm}
\hfill
\fbox{
\begin{minipage}{0.8\sourcewidth}
\centerline{\small \tt GET SOURCE CODE}
\vspace*{-8pt}
\centerline{\rule{0.91\sourcewidth}{0.7pt}}
\small DIR: {\tt #1} \\
FILE(S): {\tt #2}
\end{minipage}
} 
\end{wrapfigure} 
}

\newcommand{\sourcesC}[3]
{
 \begin{wrapfigure}[#3]{r}{\sourcewidth} 
\vspace*{-4mm}
\hfill
\fbox{
\begin{minipage}{0.8\sourcewidth}
\centerline{\small \tt GET SOURCE CODE}
\vspace*{-8pt}
\centerline{\rule{0.91\sourcewidth}{0.7pt}}
\small DIR: {\tt #1} \\
FILE(S): {\tt #2}
\end{minipage}
} 
\end{wrapfigure} 
}

\newcommand{\sourcesTOPC}[3]
{
 \begin{wrapfigure}[#3]{r}{\sourcewidth} 
\hfill
\fbox{
\begin{minipage}{0.8\sourcewidth}
\centerline{\small \tt GET SOURCE CODE}
\vspace*{-8pt}
\centerline{\rule{0.91\sourcewidth}{0.7pt}}
\small DIR: {\tt #1} \\
FILE(S): {\tt #2}
\end{minipage}
} 
\end{wrapfigure} 
}

\newcommand{\sourcesB}[2]
{
\hfill
\fbox{
\begin{minipage}[t]{0.8\sourcewidth}
\centerline{\small \tt GET SOURCE CODE}
\vspace*{-8pt}
\centerline{\rule{0.91\sourcewidth}{0.7pt}}
\small DIR: 
{\tt #1} \\
FILE(S): {\tt #2}
\end{minipage}
} 
}
\newcommand{\sourcesSOL}[2]
{
\hfill
\fbox{
\begin{minipage}[t]{0.8\sourcewidth}
\centerline{\small \tt SOLUTION SOURCE CODE}
\vspace*{-8pt}
\centerline{\rule{0.905\sourcewidth}{0.7pt}}
\small DIR: 
{\tt #1} \\
FILE(S): {\tt #2}
\end{minipage}
} 
}

 \newcommand\wrapfill{\par
  \ifx\parshape\WF@fudgeparshape
  \nobreak
  \vskip-\baselineskip
  \vskip\c@WF@wrappedlines\baselineskip
  \allowbreak
  \WFclear
  \fi
}

\usepackage{makeidx}
\usepackage{lineno}
\makeindex

\title{
Introduction to Randomness and Statistics\\
{\large excerpt from the book}\\
Practical Guide to Computer Simulations\\
{\large World Scientific 2009, ISBN 978-981-283-415-7}\\
{\large see {\tt http://www.worldscibooks.com/physics/6988.html}}\\
{\normalsize with permission by World Scientific Publishing Co. Pte. Ltd.}}
\author{Alexander K. Hartmann\\ Institute of Physics\\University of Oldenburg\\
Germany}

\begin{document}

\maketitle

\setcounter{chapter}{6}



\chapter{Randomness and Statistics\label{chap:random}\label{chap:data}}

\markboth{\booktitle}{Randomness and Statistics}  

In this chapter, we are concerned with statistics in a very broad sense.
This involves 
generation of (pseudo) random data, display/plotting 
of data and the statistical analysis of simulation results.

Frequently, a simulation
involves the explicit generation of random numbers, for instance,
 as auxiliary quantity for stochastic simulations. In this
case it is obvious that the simulation results are random as well.
Although there are many simulations which are explicitly not random,
 the resulting behavior
of the simulated systems may appear also random, for example the motion
of interacting gas atoms in a container.   
Hence, methods from
statistical data analysis are necessary for almost all analysis of
simulation results.

This chapter starts (Sec.\ \ref{sec:introProb})
  by an introduction to randomness and statistics.
In Sec.\ \ref{sec:random} 
  the generation of pseudo random numbers according to some given
probability distribution is explained. 
Basic analysis of data, i.e., the calculation of mean, variance, histograms and
corresponding error bars, is covered in Sec.\ \ref{sec:statistics}.
Next, in Sec.\
\ref{sec:plotting}, it is shown how data can be represented graphically
using suitable plotting tools, {\em gnuplot} and {\em xmgrace}.
Hypothesis testing and how to measure or ensure independence of
data is treated in Sec.\ \ref{sec:hypothesis}.
How to fit data to functions is explained in Sec.\ \ref{sec:generalEstimate}.
In the concluding section, a special technique is outlined
which allows to cope with
the limitations of simulations due to finite system sizes.

Note that some examples 
are again presented using the C programming language. 
Nevertheless, there exist very powerful 
freely available programs
like R \cite{R}, where many analysis (and plotting) tools are available
as additional packages.

\section{Introduction to probability\label{sec:introProb}}

\index{probability|(}
Here, a short introduction to concepts of probability and
randomness is given. The presentation here should be concise concerning
the subjects presented in this book. Nevertheless,
more details, in particular proofs, examples and exercises,
 can be found in standard textbooks \cite{dekking2005,lefebvre2006}.
Here often a sloppy mathematical notation is used for brevity, e.g.\
instead of writing ``a function $g:X\to Y,\, y=g(x)$'', we often write simply
``a function $g(x)$''. 

A {\em random experiment}\index{random experiment}
 is an experiment which is truly random
(like radioactive decay or quantum mechanical processes)
or at least unpredictable (like tossing a coin or predicting
the position of a certain gas atom inside a container
which holds a hot dense gas).

\begin{definition}
The {\em sample space} $\Omega$ is a set of all possible outcomes
of a random experiment.\index{sample!space|ii}
\end{definition}

For the coin example, the sample space is $\Omega=\{$head, tail$\}$.
Note that a sample space can be in principle infinite,
like the possible $x$ positions of an atom in a container. 
With infinite precision of measurement we have
$\Omega^{(x)}=[0,L_x]$,
where the container shall be a box with linear extents
$L_x$ ($L_y,L_z$ in the other directions, see below).

For a random experiment, one wants to know the probability that
certain events occur. Note that for the position of an atom in a box,
the probability to find the atom {\em precisely at} some $x$-coordinate 
${x}\in \Omega^{(x)}$ is zero if one assumes that measurements 
result in real numbers  with infinite precision. For this reason, one considers
probabilities $\Prob(A)$ of subsets $A\subset \Omega$ 
(in other words $A\in 2^\Omega$, $2^\Omega$ being the 
{\em power set}\index{power set} which is the set of all subsets of $\Omega$). 
Such a subset is called an {\em event}.\index{event}
Therefore $\Prob(A)$ is the probability that the outcome of a random 
experiment is inside $A$, i.e.\ one of the elements of $A$. More
formally:

\begin{definition}
A {\em probability function}\index{probability!function}
 $P$ is a function $P:2^\Omega \longrightarrow [0,1]$ with
 \begin{equation}\Prob(\Omega)=1
\label{eq:normalization}
\end{equation} 
and for each finite or infinite
 sequence $A_1,A_2,A_3,\ldots$ of mutual disjoint events 
($A_i \cap A_j = \emptyset$ for $i\neq j$) we have
\begin{equation}
\Prob(A_1 \cup A_2 \cup A_3 \cup \ldots) = \Prob(A_1)+\Prob(A_2)+\Prob(A_3)+\ldots
\label{eq:probII}
\end{equation}
\end{definition}

\noindent
For a fair coin, both sides would appear with the same probability, hence
one has $\Prob(\emptyset)=0$, $\Prob(\{$head$\})=0.5$, $\Prob(\{$tail$\})=0.5$, 
$\Prob(\{$head, tail$\})=1$. For the  hot gas inside the container,
we assume that no external forces act on the
atoms. Then  the atoms are  distributed uniformly.
 Thus, when measuring the $x$ position  of an atom,
the probability to find it inside the region 
$A=[x,x+\Delta x] \subset \Omega^{(x)}$ is
 $\Prob(A)=\Delta x / L_x $.

The usual set operations applies to events. The
{\em intersection}\index{intersection}
 $A\cap B$ of two events is the event which contains elements that are
both in $A$ and $B$. Hence $\Prob(A\cap B)$ is the probability that
the outcome of an experiment is contained in both events $A$ and $B$.
The {\em complement}\index{complement} 
 $A^c$ of a set is the set of all elements of $\Omega$ which are not in $A$.
Since $A^c$, $A$
are disjoint and $A\cup A^c=\Omega$, we get from Eq. (\ref{eq:probII}):
\begin{equation}
\Prob(A^c)=1-\Prob(A)\,.\label{eq:complement}
\end{equation}

Furthermore, one can show for two events $A,B\subset \Omega$:
\begin{equation}
\Prob(A\cup B)=\Prob(A)+\Prob(B)-\Prob(A\cap B)
\end{equation}
{\small \begin{proof}
$\Prob(A)=\Prob(A\cap \Omega)$$=\Prob(A\cap(B\cup B^c))$$=\Prob((A\cap B)\cup(A\cap B^c))$
$\stackrel{(\ref{eq:probII})}{=}\Prob(A\cap B)+\Prob(A\cap B^c)$. If we apply
this for $A\cup B$ instead of $A$, we get 
$\Prob(A\cup B)= \Prob((A\cup B)\cap B)+\Prob((A\cup B)\cap B^c))$ 
$=\Prob(B)+\Prob(A\cap B^c)$. Eliminating $\Prob(A\cap B^c)$ from these two equations
gives the desired result.
\end{proof}}

\noindent
Note that Eqs. (\ref{eq:probII}) and (\ref{eq:complement}) are special cases
of this equation.

If a random experiment is repeated several times, the possible
outcomes of the repeated experiment 
are tuples of outcomes of single experiments. Thus, if you throw the coin
twice, the possible outcomes are (head,head), (head,tail),
(tail,head), and (tail,tail). This means the sample space\index{sample!space}
 is a power of the single-experiment sample spaces. In general,
it is also possible to
combine different random experiments into one. 
Hence, for the general case, if
$k$ experiments with sample spaces $\Omega^{(1)},\Omega^{(2)},\ldots,
\Omega^{(k)}$
are considered, the sample space of the combined experiment is 
$\Omega=\Omega^{(1)} \times \Omega^{(2)} \times \ldots \times\Omega^{(k)}$. 
For example,
one can describe the measurement of the position of the atom in the
hot gas as a combination of the three independent random 
experiments of measuring the $x$, $y$, and $z$ coordinates, respectively.

If we assume that the different experiments are performed 
{\em independently},
then the total probability of an event for a combined random experiment
is the 
product of the single-experiment probabilities: $\Prob(A^{(1)},A^{(2)},\ldots,
A^{(k)}) = \Prob(A^{(1)}) \Prob(A^{(2)})\ldots \Prob(A^{(k)})$.

\begin{sloppypar}
For tossing the fair coin twice,
the probability of the outcome (head,tail) is
$\Prob(\{($head,head$)\})=\Prob(\{$head$\})\Prob(\{$head$\})=0.5\cdot 0.5=0.25$.
Similarly, for the experiment where all three coordinates of
 an atom inside the container are measured, one can write
 $\Prob([x,x+\Delta x]\times[y,y+\Delta y]\times [z,z+\Delta z])=
\Prob([x,x+\Delta x]) \Prob([y,y+\Delta y])\Prob([z,z+\Delta z])
= (\Delta x/L_x) (\Delta y/L_y)  (\Delta z / L_z)$
$= \Delta x \Delta y \Delta z / (L_x L_y L_z)$.
\end{sloppypar}

Often one wants to calculate probabilities which are restricted
to special events $C$ among all events, hence relative or
{\em conditioned} to $C$. For any other event $A$ we have 
$\Prob(C)$$=\Prob((A\cup A^c)\cap C)$$=\Prob(A\cap C)+\Prob(A^c\cap C)$,
which means $\frac{\Prob(A \cap C)}{\Prob(C)}
+ \frac{\Prob(A^c\cap C)}{\Prob(C)} =1$. Since $\Prob(A\cap C)$ is the probability of
an outcome in $A$ and $C$ and because $\Prob(C)$ is the probability
of an outcome in $C$, the fraction $\frac{\Prob(A \cap C)}{\Prob(C)}$ gives
the probability of an outcome $A$ and $C$ relative to $C$, i.e.
the probability of event $A$ given $C$, leading to the following 

\begin{definition}
The {\em probability of  $A$ under the condition $C$}
\index{conditional probability}\index{probability!conditional} is
\begin{equation}
\Prob(A|C) = \frac{\Prob(A \cap C)}{\Prob(C)}\,.
\label{eq:condProb}
\end{equation}
\end{definition}
\noindent
As we have seen, we have the natural normalization $\Prob(A|C)+\Prob(A^c|C)=1$.
Rewriting Eq.\ (\ref{eq:condProb}) one obtains $\Prob(A|C)\Prob(C)=\Prob(A \cap C)$.
Therefore,
 the calculation of $\Prob(A \cap C)$ can be decomposed into two parts,
which are sometimes easier to obtain.
By symmetry, we can also write  $\Prob(C|A)\Prob(A)=\Prob(A \cap C)$. Combining this
with Eq.\ (\ref{eq:condProb}), one obtains the famous {\em Bayes' rule}
\index{Bayes' rule}
\begin{equation}
\Prob(C|A)=\frac{\Prob(A|C)\Prob(C)}{\Prob(A)}\,.
\end{equation}
\noindent This means one of the conditional probabilities $\Prob(A|C)$
and $\Prob(C|A)$ can be expressed via the other, which is sometimes useful
if $\Prob(A)$ and $\Prob(C)$ are known. Note that the denominator in the
Bayes' rule is sometimes written as
$\Prob(A)=\Prob(A\cap(C\cup C^c))$$=\Prob(A\cap C)+\Prob(A\cap C^c)$
$=\Prob(A|C)\Prob(C)+\Prob(A|C^c)\Prob(C^c)$.

If an event $A$ is {\em independent}\index{independence}
 of the condition $C$, its conditional
probability should be the same as the unconditional probability, i.e.,
$\Prob(A|C)=\Prob(A)$. Using $\Prob(A \cap C)=\Prob(A|C)\Prob(C)$ we 
get $\Prob(A \cap C)=\Prob(A)\Prob(C)$,
i.e., the probabilities of independent events have to be
multiplied. This was used already above for random experiments, which are
conducted as independent subexperiments.

\index{probability|)}

\index{random variable|(}

So far, the outcomes of the random experiments can be anything like
the sides of coins, sides of a dice, colors of the eyes of randomly chosen
people or states of random systems. In mathematics, it is often easier to
handle numbers instead of arbitrary objects. For this reason one
can represent the outcomes of random experiments by numbers which
are assigned via special functions:

\begin{definition}
For a sample space $\Omega$, 
a {\em random variable} is
a function $X:\Omega \longrightarrow \mathbb{R}$.
\end{definition}
For example,
 one could use $X($head$)$=1 and $X($tail$)=0$. Hence, if one repeats
the experiments $k$ times independently, one would obtain the number
of heads by $\sum_{i=1}^k X(\omega^{(i)})$, where $\omega^{(i)}$ is the
outcome of the $i$'th experiment.

If one is interested only in the values of the random variable,
the connection to the original sample space $\Omega$ is not important
anymore. Consequently, one can consider random variables $X$ as devices,
which output a random number $x$ each time a random experiment is
performed. Note that random
variables are usually denoted by upper-case letters, while the actual
outcomes of random experiments are denoted by lower-case letters.

Using the concept of random variables, one deals only
with numbers as outcomes of random experiments. This enables
many tools from mathematics to be applied. In particular, one
can  combine random variables and functions to obtain new
random variables. This means, in the simplest case, the following:
First, one  performs a random experiment,  yielding a
random outcome $x$. Next, for a given function $g$, $y=g(x)$ 
is calculated. Then, $y$ is the
final outcome of the random experiment. This is called 
a\label{page:transformation}
{\em transformation}\index{random variable!transformation|ii}
 $Y=g(X)$ of the random variable $X$. More generally, 
one can also define a 
random variable $Y$  
by combining {\em several} random variables $X^{(1)}, X^{(2)}$,
\ldots, $X^{(k)}$ via a function $\tilde g$ such that 
\begin{equation}
Y=\tilde g\left(X^{(1)}, X^{(2)},\ldots, X^{(k)}\right)\,.
\end{equation}
 In practice,
one would perform random experiments for the random variables
$X^{(1)}, X^{(2)}$,
\ldots, $X^{(k)}$, resulting in outcomes $x^{(1)}, x^{(2)}$,
\ldots, $x^{(k)}$. The final number is obtained by
calculating $y=\tilde g(x^{(1)}, x^{(2)},\ldots, x^{(k)})$.
A simple but the most important case is the linear
combination of random variables $Y=$ 
$\alpha_1 X^{(1)}+ \alpha_2 X^{(2)}+$ \ldots $+\alpha_k X^{(k)}$,
which will be used below. For all examples  considered here,
the random variables $X^{(1)}, X^{(2)}$,
\ldots, $X^{(k)}$ have the same properties, 
which means that the same random experiment is repeated $k$ times.
Nevertheless, the most general
description which allows for different random variables will be used here.

The behavior of a random
variable is fully described by the probabilities of obtaining
outcomes smaller or equal to a given parameter $x$:

\begin{definition}
\label{def:distributionF}
The {\em distribution function}\index{distribution function}
 of a random variable $X$ is a
function $F_X: \mathbb{R}\longrightarrow [0,1]$ defined
via
\begin{equation}
F_X(x) = \Prob(X\le x)
\end{equation}
\end{definition}
\noindent
The index $X$ is omitted if no confusion arises. 
Sometimes the distribution function is also named
{\em cumulative} distribution 
function.\index{cumulative distribution function|see{distribution function}}
One also says, the distribution function
defines a {\em probability distribution}.\index{probability!distribution}
Stating 
a random variable or stating the distribution function are
fully equivalent methods to describe a random experiment.

\begin{figure}[!ht]
\includegraphics[width=0.9\textwidth]{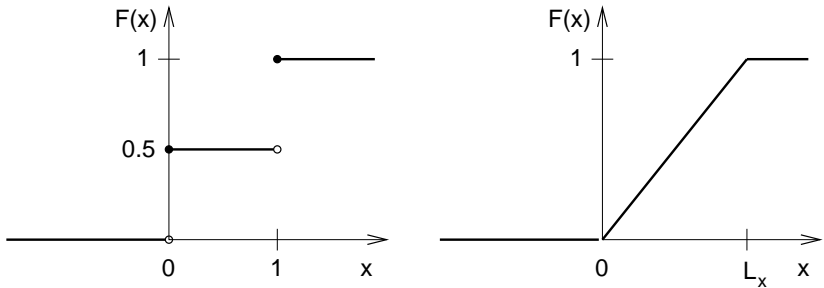}
\caption{Distribution function of the random variable for a 
fair coin (left) and for the
random $x$ position of a gas atom inside a container of length $L_x$.
\label{fig:CoinBox}}
\end{figure}

For the fair coin, we have, see left of Fig.\ \ref{fig:CoinBox}
\begin{equation}
F(x)=\begin{cases}
0 & x < 0\\
0.5 & 0\le x < 1\\
1 & x\ge 1
     \end{cases}\,.
\label{eq:F:coin}
\end{equation}

For measuring the $x$ position of an atom in the uniformly distributed gas
we obtain, see right of Fig.\ \ref{fig:CoinBox}
\begin{equation}
F(x)=\begin{cases}
0 & x < 0\\
x/L_x & 0 \le  x < L_x\\
1 & x\ge L_x
\end{cases}\,.
\label{eq:F:box}
\end{equation}

Since the outcomes of any random variable are finite,  
there are {\em no} possible outcomes $X\le x$ 
in the limit $x\to-\infty$. Also, 
 {\em all} possible outcomes fulfill $X\le x$ for 
$x\to\infty$. Consequently,
 one obtains for all random variables $\lim_{x\to-\infty} F(x)=0$ and
$\lim_{x\to+\infty} F(x)=1$. Furthermore, from Def.\ \ref{def:distributionF},
one obtains immediately:
\begin{equation}
\Prob(x_0 < X\le x_1) = F_X(x_1) - F_X(x_0)\,
\label{eq:probDistr}
\end{equation}
Therefore, one can calculate the probability to obtain a random number 
for any arbitrary interval, hence also for unions of intervals.

The distribution function, although it contains all information,
 is sometimes less convenient to handle, because
it gives information about cumulative probabilities. It is more
obvious to describe the outcomes of the random experiments directly.
For this purpose, we have to distinguish between
{\em discrete} random variables\index{random variable!discrete},
where the number of possible outcomes is denumerable or even finite,
and {\em continuous} random variables\index{random variable!continuous},
where the possible outcomes are non-denumerable. The random variable
describing the coin is discrete, while the position of an atom inside
a container is continuous.

\subsection{Discrete random variables}
\label{sec:discreteRandom}

\index{random variable!discrete|(ii}
\index{discrete random variable|(ii}
We first concentrate on discrete random variables. Here, an alternative but
equivalent description to the distribution function is to state
the probability for each possible outcome directly:

\begin{definition}
For a discrete random variable $X$, 
the {\em probability mass function}\index{probability!mass function} 
(pmf)\index{pmf|see{probability mass function}}
$p_X: \mathbb{R} \to [0,1]$ is given by
\begin{equation}
p_X(x) = \Prob(X=x)\,.
\end{equation}
\end{definition}
\noindent
Again, the index $X$ is omitted if no confusion arises. Since a discrete
random variable describes only a denumerable number of outcomes, the
probability mass function is zero almost everywhere. In the following, the
outcomes $x$ where $p_X(x)>0$ are denoted by $\tilde x_i$. Since probabilities
must sum up to one, see Eq.\ \ref{eq:normalization}, one
obtains $\sum_i p_X(\tilde x_i)=1$. Sometimes we also write 
$p_i=p_X(\tilde x_i)$. The distribution funcion $F_X(x)$
is obtained from the pmf  via summing up
all probabilities of outcomes smaller or equal to $x$:
\begin{equation}
F_X(x) = \sum_{\tilde x_i\le x} p_X(\tilde x_i)
\end{equation}

For example, the pmf of the random variable arising from the fair coin 
Eq.\ (\ref{eq:F:coin}) is given by  $p(0)=0.5$ and $p(1)=0.5$ ($p(x)=0$
elsewhere). The generalization to a possibly unfair coin,
where the outcome ``1'' arises with probability $p$,  leads to:

\begin{definition}
The {\em Bernoulli distribution with 
parameter $p$}\index{Bernoulli distribution}\index{distribution!Bernoulli} 
($0<p\le 1$) describes a discrete random variable $X$
with the following probability mass function
\begin{equation}
p_X(1)=p,\quad p_X(0)=1-p\,.
\end{equation}
\end{definition}
\noindent
Performing a Bernoulli experiment means that one throws a generalized
coin and records either ``0'' or ``1'' depending on whether one
gets head or tail.

There are a couple of important characteristic quantities describing
the pmf of a random variable. Next, we describe
the most important ones for the discrete case:

\begin{definition}
\label{def:exp:var:d}
\begin{itemize}
\item The {\em expectation value}\index{expectation value} is
\begin{equation}
\mu \equiv \E[X] = \sum_i \tilde x_i \Prob(X=\tilde x_i) = \sum_i \tilde x_i p_X(\tilde x_i)
\end{equation}

\item The {\em variance}\index{variance} is
\begin{equation}
\sigma^2 \equiv \Var[X] = \E[(X-\E[X])^2] = \sum_i (\tilde x_i-\E[X])^2 p_X(\tilde x_i) 
\label{eq:variance}
\end{equation}

\item The {\em standard deviation} \index{standard deviation}
\begin{equation}
\sigma \equiv \sqrt{\Var[X]}
\end{equation}

\end{itemize}
\end{definition}
\noindent The expectation value describes the ``average'' one would
typically obtain if the random experiment is repeated very often.
The variance is a measure for the spread of the different outcomes
of random variable.
As example, the Bernoulli distribution\index{Bernoulli distribution}%
\index{distribution!Bernoulli}  exhibits 
\begin{eqnarray}
 \E[X] & = & 0p(0)+1p(1)=p\\
 \Var[X] & = & (0-p)^2p(0)+(1-p)^2p(1)\nonumber\\
&= &p^2(1-p)+(1-p)^2p=p(1-p) \label{eq:varBernoulli}
\end{eqnarray}
One can calculate expectation values of functions $g(x)$ of random variables
$X$ via $\E[g(X)]=\sum_i g(\tilde x_i) p_X(\tilde x_i)$. 
For the calculation here, we only need that
the calculation of the
expectation value is a linear operation. Hence, for numbers $\alpha_1,\alpha_2$
and, in general, two random variables
$X_1,X_2$ one has
\begin{equation} 
E[\alpha_1 X_1+\alpha_2 X_2]=
\alpha_1 \E[X_1]+\alpha_2\E[X_2]\,.
\label{eq:expectLin}
\end{equation}

In this way, realizing that $\E[X]$ is a number,  one obtains:
\begin{eqnarray}
\sigma^2=\Var(X) &= & 
\E[(X-\E[X])^2] = \E[X^2]-2\E[X\,\E[X]]+\E[\E[X]^2] \nonumber\\
         & = & \E[X^2]-\E[X]^2 = \E[X^2]-\mu^2
\label{eq:Var:Prop} \\
\Leftrightarrow \quad \E[X^2] & =& \sigma^2+\mu^2 \label{eq:2ndMoment}
\end{eqnarray}
The variance is {\em not linear}, which can be seen when looking
at a linear combination of two  {\em independent} random variables $X_1,X_2$ 
(implying $\E[X_1X_2]=\E[X_1] \E[X_2]$ 
($\star$))
\begin{eqnarray}
\sigma^2_{\alpha_1 X_1+\alpha_2 X_2} & = & 
\Var[\alpha_1 X_1+\alpha_2 X_2] \nonumber \\
& \stackrel{(\ref{eq:Var:Prop})}{=} & 
\E[(\alpha_1 X_2+\alpha_2 X_2)^2 ] 
 -\E[\alpha_1 X_1+\alpha_2 X_2 ]^2 \nonumber \\
& \stackrel{(\ref{eq:expectLin})}{=} &
\E[\alpha_1^2 X_1^2 + 2\alpha_1\alpha_2 X_1 X_2 + \alpha_2^2X_2^2 ] 
\nonumber \\
& & - (\alpha_1 \E[X_1] + \alpha_2 \E[X_2] )^2 \nonumber \\
& \stackrel{(\ref{eq:expectLin}),(\star)}{=} &
\alpha_1^2 \E[X_1^2] + \alpha_2^2 \E[X_2^2]
- \alpha_1^2 \E[X_1]^2 + \alpha_2^2 \E[X_2]^2 \nonumber \\
& \stackrel{(\ref{eq:Var:Prop})}{=} 
& \alpha_1^2 \Var[X_1] + \alpha_2^2 \Var[X_2]
\label{eq:varLin}
\end{eqnarray}

The expectation values $E[X^n]$ are called the {\em $n$'th
moments}\index{moment!$n$'th} of the distribution. This means that
the expectation value is the first moment
and the variance can be calculated from the first and second
moments.

Next, we describe two more important distributions of discrete 
random variables. First, if one repeats a Bernoulli experiment $n$ times,
one can measure how often the result ``1'' was obtained. Formally,
this can be written as a sum of $n$ random variables $X^{(i)}$
which are Bernoulli distributed: $X=\sum_{i=1}^n X^{(i)}$ with parameter $p$.
This is a very simple example of a 
transformation\index{random variable!transformation} of
a random variable, see page \pageref{page:transformation}. 
In particular, the transformation is linear.
The probability to obtain $x$ times the result ``1'' is calculated as follows:
The probability to obtain exactly $x$ times a ``1'' is $p^x$, the
other $n-x$ experiments yield ``0'' which happens with probability 
$(1-p)^{n-x}$. Furthermore, there are ${n \choose x}=n!/(x!(n-x)!)$
different sequences with $x$ times ``1'' and $n-x$ times ``0''. Hence,
one obtains:

\begin{definition}
The {\em binomial distribution}\index{Binomial distribution}%
\index{distribution!binomial} with parameters $n\in\mathbb{N}$ and $p$ 
($0<p\le 1$) describes a random variable $X$  which has
the pmf
\begin{equation}
p_X(x)={n \choose x} p^x (1-p)^{n-k} \quad (0\le x \le n)
\end{equation}
A common notation is $X\sim B(n,p)$.
\end{definition}

\begin{figure}[!ht]
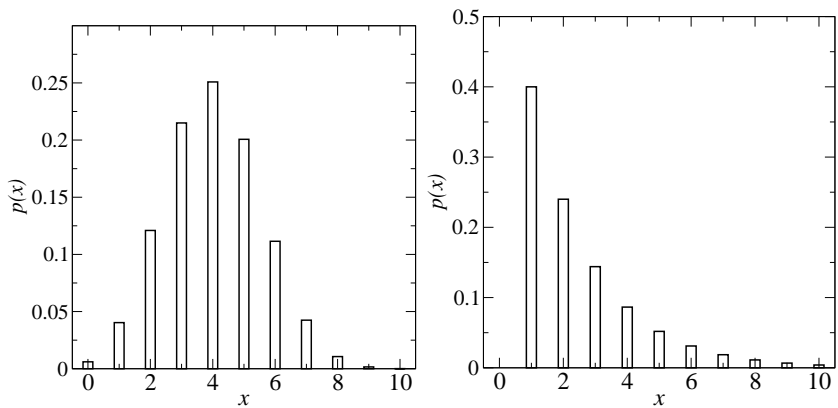

\begin{center}
\includegraphics[width=0.45\textwidth]{pic_random_small/binomial.eps}
\includegraphics[width=0.45\textwidth]{pic_random_small/geometric.eps}
\end{center}
\caption{(Left) Probability mass function of the binomial distribution
for parameters $n=10$ and $p=0.4$. (Right) 
Probability mass function of the geometric distribution
for parameter $p=0.4$.
\label{fig:binomial}}
\end{figure}

Note that the probability mass function is assumed to be zero for
argument values that are not stated.
A sample plot of the distribution for parameters $n=10$ and $p=0.4$
is shown in the left of Fig.\ \ref{fig:binomial}.
The Binomial distribution has expectation value and variance
\begin{eqnarray}
 \E[X] & = & np\\
\Var[X] & = & np(1-p) \label{eq:BinomialVar}
\end{eqnarray}
 (without proof here). The
distribution function cannot be calculated analytically in closed form.

\label{page:binomial}

In the limit of a large number  of experiments ($n\to\infty$), 
constrained such that the expectation value $\mu=np$
is kept fixed, the pmf of a Binomial distribution is well
approximated by the pmf of the {\em Poisson distribution}, which is defined
as follows:
\begin{definition}
The {\em Poisson distribution}\index{Poisson distribution|ii}%
\index{distribution!Poisson|ii} with parameter $\mu>0$ describes a random
variable $X$ with pmf 
\begin{equation}
p_X(x)=\frac{\mu^x}{x!}e^{-\mu}
\label{eq:poissonA}
\end{equation}
\end{definition}
Indeed, as required, the probabilities sum up to 1, since
$\sum_i \frac{\mu^x}{x!}$ is the Taylor series of $e^\mu$.
The Poisson distribution exhibits $\E[X]=\mu$ and $\Var[X]=\mu$. Again,
a closed form for the distribution function is not known.

Furthermore, one could repeat a Bernoulli experiment just
until the first time a ``1'' is observed,
without limit for the number of trials. If a ``1''
is observed  for the first time after
exactly $x$ times, then the first $x-1$ times 
the outcome ``0'' was observed. This happens with probability $(1-p)^{x-1}$.
At the $x$'th experiment, the outcome ``1'' is observed which has
the probability $p$. Therefore one obtains

\begin{definition}
The {\em geometric distribution} with parameter $p$ ($0<p\le 1$) describes a
random variable $X$ which
has the pmf
\begin{equation}
p_X(x)=(1-p)^{x-1}p \quad (x\in\mathbb{N})
\end{equation}
\end{definition}
A sample plot of the pmf
(up to $x=10$) is shown in the right of Fig.\ \ref{fig:binomial}.
The geometric distribution has (without proof here)
the expectation value $E[X]=1/p$,
the variance $\Var[X]=(1-p)/p^2$ and the following distribution
function:
$$
F_X(x)=\begin{cases}
0 & x<1\\
1-(1-p)^m & m\le x <m+1\quad (m\in\mathbb{N})
\end{cases}
$$
\index{random variable!discrete|)}
\index{discrete random variable|)}

\subsection{Continuous random variables}

\index{random variable!continuous|(ii}
As stated above, random variables 
are called continuous if they describe random experiments
where outcomes from a subset of the real numbers can be obtained. 
One may describe such random variables also using the 
distribution function, see Def.\ \ref{def:distributionF}. For continuous
random variables,
an alternative description 
is possible, equivalent to the pmf for
discrete random variables:
The probability density function
states the probability to obtain a certain number per unit:

\begin{definition}
For a continuous random variable $X$ with a continuous distribution
function $F_X$,
the {\em probability density function}\index{probability!density function|ii}
(pdf)\index{pdf} $p_X: \mathbb{R} \to [0,1]$ is given by
\begin{equation}
p_X(x) = \fd{F_X(x)}{x}
\end{equation}
\end{definition}

Consequently, one obtains, using the definition of a derivative and using
Eq.\ (\ref{eq:probDistr})
\begin{eqnarray}
F_X(x) & = & \int_{-\infty}^{x}d\tilde x\,p_X(\tilde x)\\
\Prob(x_0 < X\le x_1) & = & \int_{x_0}^{x_1}d\tilde x\,p_X(\tilde x)
\end{eqnarray}

Below some examples for important continuous random variables are presented.
First, we extend the definitions Def.\ \ref{def:exp:var:c}
of expectation value and variance to the continuous case:

\begin{definition}
\label{def:exp:var:c}
\begin{itemize}
\item The {\em expectation value}\index{expectation value} is
\begin{equation}
\E[X] = \int_{-\infty}^{\infty} dx\, x\, p_X(x)
\end{equation}

\item The {\em variance}\index{variance} is
\begin{equation}
\Var[X] = \E[(X-\E[X])^2] = \int_{\infty}^{-\infty} (x-\E[X])^2 p_X(x) 
\end{equation}

\end{itemize}
\end{definition}

Expectation value and variance have the same properties as for the
discrete case, i.e., Eqs.\ (\ref{eq:expectLin}), (\ref{eq:Var:Prop}),  and
(\ref{eq:varLin}) hold as well.
Also the definition of the n'th moment\index{moment!$n$'th} 
of a continuous distribution is the same.

Another quantity of interest is the {\em median}, which describes the central
point of the distribution. It is given by  the point
 such that the cumulative probabilities left and right
of this point are both equal to 0.5:
\begin{definition}
\label{def:median}
 The {\em median}\index{median}  $x_{\rm med} = \Med[X]$
is defined via
\begin{equation}
 F(x_{\rm med}) = 0.5 
\end{equation}
\end{definition}

The simplest distribution is the  uniform distribution, where
the probability density function is nonzero and constant
  in some interval $[a,b)$:
\begin{definition}
The {\em uniform distribution},\index{uniform distribution}%
\index{distribution!uniform}
with real-valued parameters $a<b$,  
 describes a random variable $X$  which has
the pdf
\begin{equation}
p_X(x)= \begin{cases}
0 & x < a \\
\frac 1 {b-a} & x\le x < b\\
0 & x \ge 0
\end{cases}
\end{equation}
One writes $X\sim U(a,b)$.
\end{definition}
The distribution function simply rises linearly from zero,
starting at $x=a$, till it reaches 1 at $x=b$, see for example
Eq.\ \ref{eq:F:box} for the case $a=0$ and $b=L_x$.
The uniform distribution exhibits the expectation value 
$\E[X]=(a+b)/2$ and variance $\Var[X]=(b-a)^2/12$. Note that
via the linear transformation $g(X)=(b-a)*X+a$ one obtains $g(X) \sim U(a,b)$
if $X\sim U(0,1)$.
The uniform distribution serves as a basis for the generation of
(pseudo) random numbers in a computer, see Sec.\ \ref{sec:generators}.
All distributions can be in some way obtained via transformations
from one or several uniform distributions, see 
Secs.\ \ref{sec:drawDiscrete}--\ref{sec:gauss}.

\begin{figure}[!ht]
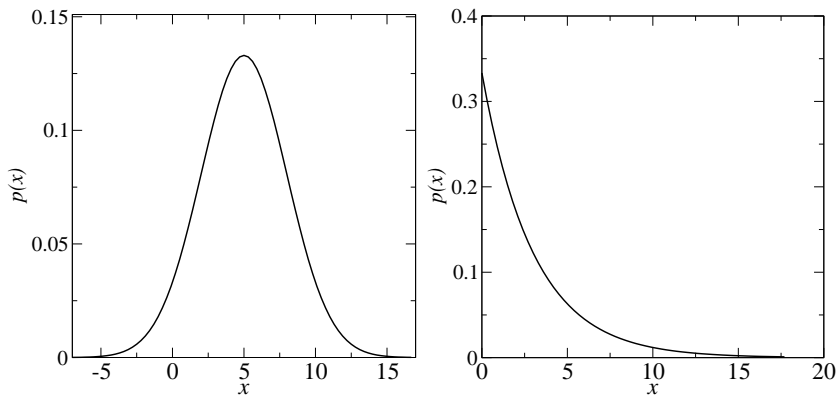

\begin{center}
\includegraphics[width=0.45\textwidth]{pic_random_small/gauss.eps}
\includegraphics[width=0.45\textwidth]{pic_random_small/expo_distr.eps}
\end{center}
\caption{(Left) Probability density function of the Gaussian distribution
for parameters $\mu=5$ and $\sigma=3$. (Right) 
Probability density function of the exponential distribution
for parameter $\mu=3$.
\label{fig:gauss}}
\end{figure}

Probably the most important continuous distribution  in the context of
simulations  is the Gaussian
distribution:

\begin{definition}
The {\em Gaussian distribution},\index{Gaussian distribution}%
\index{distribution!Gaussian} also called 
{\em normal distribution},%
\index{normal distribution|see{Gaussian distribution}}%
\index{distribution!normal|see{Gaussian distribution}}
with real-valued parameters $\mu$ and $\sigma>0$, 
 describes a random variable $X$  which has
the pdf
\begin{equation}
p_X(x)= \frac{1}{\sqrt{2\pi\sigma^2}}
\exp\left( -\frac{(x-\mu)^2}{2\sigma^2}\right)
\label{eq:Gauss}
\end{equation}
One writes $X\sim N(\mu,\sigma^2)$.
\end{definition}
\noindent
The Gaussian distribution has expectation value $\E[X]=\mu$
and variance $\Var[X]=\sigma^2$.
A sample plot of the distribution for parameters $\mu=5$ and $\sigma=3$
is shown in the left of Fig.\ \ref{fig:gauss}. 
The Gaussian distribution for $\mu=0$  and $\sigma=1$ is called
{\em standard normal distribution}\index{standard normal distribution}%
\index{distribution!standard normal} $N(0,1)$. One can obtain any Gaussian
distribution from $X_0\sim N(0,1)$ by applying the 
transformation $g(X_0)=\sigma X_0+\mu$.
Note that the
distribution function for the Gaussian distribution
cannot be calculated analytically. Thus, one
uses usually numerical integration or tabulated values of $N(0,1)$

The {\em central limit theorem}\index{central limit theorem}
describes how the Gaussian distribution arises from a sum
of random variables:

\begin{theorem}
\label{theo:centralLimit}
Let $X^{(1)}, X^{(2)}$,
\ldots, $X^{(n)}$ be independent random variables, which follow all
the same distribution exhibiting expectation value $\mu$ and
variance $\sigma^2$. Then 
\begin{equation}
X=\sum_{i=1}^n  X^{(i)}
\end{equation}
is in the limit of large $n$ approximately Gaussian distributed with
mean $n\mu$ and variance $n\sigma^2$, i.e.\ $X\sim N(n\mu,n\sigma^2)$.

Equivalently, the suitably normalized sum
\begin{equation}
Z=\frac{\frac{1}{n}\sum_{i=1}^n  X^{(i)}-\mu}{\sigma/\sqrt{n}}
\end{equation}
is approximately standard normal distributed $Z\sim N(0,1)$.
\end{theorem}
For a proof, please refer to standard text books on probability.
Since sums of random processes arise very often in nature, the
Gaussian distribution is ubiquitous. For instance, the movement of a
``large'' particle swimming in a liquid called
{\em Brownian motion}\index{Brownian motion} is described by a Gaussian
distribution.

Another common probability distribution is the exponential
distribution. 
\begin{definition}
The {\em exponential distribution},\index{exponential distribution|ii}%
\index{distribution!exponential|ii}
with real-valued parameter $\mu>0$, 
 describes a random variable $X$  which has
the pdf
\begin{equation}
\label{eq:exponentialDistr}
p_X(x)= \frac{1}{\mu} \exp\left( -x/\mu \right)
\end{equation}
\end{definition}
\noindent
A sample plot of the distribution for parameter $\mu=3$ 
is shown in the right of Fig.\ \ref{fig:gauss}. 
The exponential distribution has expectation value $\E[X]=\mu$
and variance $\Var[X]=\mu^2$. The distribution function
can be obtained analytically and is given by \label{exp:distr}
\begin{equation}
F_X(x)= 1- \exp\left( -x/\mu \right)
\end{equation}

The exponential distribution
 arises under circumstances where  processes happen
with certain {\em rates}, i.e., with a constant probability per time
unit. Very often, waiting queues or the decay of radioactive atoms
are modeled by such random variables.
Then the time duration till the first event (or between
two events if the experiment is repeated several times) 
 follows  Eq.\ (\ref{eq:exponentialDistr}).

\begin{figure}[!ht]
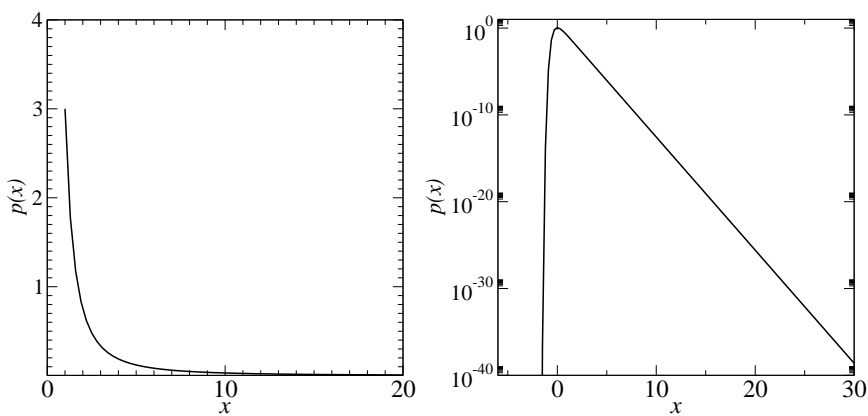

\begin{center}
\includegraphics[width=0.45\textwidth]{pic_random_small/power_law.eps}
\includegraphics[width=0.485\textwidth]{pic_random_small/fisher_tippett.eps}
\end{center}
\caption{(Left) Probability density function of the power-law distribution
for parameters $\gamma=3$ and $\kappa=1$. (Right) 
Probability density function of the Fisher-Tippett distribution
for parameter $\lambda=3$ with logarithmically scaled $y$-axis.
\label{fig:powerLaw}}
\end{figure}

Next, we discuss a distribution, which has attracted recently
\cite{newman2003,newman2006}
much attention in various disciplines like sociology, physics and
computer science. Its probability distribution is a power law:

\begin{definition}
The {\em power-law distribution},\index{power-law distribution}%
\index{distribution!power-law} also called 
{\em Pareto distribution},%
\index{Pareto distribution|see{power-law distribution}}%
\index{distribution!Pareto|see{power-law distribution}}
with real-valued parameters $\gamma>0$ and $\kappa>0$, 
 describes a random variable $X$  which has
the pdf
\begin{equation}
\label{eq:powerLawDistr}
p_X(x)= \begin{cases} 
0 & x<1 \\
\frac \gamma \kappa  (x/\kappa)^{-\gamma+1} & x\ge 1
\end{cases}
\end{equation}
\end{definition}
\noindent
A sample power-law distribution is shown in Fig.\ \ref{fig:powerLaw}.
When plotting a power-law distribution with double-logarithmic scale,
one sees just a straight line.

A discretized version of the
 power-law distribution appears  for example in empirical social networks.
The probability that a person  has $x$ ``close friends'' follows
a power-law distribution. The same is observed for  computer
networks for the probability that a computer is connected to $x$ other
computers. The power-law distribution has a finite expectation value
only if $\gamma>1$, i.e.\ if it falls off quickly enough. In that
case one obtains $\E[X]=\gamma\kappa/(\gamma-1)$. Similarly,
it exhibits a finite variance only for $\gamma>2$: 
$\Var[X]=\frac{\kappa^2\gamma}{(\gamma-1)^2(\gamma-2)}$. The
distribution function can be calculated analytically:
\begin{equation}
F_X(x)=1- (x/\kappa)^{-\gamma} \quad (x\ge 1)
\end{equation}

In the context of extreme-value statistics, the 
Fisher-Tippett distribution 
 (also called log-Weibull distribution) plays an important role.

\begin{definition}
The {\em Fisher-Tippett distribution}, \index{Fisher-Tippett distribution|ii}%
\index{distribution!Fisher-Tippett|ii}
with real-valued parameters $\lambda>0, x_0$, 
 describes a random variable $X$  which has
the pdf
\begin{equation}
\label{eq:FiherTippet}
p_X(x)= \lambda e^{-\lambda x} e^{-e^{-\lambda x}}
\end{equation}
In the special case of $\lambda=1$, the Fisher-Tippett distribution
is also called {\em Gumbel} distribution. \index{Gumbel distribution}%
\index{distribution!Gumbel}
\end{definition}
\noindent
A sample Fisher-Tippett distribution is shown in the right part of 
Fig.\ \ref{fig:powerLaw}.
The function exhibits a maximum at $x=0$. This can be shifted to any
value $\mu$ by replacing $x$ by $x-\mu$. The expectation value is 
$\E[X]=\nu/\lambda$, where $\nu \equiv 0.57721\ldots$ is
the {\em Euler-Mascheroni constant}.\index{Euler-Mascheroni constant}
The distribution exhibits a variance of $\Var[X]=\frac{\pi}{\sqrt{6}\lambda}$.
Also, the distribution function is known analytically:
\begin{equation}
F_X(x)=e^{-e^{-\lambda x}}
\end{equation}

\begin{sloppypar}
Mathematically, one can obtain  a 
Gumbel ($\lambda=1$) distributed random variable
from $n$ standard normal $N(0,1)$ distributed variables 
$X^{(i)}$ by taking
the maximum of them and performing the limit $n\to\infty$, i.e.
$X=\lim_{n\to\infty} \max\left\{X^{(1)}, X^{(2)},
\ldots, X^{(n)}\right\}$. This is also true for some other
``well-behaved'' random variables like exponential distributed ones,
if they are normalized such that they have zero mean and variance one.
The Fisher-Tippett distribution can be obtained from the Gumbel
distribution via a linear transformation.
\end{sloppypar}

For the estimation of confidence intervals (see Secs.\ \ref{sec:confidence}
and \ref{sec:histogram}) one needs the chi-squared distribution and
the $F$ distribution, which are presented next for completeness.

\begin{definition}
The {\em chi-squared distribution},%
\index{chisquareddistribution@chi-squared distribution}%
\index{distribution!chi2@chi-squared}
with $\nu>0$ {\em degrees of freedom} 
 describes a random variable $X$  which has
the probability density function (using the 
{\em Gamma function}\index{Gamma function} $\Gamma(x)=\int_0^\infty t^{x-1}
e^{-t}\,dt$)
\begin{equation}
\label{eq:chi2}
p_X(x)= \frac{1}{2^{\nu/2}\Gamma(\nu/2)}x^{\frac{\nu-2}{2}}e^{-\frac{x}{2}}
\quad (x>0)
\end{equation}
and $p_X(x)=0$ for $x\le 0$.
\end{definition}
\label{page:chi2}\noindent
Distribution function, mean
and variance are not stated here.  A
chi-squared distributed random variable can be obtained
from a sum of $\nu$ squared standard normal distributed random
variables $X_i$: $X=\sum_{i=1}^\nu X_i^2.$
 The chi-squared distribution
is implemented in the {\em GNU scientific library}
\index{GNU scientific library}
(see Sec.\ \ref{sec:gsl}).

\begin{definition}
\label{def:Fdistr}
The {\em F distribution},%
\index{F distribution}%
\index{distribution!F}
with $d_1,d_2>0$ {\em degrees of freedom} 
 describes a random variable $X$  which has
the pdf 
\begin{equation}
\label{eq:F}
p_X(x)= d_1^{d_1/2} d_2^{d_2/2} 
\frac{\Gamma(d_1/2+d_2/2)}{\Gamma(d_1/2)\Gamma(d_2/2)}
\frac{x^{d_1/2-1}}{(d_1x+d_2)^{d_1/2+d_2/2}}
\quad (x>0)
\end{equation}
and $p_X(x)=0$ for $x\le 0$.
\end{definition}
\label{page:F}\noindent
Distribution function, mean
and variance are not stated here. An
F distributed random variable can be obtained
from  a chi-squared distributed random variable $Y_1$ with $d_1$ degrees
of freedom and a chi-squared 
distributed random variable $Y_2$ with $d_2$ degrees
of freedom via $X=\frac{Y_1/d_1}{Y_2/d_2}$.
 The F distribution
is implemented in the {\em GNU scientific library}
\index{GNU scientific library}
(see Sec.\ \ref{sec:gsl}).

Finally, note that also discrete random variables can be described using
probability density functions if one applies the so-called
{\em delta function} $\delta(x-x_0)$. 
For the purpose of computer simulations this
is not necessary. Consequently, no further details are presented here.

\index{random variable!continuous|)}
\index{random variable|)}

\section{Generating (pseudo) random numbers}
\label{sec:random}

For many simulations in science, economy or social sciences, 
random numbers are necessary. Quite
often the model itself exhibits random parameters which remain  fixed 
throughout the simulation; one speaks of {\em quenched
  disorder}\index{quenched disorder}\index{disorder!quenched}. 
A famous example in the field
of condensed matter physics are {\em spin glasses},
which are random alloys of magetic and non-magnetic 
materials\index{spin glass|ii}. In this case, when one performs
simulations of small systems, one has
to perform an average over different disorder realizations to
obtain physical quantities. Each realization of the
disorder consists of randomly chosen positions of the magnetic
and non-magnetic particles. To generate a disorder realization within the
simulations, random numbers are required.

But even when the simulated system is not inherently random, very often
random numbers are required by the algorithms, e.g., to realize a
finite-temperature ensemble or when using randomized algorithms. 
In summary, the application of random numbers in computer simulations
is ubiquitous.

In this section an introduction to the generation of random
numbers is given. First it is explained how they can be generated at all on a
computer. Then, different methods  are presented for obtaining numbers 
which obey a target distribution: the {\em inversion method}\/, the 
{\em rejection method}\/ and {\em Box-M\"uller
method}\/. More comprehensive 
information about these and similar techniques can be found in
Refs.\ \cite{PRA-morgan1984,devroye1986,PRA-numrec1995}. 
In this section it is assumed that you are familiar with the basic
concepts of probability theory and statistics, as presented
in Sec.\ \ref{sec:introProb}.

\subsection{Uniform (pseudo) random numbers\label{sec:generators}}

\index{random number generator|(ii}

First, it should be pointed out that standard computers are
deterministic machines. Thus, it is completely impossible  to generate
true random numbers directly. One could, for example, include
interaction with the user. It is,
for example, possible to measure the time interval between successive
keystrokes, which is randomly distributed by nature. But the
resulting time intervals depend
heavily on the current user which means the statistical properties
cannot be controlled.
 On the other hand, there
are external devices, which have a true random physical process
built in and which can be attached to a computer \cite{quantis,westphal}
or used via the internet \cite{hotbits}.
Nevertheless, since these numbers are really random, they do
not allow to perform stochastic simulations in a controlled and reproducible 
way. This is important in a scientific context, because spectacular
or unexpected results are often tried to be reproduced by other
research groups. Also, some program bugs turn up only
for certain random numbers. Hence, for debugging purposes it is important
to be able to run exactly the same simulation again.
Furthermore, for the true random numbers, 
either the speed of random number generation is limited if the
true random numbers are cheap, or otherwise the generators are expensive.

This is the reason why
{\em pseudo random numbers}\/ are usually taken. They are generated by
deterministic rules. As basis serves a
number generator function {\tt{rand()}} for a uniform distribution.
Each time {\tt{}rand()} is called, a new
(pseudo) random number is returned. (Now the ``pseudo'' is omitted
for convenience)
These random numbers 
should ``look like'' true random numbers
and should have many of the properties of them. One
says they should be ``good''. 
What ``look like'' and ``good'' means, has to be specified:
One would like to have a random number generator
 such that each possible number has indeed the same
probability of occurrence. 
Additionally, if two generated numbers $r_i,r_k$ differ
only slightly, the random numbers $r_{i+1},r_{k+1}$ returned by the
respective subsequent calls should differ sustancially, hence
consecutive numbers should have a low
correlation. There are many ways to specify a correlation, hence
there is no unique criterion. Below, the simplest one will be discussed.

The simplest methods to generate pseudo random numbers are 
{\em linear congruential generators}\/. 
\index{linear congruential generators} They generate a sequence
$x_1,x_2,\ldots$ of integer numbers between 0 and $m-1$ by a recursive
rule:
\begin{equation}
  x_{n+1} = (ax_n+c) \mbox{mod}\, m\,.
\end{equation}

The initial value $x_0$ is called {\em seed}\index{seed}%
\index{random number generator!seed}.
Here we show a simple C implementation \verb!lin_con()!. 
It stores the current number
in the local variable \verb!x! which is declared as \verb!static!, such that
it is remembered, even when the function is terminated (see
Sec.\ \ref{sec:functions}).
There are two arguments. The first

\sources{randomness}{rng.c}
\noindent 
argument \verb!set_seed! indicates
whether one wants to set a seed. If yes, the new seed should be passed
as second argument, otherwise the value of the second argument is ignored. 
The function returns the seed if it is changed, or the new random number.
Note that the constants $a$ and $c$ are defined inside the function,
while the modulus $M$ is implemented via a macro 
\verb!RNG_MODULUS! to make it visible
outside \verb!lin_con()!:
\index{lincon@{\tt lin\_con()}|)}

{\small 
\linenumbers[1]
\begin{verbatim}
#define RNG_MODULUS  32768                  /* modulus */

int lin_con(int set_seed, int seed)
{
  static int x = 1000;        /* current random number */
  const int a = 12351;                   /* multiplier */
  const int c = 1;                            /* shift */

\end{verbatim}
\begin{verbatim}
  if(set_seed)                           /* new seed ? */
    x = seed;
  else                          /* new random number ? */
    x = (a*x+c) % RNG_MODULUS;

  return(x);
}
\end{verbatim}
\nolinenumbers}

If you just want to obtain the next random number, you do not care about the
seed. Hence, we use for convenience 
\verb!rn_lin_con()! to call \verb!lin_con()!
with the first argument being 0:
\index{randlincon@{\tt rand\_lin\_con()}|)}

{\small 
\linenumbers[1]
\begin{verbatim}
int rand_lin_con()
{
  return(lin_con(0,0));
}
\end{verbatim}
\nolinenumbers} 

If we want to set the seed, we also use for convenience a special trivial
function \verb!seed_lin_con()!:
\index{srandlincon@{\tt srand\_lin\_con()}|)}

{\small 
\linenumbers[1]
\begin{verbatim}
void srand_lin_con(int seed)
{
  lin_con(1, seed);
}
\end{verbatim}
\nolinenumbers}

To generate random numbers $r$ distributed in the interval $[0,1)$ one
has to divide the current random number by the modulus $m$. It is desirable to
obtain equally distributed outcomes in the interval, i.e.\ a uniform
distribution. Random numbers generated from this 
 distribution can be used as
input to generate random numbers distributed according to other,
basically arbitrary, distributions.  Below, you will see
how random numbers obeying other distributions can be generated.
 The following simple C function
generates random numbers in $[0,1)$ using the macro \verb!RNG_MODULUS!
defined above:

{\small 
\linenumbers[1]
\begin{verbatim}
double drand_lin_con()
{
  return( (double) lin_con(0,0) / RNG_MODULUS);
}
\end{verbatim}
\nolinenumbers}

One has to
choose the parameters $a,c,m$ in a way that ``good'' random numbers are
obtained, where ``good'' means ``with less correlations''. Note that
in the past several results from simulations have been proven
to be wrong because of the application of bad random number
generators \cite{ferrenberg1992,vattulainen1994}. 

\begin{example}
  To see what ``bad generator'' means, consider as an example the
parameters $a=12351,c=1,m=2^{15}$ and the seed value $I_0=1000$. 
10000 random numbers are generated by dividing each of
them by $m$. They are distributed in the interval $[0,1)$. In
Fig.\ \ref{fig:randomDistr} the distribution of the random numbers is shown.
\begin{figure}[th]
\centerline{\psfig{file=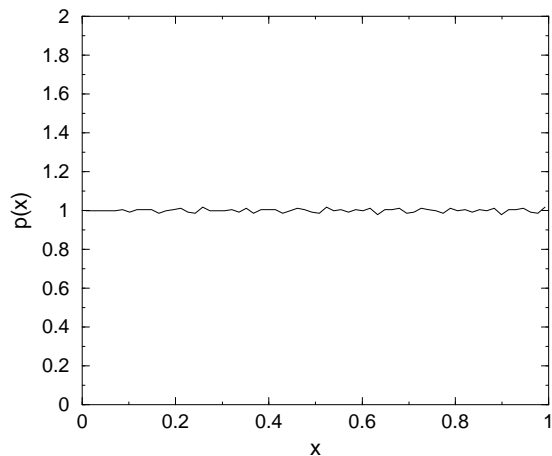,width=0.6\textwidth}}
\caption{Distribution of random numbers in the interval $[0,1)$ obtained
from converting a histogram into a pdf, see Sec.\ \ref{sec:histogram}. The
random numbers  are generated using a linear congruential generator with the
  parameters $a=12351,c=1,m=2^{15}$.
\label{fig:randomDistr}}
\end{figure}

The distribution looks rather flat, but by taking a closer look some
regularities can be observed. These regularities can be studied
 by recording $k$-tuples of $k$ successive random numbers
$(x_i,x_{i+1},\ldots, x_{i+k-1})$. A good random number generator, 
exhibiting no correlations,
would fill up the $k$-dimensional space uniformly. 
Unfortunately, for linear congruential generators, 
instead the points lie on $(k-1)$-dimensional planes. It can be shown
that there are {\em at most}\/ of the order $m^{1/k}$ such planes. A bad
generator has much fewer planes. This is the case for the example
studied above, see top part of Fig.\ \ref{fig:randomCorr}

\begin{figure}[!ht]
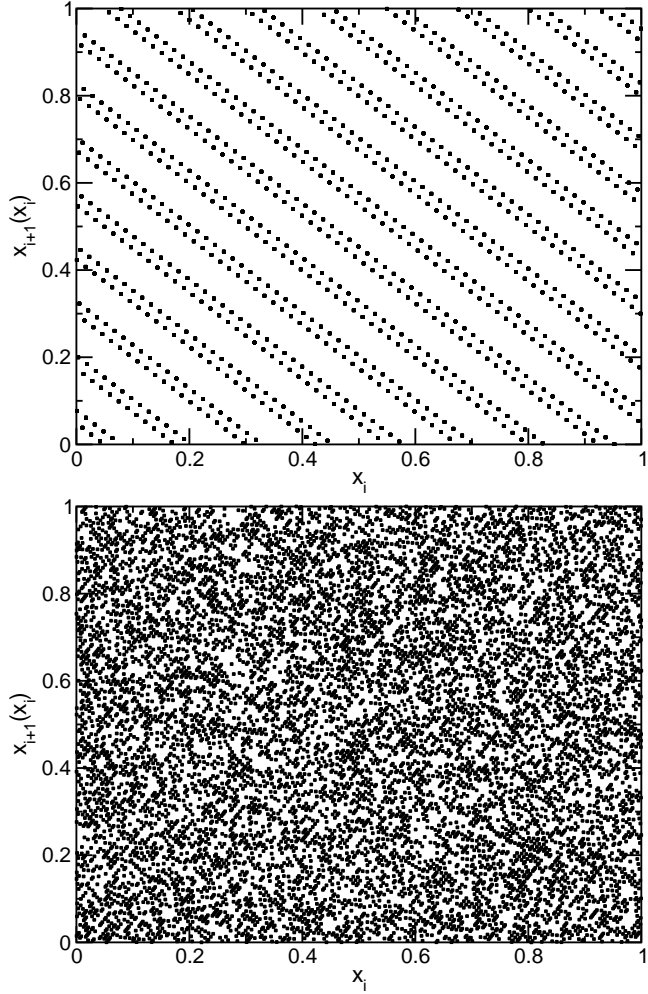

\centerline{\psfig{file=pic_random_small/random_corr51.eps,width=0.7\textwidth}}
\centerline{\psfig{file=pic_random_small/random_corr49.eps,width=0.7\textwidth}}
\caption{Two point correlations $x_{i+1}(x_i)$ 
between successive random numbers $x_i,x_{i+1}$. The top case
  is generated using a linear congruential generator with the
  parameters $a=12351,c=1,m=2^{15}$, the bottom case has instead $a=12349$.
\label{fig:randomCorr}}
\end{figure}

The result for $a=123450$ is even worse: only 15 different ``random''
numbers are generated (with seed 1000), then the iteration reaches a
fixed point (not shown in a figure).

If instead $a=12349$ is chosen, the two-point correlations look like
that shown in the bottom half of 
Fig.\ \ref{fig:randomCorr}. Obviously, the behavior is much more
irregular, but poor correlations may become visible for  higher $k$-tuples.
\end{example}

A generator which has passed several empirical tests is
$a=7^5=16807$, $m=2^{31}-1$, $c=0$. When implementing this 
generator you have to
be careful, because during the calculation numbers are generated which
do not fit into 32 bit. A clever implementation is presented in
Ref. \cite{PRA-numrec1995}. Finally, it should be stressed that this
generator, like all linear congruential generators, has the low-order
bits much less random than the high-order bits. For that reason, when
you want to generate integer numbers in an interval [1,N], you should
use

{\small \begin{verbatim}
 r = 1+(int) (N*x_n/m);

\end{verbatim}}
\vspace*{-5mm}
 
\noindent
 instead of using the modulo operation as with {\tt r=1+(x\_n \% N)}.

In standard C, there is a simple built-in random number generator called
\verb!rand()! (see corresponding documentation), which has
a modulus $m=2^{15}$, which is very poor. On most operating systems,
also \verb!drand48()!\index{drand48@{\tt drand48()}}
 is available, which is based on $m=2^{48}$
($a=$, $c=11$) and  needs also special arithmetics. It 
is already sufficient for simulations which no not need many random numbers
and do not required highest statistical quality. In recent years,
several high-standard random number generators have been developed.
Several very good ones are included in the 
freely availabe {\em GNU scientific library}
\index{GNU scientific library} (see Sec.\ \ref{sec:gsl}). Hence, you do
not have to implement them yourself.

So far, it has been shown how random numbers can be
generated which are distributed uniformly in the interval $[0,1)$. In
general, one is interested in obtaining random numbers which are
distributed according to a given probability distribution with some density
$p(x)$. In the next sections, several techniques suitable for this task
are presented.

\subsection{Discrete random variables\label{sec:drawDiscrete}}

In case of discrete distributions with finite number
of possible outcomes, one can  create
a table of the possible outcomes together with their probabilities $p_X(x_i)$
($i=1,\ldots,i_{\max}$), assuming that the $x_i$ are sorted in ascending order.
To draw a number, one has to draw a  
random number $u$ which is uniformly distributed in $[0,1)$  
and take the entry $j$ of the table such that the
sum $s_j \equiv \sum_{i=1}^j p_X(x_i)$ 
of the probabilities is larger than
$u$, but $s_{j-1}\equiv \sum_{i=1}^{j-1}p_X(_i)<u$. Note that
one can search the array quickly by 
{\em bisection search}:\index{bisection search} The array is 
iteratively divided it into two halves
and each time continued   in that half where the corresponding entry
$j$ is contained. In this way,
generating a random number has a time complexity which grows
only logarithmically with the number $i_{\max}$ of possible outcomes.
This pays off if the number of possible outcomes is very large.

In exercise (\ref{ex:samplingDiscrete}) you are asked to write
a function to sample from the probability distribution of a discrete
variable, in particular for a Poisson distribution.

In the following, we concentrate on techniques for 
generating continuous random variables.

\subsection{Inversion Method\index{inversion method|(}}

Given is a random number generator \verb!drand()! which is assumed to
generate random numbers $U$ which are distributed uniformly in $[0,1)$. The
aim is to generate random numbers $Z$ with probability density
$p_Z(z)$. The  corresponding distribution function is

\begin{equation}
  F_Z(z)\equiv \Prob(Z\le z) \equiv
\int_{-\infty}^z dz^{\prime} p_Z(z^{\prime}) \label{eq:pra:transform}
\end{equation}

The target is to find a function $g(u)$, such that after the 
transformation $Z=g(U)$
the outcomes of $Z$ are distributed according to
(\ref{eq:pra:transform}). It is assumed that $g$ can be inverted and is
strongly monotonically increasing. Then one obtains
\begin{equation}
  F_Z(z)=\Prob(Z\le z)=\Prob(g(U)\le z) = \Prob(U\le g^{-1}(z))
\end{equation}
Since the distribution function $F_U(u)=\Prob(U\le u)$ for a uniformly 
distributed variable is just $F_U(u)=u$ ($u\in[0,1]$), one obtains
$F_Z(z)=g^{-1}(z)$. Thus, one just has to choose $g(z)=F_Z^{-1}(z)$ for the
transformation function in order to obtain random numbers, which are
distributed according to the probability distribution $F_Z(z)$. Of course,
this only works if $F_Z$ can be inverted. If this is not possible,
you may use the methods presented in the subsequent sections, or you
could generate a table of the distribution function, which is in fact
a discretized approximation of the distribution function, 
and use the methods for generating discrete random numbers as shown
in Sec.\ \ref{sec:drawDiscrete}. This can be even refined by using
a linearized approximation of the distribution function. Here,
we do not go into further details, but present an example where 
the distribution function can be indeed inverted.

\begin{example}
\index{exponential distribution} \index{distribution!exponential}
  Let us consider the exponential distribution with parameter $\mu$,
  with distribution function $F_Z(z)=1-\exp(-z/\mu)$,
see page \pageref{exp:distr}. Therefore, one can
obtain exponentially distributed random numbers $Z$ by generating
uniform distributed random numbers $u$ and choosing
$z=-\mu\ln(1-u)$.

\begin{figure}[ht]
\centerline{\psfig{file=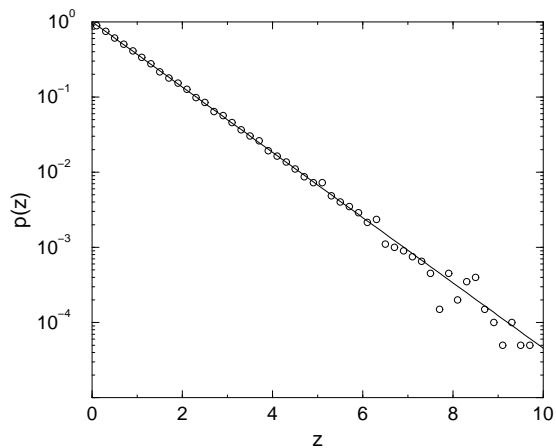,width=0.6\textwidth}}
\caption{Histogram pdf (see page \pageref{page:histoPdf}) of  random numbers
  generated according to an exponential distribution
  ($\mu=1$) compared with the probability density function (straight line) in
  a logarithmic plot.
\label{fig:exponential}}
\end{figure}

\sources{random}{expo.c}
The following simple 
C function generates a random number which is exponentially
distributed. The parameter $\mu$ of the distribution 
is passed as argument.

\newpage
{\small 
\linenumbers[1]
\begin{verbatim}
double rand_expo(double mu)
{
  double randnum;                  /* random number U(0,1) */
  randnum = drand48();

  return(-mu*log(1-randnum));
}
\end{verbatim}
\nolinenumbers} 
\noindent
Note that we use in line 4 the simple \verb!drand48()!
random number generator, which is included in the C standard library
and works well for applications with moderate statistical requirements.
For more sophisticated generates, see e.g.\ the  {\em GNU scientific library}
\index{GNU scientific library} (see Sec.\ \ref{sec:gsl}).

In Fig.\ \ref{fig:exponential} a histogram pdf (see page 
\pageref{page:histoPdf})
for $10^5$ random numbers
generated in this way and the exponential probability function for
$\mu=1$ are shown with a logarithmically scaled $y$-axis. Only for
larger values are deviations  visible. They are due to statistical
fluctuations since $p_Z(z)$ is very small there.

\end{example}
\index{inversion method|)}

\subsection{Rejection Method\label{sec:reject}}
\index{rejection method|(}

As mentioned above, the inversion method works only when the
distribution function $P$ can be inverted analytically. For distributions not
fulfilling this condition, sometimes this problem can
be overcome by drawing several random numbers and combining them in a
clever way.

\begin{figure}[th]
\centerline{\psfig{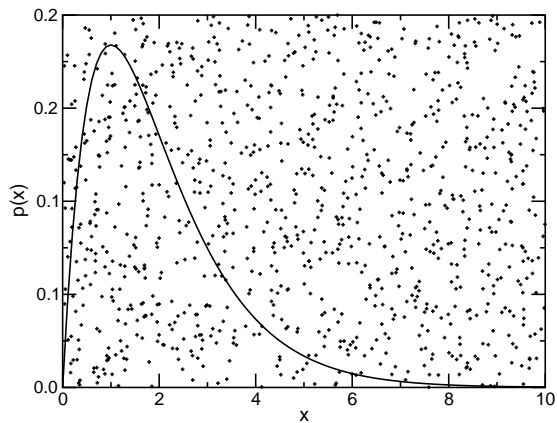}}
\caption{The rejection method: Points ($x,y$) are scattered uniformly over a
  bounded rectangle. The probability that $y\le p(x)$ is proportional
  to $p(x)$.
\label{fig:rejection}}
\end{figure}

The {\em rejection method}\/
 works for random variables where the pdf 
 $p(x)$ fits into a box $[x_0,x_1)\times [0,y_{\max})$,
i.e., $p(x)=0$ for $x\not\in [x_0,x_1]$ and $p(x)\le y_{\max}$. The
basic idea of generating a random number distributed according to $p(x)$ is
to generate random pairs ($x,y$), which are distributed uniformly in
$[x_0,x_1]\times [0,y_{\max}]$ and accept only those numbers $x$ where 
$y\le p(x)$ holds, i.e., the pairs which are located below $p(x)$, see
Fig.\ \ref{fig:rejection}. Therefore, the probability that $x$ is drawn is
proportional to $p(x)$, as desired.

\sources{randomness}{reject.c}
The following C function realizes the rejection method for an arbitrary pdf.
It takes as arguments the boundaries of the box \verb!y_max!, \verb!x0!
and \verb!x1! as well as a pointer \verb!pdf! 
to the function realizing the pdf. For an explanation of function pointers, 
see Sec.\ \ref{sec:pointers}.

{\small 
\linenumbers[1]
\begin{verbatim}
double reject(double y_max, double x0, double x1, 
              double (* pdf)(double))
{
  int found;               /* flag if valid number has been found */
  double x,y;               /* random points in [x0,x1]x[0,p_max] */
  found = 0;
\end{verbatim}
\begin{verbatim}
  while(!found)                 /* loop until number is generated */
  {
    x = x0 + (x1-x0)*drand48();           /* uniformly on [x0,x1] */
    y = y_max *drand48();               /* uniformly in [0,p_max] */
    if(y <= pdf(x))                                   /* accept ? */
      found = 1;
  }
  return(x);
}
\end{verbatim}
\nolinenumbers} 
\noindent
In lines 9--10 the random point, which is uniformly distributed  in the box,
is generated. Lines 11--12 contain the check whether a point below the
pdf curve has been found. The search in the loop (lines 7--13)
continues until a random number has been accepted, 
which is returned in line 14.

\begin{example}
The rejection method is applied to a pdf, which has density 1 in $[0,0.5)$
and rises linearly from 0 to 4 in $[1,1.5)$. Everywhere else it is zero.
This pdf is realized by the following C function:

{\small 
\linenumbers[1]
\begin{verbatim}
double pdf(double x)
{
  if( (x<0)||
      ((x>=0.5)&&(x<1))||
      (x>1.5) )
      return(0.0);
  else if((x>=0)&&(x<0.5))
      return(1.0);
  else
      return(4.0*(x-1));
}
\end{verbatim}
\nolinenumbers} 
\noindent
The resulting empirical histogram pdf is shown in Fig.\ \ref{fig:rejectPdf}.

\begin{figure}[!ht]
\begin{center}
\includegraphics[width=0.6\textwidth]{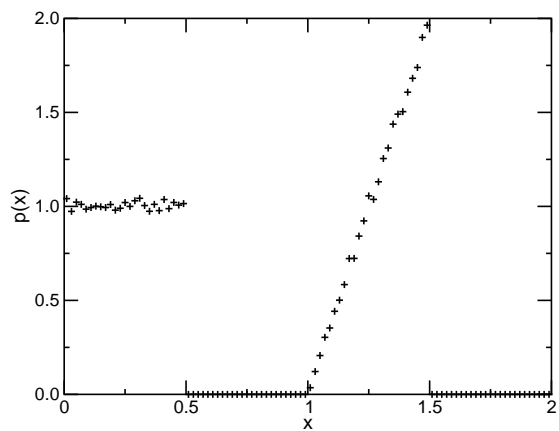}
\end{center}
\caption{Histogram pdf (see page \pageref{page:histoPdf}) of  $10^5$ 
random numbers
  generated using the rejection method for an artificial pdf.
\label{fig:rejectPdf}}
\end{figure}

\end{example}

The rejection method can always be applied  
if the probability density is boxed,
but it has the drawback that more random numbers have to be generated
than can be used: If $A=(x_1-x_0)y_{\max}$ is the area of the box,
one has on average to generate $2A$ auxiliary 
random numbers to obtain one random number of the desired distribution.
If this leads to a very poor efficiency, you can consider to use
several boxes for different parts of the pdf.

\index{rejection method|)}

\subsection{The Gaussian Distribution}
\label{sec:gauss}
\index{Gaussian distribution|(ii}
\index{distribution!Gaussian|(ii}

In case neither the distribution function can be inverted nor the
probability fits into a box, special methods have to be applied. As an
example,  a method for generating random numbers distributed
according to a Gaussian distribution is considered. Other methods and
examples of how different techniques can be combined are collected in
 \cite{PRA-morgan1984}.

\begin{figure}[th]
\centerline{\psfig{file=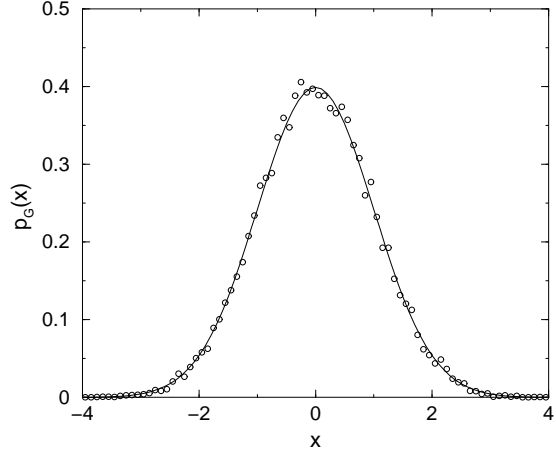,width=0.6\textwidth}}
\caption{Gaussian distribution with zero mean and unit width. The
  circles represent a histogram pdf (see page \pageref{page:histoPdf})
obtained from $10^4$ numbers drawn with
  the Box-M\"uller method.
\label{fig:GaussGenerate}}
\end{figure}

The probability density function
for the Gaussian distribution with mean $\mu$ and variance
$\sigma^2$ is shown in Eq. (\ref{eq:Gauss}), 
see also Fig.\ \ref{fig:GaussGenerate}.
It is, apart from uniform distributions, the most
common distribution occurring in simulations.

Here, the case of a standard Gaussian distribution 
($\mu=0,\;\sigma=1$) 
is considered. If you want to realize the general case, you
have to draw a standard Gaussian distributed number $z$ and then use $\sigma
z+\mu$ which is distributed as desired.

Since the Gaussian distribution extends over an infinite interval and
because the distribution function 
cannot be inverted, the methods from above are not applicable.
The simplest technique to generate random numbers distributed
according to a Gaussian distribution makes use of the central limit
theorem \ref{theo:centralLimit}.\index{central limit theorem}%
 \index{theorem!central limit}
It tells us that any sum of $K$ independently distributed
random variables $U_i$ (with mean $\mu$ and variance $v$) will converge
to a Gaussian distribution with mean $K\mu$ and variance $Kv$. If 
again $U_i$ is taken  to be uniformly distributed in $[0,1)$ (which has
mean $\mu=0.5$ and variance $v=1/12$), one can choose $K=12$ and
the random variable $Z=\sum_{i=1}^{K} U_i-6$ will be distributed approximately
according to a standard Gaussian distribution.
 The drawbacks of this method are that 12 random numbers are
needed to generate one final random number and that numbers
larger than 6 or smaller than -6 will never appear. 

In contrast to this technique the {\em Box-M\"uller method}\/ is
exact. You need two random variables $U_1,U_2$ uniformly distributed  
in $[0,1)$ to generate two independent Gaussian variables
$N_1,N_2$.  This can be achieved by generating $u_1,u_2$ 
from $U_1,U_2$ and assigning
\begin{eqnarray*}
  n_1 & = & \sqrt{-2 \log (1-u_1)} \cos(2\pi u_2) \\
  n_2 & = & \sqrt{-2 \log (1-u_1)} \sin(2\pi u_2) 
\end{eqnarray*}
A proof that $n_1$ and $n_2$ are indeed distributed according to
(\ref{eq:Gauss}) can be found e.g.\ in  \cite{PRA-numrec1995,PRA-morgan1984},
where also other methods for generating Gaussian random numbers, some even
more efficient, are
explained. A method which is based on the simulation of particles in a
box is explained in \cite{PRA-fernandez1999}. 
In Fig.\ \ref{fig:GaussGenerate} a histogram pdf  of $10^4$ random
numbers drawn with the Box-M\"uller method is shown. Note that
you can find an implementation of the Box-M\"uller method
in the solution of Exercise (\ref{ex:variance}).
\index{random number generator|)}
\index{Gaussian distribution|)}
\index{distribution!Gaussian|)}


        




\section{Basic data analysis\label{sec:statistics}}
\index{analyzing data|(}
\index{data!analysis|(}

The starting point is a {\em sample}\index{sample|ii} of
 $n$ measured points $\{x_0,x_1,$ \ldots, $x_{n-1}\}$ 
of some quantity, as obtained from a simulation. Examples
are the density of a gas, the transition time between
two conformations of a molecule, or the price of a stock.
We assume that formally all measurements can be described
by random variables $X_i$ representing 
the same random variable $X$ and that all measurements are statistically
independent of each other (treating statistical dependencies
is treated in Sec.\ \ref{sec:hypothesis}). Usually, one
does not know the underlying probability distribution $F(x)$, having density
$p(x)$, which describes $X$.

\subsection{Estimators\label{sec:estimators}}

Thus, one wants to obtain information about $X$ by looking
at the sample $\{x_0,x_1,$ \ldots, $x_{n-1}\}$. In principle, one does
this by considering {\em estimators}\index{estimator}\label{page:estimator}
$h=h(x_0,x_1, \ldots, x_{n-1})$. Since the measured points are
obtained from random variables, $H=h(X_0,X_1,\ldots,X_{n-1})$ is
a random variable itself. Estimators are often used to estimate
parameters $\theta$ of random variables, e.g.\ moments of distributions.
The most fundamental estimators are:
\begin{itemize}
\item The {\em mean}\index{mean}
\begin{equation}
\overline{x} \equiv \frac{1}{n} \sum_{i=0}^{n-1} x_i
\label{eq:sampleMean}
\end{equation}
\item The {\em sample variance}\index{sample!variance}\index{variance!sample}
\begin{equation}
s^2 \equiv \frac{1}{n}\sum_{i=0}^{n-1} (x_i-\overline{x})^2
\label{eq:sampleVariance}
\end{equation}
The sample standard deviation\index{sample!standard deviation}%
\index{standard deviation!sample} is $s\equiv \sqrt{s^2}$.

\end{itemize}

\sources{randomness}{mean.c} 
As example, next a simple C function is shown, which calculates the
mean of $n$ data points. The function obtains the number $n$ of data
points and an array containing the data as arguments. It returns
the average:

{\small 
\linenumbers[1]
\begin{verbatim}
double mean(int n, double *x)
{
  double sum = 0.0;                       /* sum of values */
  int i;                                        /* counter */
  
  for(i=0; i<n; i++)          /* loop over all data points */
    sum += x[i];
  return(sum/n);
}

\end{verbatim}
\nolinenumbers} 
\noindent
You are asked to write a similar function for calculating
the variance in exercise
(\ref{ex:variance}).

The sample mean can be used to estimate the expectation value
$\mu\equiv \E[X]$ of the distribution. This estimate is 
{\em unbiased}\index{unbiased estimator},
which means that the expectation value of the mean, for any
sample sizes $n$, is indeed the expectation value of the random variable. 
This can be shown
quite easily. Note that formally the random variable
from which the sample mean $\overline{x}$ is drawn is
$\overline{X}=\frac{1}{n} \sum_{i=0}^{n-1} X_i$:

\begin{equation}
\mu_{\overline{X}} \equiv 
\E[\overline{X}] = \E\left[\frac{1}{n} \sum_{i=0}^{n-1} X_i\right]
= \frac{1}{n} \sum_{i=0}^{n-1} \E[X_i] = \frac{1}{n} n \E[X] = \E[X]=\mu   
\label{eq:expectMean}
\end{equation}
\indent Here again the linearity of the
expectation value was used. The fact that the estimator is unbiased 
means that if you repeat the estimation of the expectation value
via the mean several times, on average the correct value is obtained.
This is independent of the sample size.
In general, the estimator $h$ for a parameter $\theta$ is
called unbiased\index{unbiased estimator}\index{estimator!unbiased} 
if $\E[h]=\theta$.

Contrary to what you might expect due to the
symmetry between Eqs.\ (\ref{eq:variance}) and (\ref{eq:sampleVariance}), the
sample variance is {\em not} an unbiased estimator for the
variance $\sigma^2\equiv \Var[X]$ 
of the distribution, but is {\em biased}\label{biased estimator}.
The fundamental reason is, as mentioned above, that $\overline{X}$
is itself a random variable which is described by a distribution 
$P_{\overline{X}}$. As shown in Eq.\ (\ref{eq:expectMean}), this
distribution has mean $\mu$, independent of the sample size.
On the other hand, the distribution has the variance
\begin{eqnarray}
\sigma^2_{\overline{X}} & \equiv & 
\Var[\overline{X}] = \Var\left[ \frac{1}{n} \sum_{i=0}^{n-1} X_i\right]
\stackrel{(\ref{eq:varLin})}{=}
\frac{1}{n^2} \sum_{i=0}^{n-1} \Var[X_i] \nonumber \\
& = &\frac{1}{n^2} n \Var[X] = 
\frac{\sigma^2}{n} \label{eq:sigmaMean}
\end{eqnarray}
Thus, the distribution of $\overline{X}$ gets narrower with increasing 
sample size $n$. This has the following consequence for the
expectation value of the sample variance which is
described by the random variable 
$S^2=\frac{1}{n} \sum_{i=0}^{n-1} (X_i-\overline{X})^2$:
\begin{eqnarray}
\E[S^2] & = & 
\E\left[ \frac{1}{n} \sum_{i=0}^{n-1} (X_i-\overline{X})^2\right]
= \E\left[ \frac{1}{n} \sum_{i=0}^{n-1} (X_i^2-2X_i \overline{X}
+\overline{X}^2)\right]
\nonumber \\
& = & \frac{1}{n} \left(\sum_{i=0}^{n-1} \E[X_i^2]-n \E[\overline{X}^2]\right)
\stackrel{(\ref{eq:2ndMoment})}{=} \frac{1}{n}
\left(n(\sigma^2+\mu^2)- n(\sigma^2_{\overline{X}}+\mu_{\overline{X}}^2) 
\right)
\nonumber \\
& \stackrel{(\ref{eq:sigmaMean})}{=} & 
\frac{1}{n} \left(n\sigma^2+n\mu^2- 
n\frac{\sigma^2}{n}-n\mu^2 \right) = \frac{n-1}{n} \sigma^2
\label{eq:varianceBiased}
\end{eqnarray}
This means that, although $s^2$ is biased,
 $\frac{n}{n-1}s^2$ is an unbiased estimator
for the variance of the underlying distribution of $X$. Nevertheless, 
 $s^2$ also becomes unbiased  for $n\to\infty$.\footnote{Sometimes
the sample variance is defined as $S^\star= 
\frac{1}{n-1}\sum_{i=0}^{n-1} (x_i-\overline{x})^2$ to make it
an unbiased estimator of the variance.}

For some distributions, for instance a power-law distribution
Eq.\ (\ref{eq:powerLawDistr}) with exponent $\gamma\le 2$, the
variance does not exist. Numerically, when calculating
$s^2$ according Eq.\ (\ref{eq:sampleVariance}), one
observes that it will not converge to a finite value when increasing
the sample size $n$. Instead one will observe occasionally jumps
to higher and higher values. One says the estimator
is {\em not robust}\index{estimator!robust}%
\index{robust estimator}. To get still an impression of
the spread of the data points, one can instead calculate 
the {\em average deviation}
\begin{equation}
D \equiv \frac{1}{n} \sum_{i=0}^{n-1} |x_i-\overline{x}|
\end{equation}
In general, an estimator is the less robust, the higher the involved
moments
are. Even the sample mean may not be robust,
for instance for a power-law distribution with $\gamma \le 1$.
In this case one can use the 
{\em sample median}\index{median!sample}\index{sample!median},
which is the value $x_m$ such that $x_i\le x_m$ for half the
sample points, i.e.\ $x_m$ is the $(n+1)/2$'th sample point if they are
sorted in ascending order.\footnote{If $n$ is even, one can take
the average between the $n/2$'th and the $(n+1)/2$'th sample point in
ascending order.} The sample median is clearly 
an estimator of the median (see Def.\ \ref{def:median}). It is
more robust, because it is less influenced by the sample
points in the tail. The simplest way to calculate the median
is to sort all sample points in ascending order and take
the sample point at the $(n/2+1)$'th position. This process
takes a running time ${\cal O}(n \log n)$. Nevertheless,
there is an algorithm \cite{PRA-numrec1995,ALG-cormen2001}
which calculates the median even in linear
running time ${\cal O}(n)$. 

\subsection{Confidence intervals\label{sec:confidence}}

In the previous section, we have studied estimators for parameters
of a random variable $X$
using a sample
obtained from a series of independent random experiments.
This is a so-called {\em point estimator}\index{point estimator},
because just one number is estimated.

Since each estimator is itself a random variable, each 
estimated value will be usually off the true value $\theta$. Consequently,
one wants to obtain an impression of how far off the estimate
  might be from the real value $\theta$. This can be obtained for instance
from:

\begin{sloppypar}
\begin{definition}
 The {\em mean squared error}\index{mean-squared error} 
of a point estimator $H=h(X_0,X_1, \ldots, X_{n-1})$ 
for a parameter $\theta$ is
\begin{eqnarray}
\MSE(H) &\equiv & \E[ (H-\theta)^2] =  \E[ (H-\E[H]+\E[H]-\theta)^2] \nonumber \\
& = & \E[ (H-\E[H])^2] + \E[ 2(H-\E[H])(\E[H]-\theta) ]
+ \E[ (\E[H]-\theta)^2 ] \nonumber \\
& = & \E[ (H-\E[H])^2] + 2\underbrace{(\E[H]-\E[H])}_{=0}(\E[H]-\theta)
+ (\E[H]-\theta)^2 \nonumber \\
& = & \Var[H] + (\E[H]-\theta)^2
\end{eqnarray}
\end{definition}
\end{sloppypar}

If an estimator is unbiased, i.e., if $\E[H]=\theta$, the
mean squared error is given by the variance of the estimator. Hence,
if for independent samples (each consisting of $n$ sample points) 
the estimated values are  close
to each other, the estimate is quite accurate. Unfortunately,
usually only {\em one} sample is available (how
to circumvent this problem rather ingeniously, see Sec.\ \ref{sec:bootstrap}).
  Also the mean squared error does not immediately provide a probabilistic
interpretation of how far the estimate is away from the true value $\theta$.

Nevertheless, one can obtain an estimate of the
error in a probabilistic sense.
Here we want to calculate a so-called 
 {\em confidence interval}\index{confidence interval} also sometimes
named {\em error bar}\index{error bar}.

\begin{definition}
For a parameter $\theta$ describing a random variable, two estimators
$l_\alpha=l_\alpha(x_0,x_1, \ldots, x_{n-1})$ and  
$u=u_\alpha(x_0,x_1, \ldots, x_{n-1})$
which are obtained from a sample $\{x_0,x_1,$ \ldots, $x_{n-1}\}$ provide
a {\em confidence interval}\index{confidence interval} if,
for given {\em confidence level}\index{confidence level} $1-\alpha\in (0,1)$
we have
\begin{equation}
\Prob(l_\alpha < \theta < u_\alpha)= 1-\alpha
\end{equation}
The value $\alpha\in(0,1)$ is called conversely 
{\em significance level}\index{significance level}.
\end{definition}
 This means, the true but unknown value $\theta$ is contained in
the interval $(l,u)$, which is itself a random variable as well,
with probability $1-\alpha$. Typical values of the confidence level
are 0.68, 0.95 and 0.99 ($\alpha=0.32$, $0.05$, $0.01$, respectively), 
providing increasing confidence.
The more one wants to be sure that
the interval really contains the true parameter, i.e.\ the smaller
the value of $\alpha$, the larger the confidence interval will be.

\begin{figure}[!ht]
\begin{center}
\includegraphics[width=0.7\textwidth]{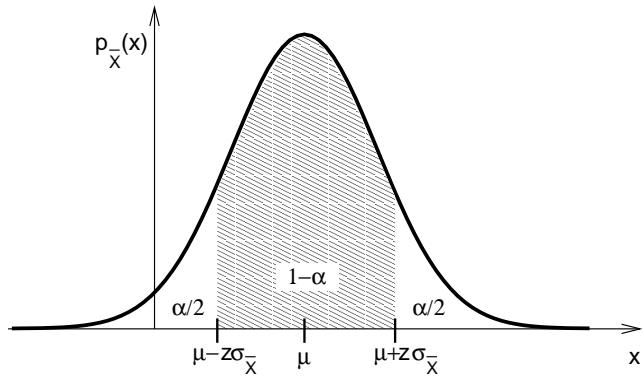}
\end{center}
\caption{Probability density function of the sample mean $\overline{X}$
for large enough sample sizes $n$ where the distribution becomes Gaussian. 
The true expectation value is denoted by 
$\mu$ and $\sigma_{\overline{X}}$ is the variance of the sample mean. 
The probability that a random number drawn from this distribution falls
outside the symmetric 
interval $[\mu-z\sigma_{\overline{X}},\mu+z\sigma_{\overline{X}}]$
is $\alpha$.
\label{fig:errorA}}
\end{figure}

Next, it is quickly
outlined how one arrives at the confidence interval 
for the mean, for details please
consult the specialized literature. First we recall that according
to its definition
the mean is a sum of independent random variables. For computer
simulations, one can assume that usually (see below
for a counterexample) a sufficiently large
number of experiments is performed.\footnote{This is different for
many empirical experiments, for example, when testing new treatments
in medical sciences, where often only a very restricted number
of experiments can be performed. In this case, one has to consider
special distributions, like the {\em Student distribution}.}
 Therefore,
according to the central limit theorem\index{central limit theorem}
 \ref{theo:centralLimit} 
$\overline{X}$ should exhibit (approximately) a pdf 
$f_{\overline{X}}$ which is Gaussian\index{Gaussian distribution}%
\index{distribution!Gaussian}
with an expectation value $\mu$ and some variance 
$\sigma^2_{\overline{X}}=\sigma^2/n$.
This means, the probability $\alpha$ 
that the sample means fall {\em outside} an
interval $I=[\mu-z\sigma_{\overline{X}},\mu+z\sigma_{\overline{X}}]$
can be easily obtained from the standard normal distribution.
 This
situation is shown in the Fig.\ \ref{fig:errorA}.
Note that the interval is symmetric about the mean $\mu$ and that
its width is stated in multiples $z=z(\alpha)$ of the standard deviation 
$\sigma_{\overline{X}}$. 
The relation between significance level $\alpha$ and half
interval width $z$ is just $\int_{-z}^{z}dx\,f_{\overline{X}}(x)=1-\alpha$. 
Hence,
the weight of the standard normal distribution {\em outside} 
the interval $[-z,z]$ is $\alpha$. This relation
can be obtained from any table
of the standard Gaussian distribution or from the function
\verb!gsl_cdf_gaussian_P()! of the {\em GNU scientific library}
\index{GNU scientific library}
(see Sec.\ \ref{sec:gsl}).
Usually, one considers integer values $z=1,2,3$
which correspond to significance levels $\alpha=0.32$, $0.05$,
and $0.003$, respectively. 
So far, the confidence interval $I$ contains the
unknown expectation value $\mu$ and the unknown 
variance $\sigma_{\overline{X}}$.  First, one can rewrite
\begin{eqnarray*}
1-\alpha & = &
\Prob(\mu-z\sigma_{\overline{X}}\le\overline{X}\le\mu+z\sigma_{\overline{X}})\\
& = & 
\Prob(-z\sigma_{\overline{X}}\le\overline{X}-\mu\le z\sigma_{\overline{X}})\\ 
& = & 
\Prob(-\overline{X}-z\sigma_{\overline{X}}\le-\mu\le 
-\overline{X}z\sigma_{\overline{X}})\\ 
& = & \Prob(\overline{X}-z\sigma_{\overline{X}} \le\mu\le
\overline{X}+z\sigma_{\overline{X}})
\,.
\end{eqnarray*}
This now states the probability that the true value, which is estimated
by the sample mean $\overline{x}$, lies within an interval which
is symmetric about the estimate $\overline{x}$. 
Note that the width $2z\sigma_{\overline{X}}$ is basically given
by  $\sigma_{\overline{X}}=\sqrt{\Var[\overline{X}]}$. This explains
why the mean squared error $\MSE(H)=\Var[H]$,\label{page:applyMSE} 
as presented in the beginning of this section, 
 is a good measure for the statistical error made by the estimator.
This will be used in Sec.\ \ref{sec:bootstrap}.

To finish, we  estimate
the true variance $\sigma^2$ using $\frac{n}{n-1}s^2$, hence
we get $\sigma_{\overline{X}}=\frac{\sigma}{\sqrt{n}}
\approx\frac{S}{\sqrt{n-1}}$. To summarize we 
get\index{expectation value!confidence interval}%
\index{confidence interval!expectation value}:

\begin{center}
\fbox{
\begin{minipage}[t]{0.9\textwidth}
\begin{equation}
\Prob \left(
\overline{X}-z\frac{S}{\sqrt{n-1}}  \le\mu\le 
\overline{X}+z\frac{S}{\sqrt{n-1}}
\right)\approx 1-\alpha \label{eq:confidence}
\end{equation}
\vspace*{2pt}
\end{minipage}
}
\end{center}
\noindent
Note that this confidence interval, with 
$l_\alpha=\overline{x}-z(\alpha)S/\sqrt{n-1}$
and $u_\alpha=\overline{x}+z(\alpha)S/\sqrt{n-1}$, is 
symmetric about $\overline{x}$,
which is not necessarily the case for other confidence intervals.
Very often in scientific publications, to state the estimate for $\mu$ including
 the confidence interval,  one gives the range where the true mean
is located in 68\% of all cases ($z=1$) i.e.\ $\overline{x}
\pm \frac{S}{\sqrt{n-1}}$, this is called the
{\em standard Gaussian error bar} or {\em one $\sigma$  
error bar}\index{error bar}. Thus, the sample variance and the
sample size determine the error bar/ confidence interval.

For the variance,\index{variance!confidence interval}%
\index{confidence interval!variance} 
the situation is more complicated, because it
is not simply a sum of statistically independent
 sample points $\{x_0,x_1,$ \ldots, $x_{n-1}\}$.
Without going into the details, here only
the result from the corresponding statistics literature 
\cite{dekking2005,lefebvre2006}
is cited:
The confidence interval where with probability $1-\alpha$ the
true variance is located is given by $[\sigma^2_l,\sigma^2_u]$ where

\begin{center}
\fbox{
\begin{minipage}[t]{0.7\textwidth}
  \begin{eqnarray}
    \label{eq:confVariance}
 \sigma^2_l & = & \frac{ns^2}{\chi^2(1-\alpha/2, n-1)} \nonumber \\    
 \sigma^2_u & = & \frac{ns^2}{\chi^2(\alpha/2, n-1)}\,.    
  \end{eqnarray}
\vspace*{2pt}
\end{minipage}
}
\end{center}
\noindent
Here, $\chi^2(\beta,\nu)$ is the inverse of the cumulative chi-squared
distribution with $\nu$ degrees of freedom.
It states the value where $F(\chi^2,\nu)=\beta$,
see page \pageref{page:chi2}. This chi-squared function
is implemented in the {\em GNU scientific library}
\index{GNU scientific library}
(see Sec.\ \ref{sec:gsl}) in the function
\verb!gsl_cdf_chisq_Pinv()!.

Note that as one alternative, you could regard $y_i\equiv (x_i-\overline{x})$
approximately as independent data points and use the above standard error
estimate described for the mean of the sample $\{y_i\}$. Also, one can use the
bootstrap method as explained below (Sec.\ \ref{sec:bootstrap}), 
which allows to calculate
confidence intervals for arbitrary estimators.

\subsection{Histograms\label{sec:histogram}}
\index{histogram|(}
Sometimes, you do not only want to estimate moments of an underlying
 distribution,
but you want to get an impression of the full distribution.
In this case you can use {\em histograms}.

\begin{definition}
 A histogram is given
by a set 
of disjoint intervals  
\begin{equation}
B_k=[l_k,u_k)\,,
\end{equation}
 which are
called {\em bins}\index{bin} and a counter $h_k$ for each bin. For a given
{\em sample}\index{sample} of
 $n$ measured points $\{x_0,x_1,$ \ldots, $x_{n-1}\}$, bin $h_k$
contains the number of sample points $x_i$ which are contained in $B_k$. 
\end{definition}

\begin{example}
\label{example:histo}
For the sample
\begin{eqnarray*}
\{x_i\} & = & \{ 1.2,\, 1.5,\, 1.0,\, 0.7,\, 1.4,\, 2.0,\,\\
 & & \;\; 1.5,\, 1.1,\, 0.9,\, 1.9,\, 1.2,\, 0.8\}   
\end{eqnarray*}
\noindent
the bins
$$
[0,0.5),\,[0.5,1.0)\,[1.0,1.5),\,[1.5,2.0),\,[2.0,2.5)\,[2.5,3.0)\,,
$$
are used, resulting in

$$
h_1 = 0,\, h_2 =3,\, h_3=5,\, h_4=3,\, h_5=1,\, h_6 = 0\,
$$
which is depicted in Fig.\ \ref{fig:histo}.

\begin{figure}[!ht]
\begin{center}
\includegraphics[width=0.5\textwidth]{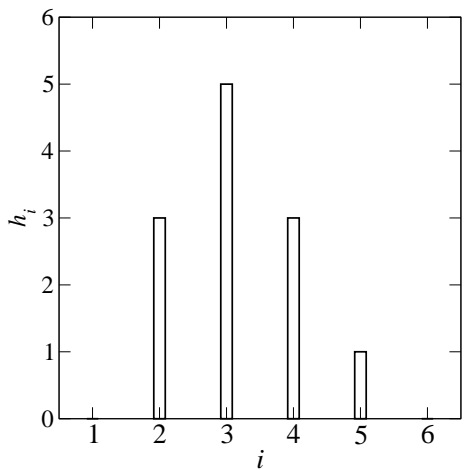}
\end{center}
\caption{Histogram for the data shown in Ex.\ \ref{example:histo}.
\label{fig:histo}}
\end{figure}

\end{example}

In principle, the bins can be chosen arbitrarily. You should take care
that the union of all intervals covers all (possible or actual) sample
points. Here, it is assumed that the bins are properly chosen.
Note also that the width $b_k=u_k-l_k$ of each bin can be different. 
Nevertheless,
often bins with uniform width are used. 
Furthermore, for many applications, for instance, when 
assigning different weights to different sample points\footnote{This 
occurs for some advanced simulation techniques.}, it is useful
to consider the counters as real-valued variables.
A simple (fixed-bin width)
C implementation of histograms is described in Sec.\ \ref{sec:ooCExample}.
The {\em GNU scientific library}
\index{GNU scientific library} (see Sec.\ \ref{sec:gsl}) contains
data structures and functions which implement histograms allowing for
variable bin width.

Formally, for a given random variable $X$, the count $h_k$ in bin $k$ can be 
seen as a result of  
a random experiment for the binomial random variable 
$H_k\sim B(n,p_k)$ with parameters $n$ and $p_k$, where 
$p_k=\Prob(X\in B_k)$ is the
probability that a random experiment for $X$ results in a value which is
contained in bin $B_k$. This means that
confidence intervals\index{confidence interval!histogram}%
\index{histogram!confidence interval} for a histogram bin can be obtained
in principle from a binomial distribution. Nevertheless, for
each sample the true value for a value $p_k$ is unknown and can
only be estimated by $q_k\equiv h_k/n$. Hence, the true binomial 
distribution is unknown. On the other hand, a binomial random variable
is a sum of $n$ Bernoulli random variables with parameter $p_k$. 
Thus, the estimator $q_k$
is nothing else than a sample mean for a Bernoulli random variable.
If the number of sample points $n$ is ``large'' (see below), 
from the central limit
theorem \ref{theo:centralLimit} and as discussed in 
Sec.\ \ref {sec:confidence}, the distribution of the sample mean
(being binomial in fact)
is approximately Gaussian. Therefore, 
one can use the standard confidence interval
Eq.\ (\ref{eq:confidence}), in this case

\begin{eqnarray}
\Prob \left(
q_k-z\frac{S}{\sqrt{n-1}} \le p_k\le
q_k+z\frac{S}{\sqrt{n-1}}
\right)\approx 1-\alpha \label{eq:confidenceHisto}
\end{eqnarray}
Here, according 
Eq.\ (\ref{eq:varBernoulli}), the Bernoulli random variable
exhibits   a 
sample variance $s^2=q_k(1-q_k)=
(h_k/n)(1-h_k/n)$. 
Again, $z=z(\alpha)$ denotes the half width
of an interval $[-z,z]$ such that the weight of the standard normal
distribution outside the interval equals $\alpha$. Hence,
the estimate with standard error bar ($z=1$) is 
$q_k \pm \sqrt{q_k(1-q_k)/(n-1)}$.

The question remains: What is ``large'' such that you can trust this 
``Gaussian'' confidence
interval? Consider that you measure for example no point at all 
for a certain bin $B_k$.
This can happen easily in the regions where $p_k$ is smaller than $1/n$ but 
non-zero,
i.e.\ in regions of  the histogram which are used to sample the tails
of a probability density function. In this case
the estimated fraction can easily be $q_k=0$ resulting also in a zero-width
confidence interval, which is certainly wrong. This means,
the number of samples $n$ needed to have a reliable confidence interval
for a bin $B_k$ depends on the number of bin entries. A rule of thumb
from the statistics literature 
is that  $nq_k(1-q_k)>9$ should hold. If this condition is
not fulfilled, the correct confidence interval $[q_{i,l},q_{i,u}]$
for $q_k$ has to be obtained
from the binomial distribution and it is quite complicated,
since it uses the  {\em F distribution} (see Def.\ \ref{def:Fdistr}
on page \pageref{page:F}) 

  \begin{eqnarray} 
    \label{eq:confHistoB}
 q_{i,l} & = & \frac{h_k}{h_k+(n-h_k+1)F_1} \nonumber \\    
 q_{i,u} & = & \frac{(h_k+1)F_2}{(h_k+1)F_2+n-h_k}\,,\\
 \mbox{where}\;\; F_1 & = & F(1-\alpha/2;\, 2n-2h_k+2,\, 2h_k) \nonumber\\
F_2 & = & F(1-\alpha/2;\, 2h_k+2,\, 2n-2h_k) \nonumber   
  \end{eqnarray}
The value $F(\beta;\,r_1,\,r_2)$ states the $x$ value such that
the distribution function for the F distribution with
number of degrees $r_1$ and $r_2$ reaches the value $\beta$. This inverse
distribution function is implemented in the {\em GNU scientific library}
\index{GNU scientific library}
(see Sec.\ \ref{sec:gsl}). If you always use these confidence intervals,
which are usually {\em not} symmetric about $q_k$, then you cannot go wrong.
Nevertheless, for most applications the standard Gaussian error bars are fine.

\
Finally, in case you want to use a histogram to represent
a sample from a continuous random variable, you can easily interpret 
a histogram as a sample for a 
probability density function\index{probability!density function}\index{pdf},
which can be represented as a set of points $\{(\tilde x_k,p(\tilde x_k))\}$.
This is called the \label{page:histoPdf} 
{\em histogram pdf}\index{histogram!pdf}\index{pdf!histogram} 
or the {\em sample pdf}.\index{sample!pdf}\index{pdf!sample} 
For simplicity, it is assumed that the interval mid points of the intervals
are used as $x$-coordinate. For the normalization, we have to divide
 by the total number of counts, as for $q_k=h_k/n$ 
and to divide by the bin width.
This ensures that the integral of the sample pdf, approximated
by a sum, gives just unity.  Therefore, we get
\begin{eqnarray}
 \tilde x_k & \equiv &(l_k+u_k)/2  \nonumber \\
  p(\tilde x_k) & \equiv & h_k/(n b_k)\,.
\end{eqnarray}
The confidence interval,
whatever type you choose, has to be normalized in the same way.
A function which prints a histogram as pdf, with simple
Gaussian error bars, is shown in Sec. \ref{sec:ooCExample}.

For discrete random variables, the histogram can be used
to estimate the pmf.\footnote{For discrete random variables,
the  $q_k$ values are already suitably normalized}.
 In this case the choice of the bins, in 
particular the bin widths, is easy, since usually all possible
outcomes of the random experiments are known.
For a histogram pdf, which is used to describe approximately
a continuous random variable, the choice of the bin width is important.
Basically,   you have to adjust the width manually, such that
the sample data is respresented ``best''. Thus, the bin width
should not be too small nor too large. Sometimes a non-uniform
bin width is the best choice. In this case no general advice can be given,
except that the bin width should be large where few data points have 
been sampled. This means that 
each bin should contain roughly the same number of sample points. 
Several different
 rules of thumb exist for uniform bin widths. For 
example $b=3.49 S n^{-1/3}$ \cite {scott1979},
which comes from minimizing the mean integrated squared difference
between a Gaussian pdf and  a sample drawn from this Gaussian distribution.
Hence, the larger the variance $S$ of the sample, the larger the bin width,
while increasing the number of sample points enables the bin width to be 
reduced.

In any case, you should be aware that the histogram pdf can be
only an approximation of the real pdf, due to the finite number
of data points and due to 
the underlying discrete nature resulting from the bins.
The latter problem has been addressed in recent years by so-called
{\em kernel estimators}\index{kernel!estimator}\index{estimator!kernel} 
\cite{dekking2005}. Here,  
each sample point $x_i$ is represented by 
a so-called {\em kernel 
function}.\index{kernel!function}\index{function!kernel}
A kernel function $k(x)$ is a peaked function, formally exhibiting
 the following properties:
  \begin{itemize}
  \item It has a maximum at 0.
\item It falls off to zero over some
some distance $h$.
\item Its integral $\int k(x)\,dx$ is normalized to one.
  \end{itemize}
Often used kernel functions are, e.g., a triangle, a cut upside-down parabola
or a Gaussian function.
Each sample point $x_i$ is represented such that
a kernel function is shifted having the
 maximum at $x_i$.
 The estimator $\hat p(x)$ for the pdf is the suitably normalized
sum (factor $1/n$) 
of all these kernel functions, one for each sample point: 
\begin{equation}
\hat p(x) = \frac 1 n \sum_i k(x-x_i)
\end{equation}
The advantages of these kernel estimators are that they result usually
in a smooth function $\hat p$ and that for a value $\hat p(x)$ 
also sample points more distant from $x$  may contribute, with decreasing
weight for increasing distance. The most
important parameter is the width $h$, because  too small a value of
$h$ will result
in many distinguishable peaks, one for each sample point, while  too
large value a of $h$ leads to a loss of important details. This is of similar
importance as the choice of the bin width for histograms.
The choice of the kernel function (e.g.\ a triangle, an upside-down parabola
or a Gaussian function) seems to be less important.

\index{histogram|)}

\subsection{Resampling using Bootstrap\label{sec:bootstrap}
}
\index{bootstrap approach|(ii}
\index{resampling|(}

\begin{sloppypar}
As pointed out, an estimator for some parameter $\theta$, given by a function
$h(x_0,x_1, \ldots, x_{n-1})$, is in fact a random variable
$H=h(X_0,X_1,\ldots,X_{n-1})$. Consequently, 
to get an impression of how much an estimate differs from the true value
of the parameter,
one needs in principle to know the distribution of the
estimator, e.g.\ via the pdf $p_H$ or the distribution function $F_H$.
In the previous chapter, the distribution was known for few estimators,
in particular if the sample size $n$ is large. For instance, the distribution
of the sample mean converges to a Gaussian distribution, irrespectively
of the distribution function $F_X$ describing the sample points $\{x_i\}$.
\end{sloppypar}

For the case of a general estimator $H$, in particular if $F_X$ is not known,
 one may not know anything about the distribution of  $H$.
In this case one can approximate $F_X$ by the sample distribution
function:

\begin{definition}
For a sample\index{sample} $\{x_0,x_1, \ldots, x_{n-1}\}$, the 
 {\em sample distribution function}\index{sample!distribution function|ii}%
\index{distribution function!sample}
(also called {\em empirical distribution function})%
\index{empirical distribution function}\index{distribution function!empirical}
is 
\begin{equation}
F_{\hat X}(x) \equiv 
\frac{\mbox{number of sample points } x_i 
\mbox{ smaller than or equal to } x }{n}
\label{eq:FhatX}
\end{equation}
\end{definition}
\noindent
Note that this distribution function  
describes in fact a  discrete 
random variable (called $\hat X$ here),
but is usually (but not always) used  to approximate a continuous
distribution function.

The {\em bootstrap principle} is to use $F_{\hat X}$ instead of $F_X$.
The name of this principle was made popular by
B. Efron \cite{efron1979,efron1994} and
comes from the fairy tale of Baron M\"unchhausen, who dragged
himself out of a swamp by pulling on the strap of his boot.\footnote{In the
European version, he dragged himself out by pulling his hair.}
Since the distribution function  $F_X$ is replaced by
the empirical sample distribution function, the approach
is also called {\em empirical bootstrap}, for a variant
called {parametric bootstrap} see below.

Now, having $F_{\hat X}$ one could in principle calculate 
the distribution function $F_{\hat H}$ for the random variable
$\hat H = h(\hat X_0, \hat X_1,\ldots, \hat X_{n-1})$ exactly,
which then is an approximation of $F_H$.
Usually, this is to cumbersome and one uses a second approximation:
One draws so-called {\em bootstrap samples}
$\{\hat x_0, \hat x_1, \ldots, \hat x_{n-1}\}$ from the random variable 
$\hat X$. This is called {\em resampling}\index{resampling}.
This can be done quite simply by $n$ times selecting ({\em with
replacement}\/) one
of the data points of the original sample $\{x_i\}$, each one with the
same probability $1/n$. This means that some sample points
from $\{x_i\}$ may appear several times in $\{\hat x_i\}$, some may
not appear at all.\footnote{The probability for a sample point not
to be selected is $(1-1/n)^n=\exp(n \log(1-1/n)) \to
\exp(n (-1/n))=\exp(-1)\approx 0.367$ for $n\to\infty$.}
Now, one can calculate the estimator value 
$h^*=h(\hat x_0, \hat x_1, \ldots, \hat x_{n-1})$ for each bootstrap 
sample. This is repeated
$K$ times for different bootstrap samples
resulting in $K$ values $h_k^*$ ($k=1,\ldots,K$)  of the estimator. 
The sample distribution function $F_{H^*}$
 of this sample $\{h_k^*\}$ is the final result,
which is an approximation of the desired distribution function $F_H$.
Note that the second approximation, replacing $F_{\hat H}$ by
 $F_{H^*}$ can be made arbitrarily accurate by making $K$ as large as
desired, which is computationally cheap.

You may ask: Does this work at all, i.e., is $F_{H^*}$
a good approximation of $F_{H}$? For the general case, there
is no answer. But for some cases there are mathematical proofs. For example
for the mean $H=\overline{X}$ the distribution function  $F_{\overline{X}^*}$
in fact converges to $F_{\overline{X}}$. Here, only the subtlety arises
that one has to consider in fact the {\em normalized} 
distributions of $\overline{X}-\mu$ ($\mu=\E[X]$) and $\hat X -\overline{x}$
($\overline{x}=\sum_{i=0}^{n-1}x_i/n$). Thus, the
random variables are just shifted by constant values. For other cases, like 
for estimating the median  or the variance, one has
to normalize in a different way, i.e., by subtracting the (empirical)
median or by dividing by the (empirical) variance. Nevertheless,
for the characteristics of $F_{H}$ we are interested in,
in particular in the variance, see below, normalizations like
shifting and stretching
are not relevant, hence they are ignored in the following.
Note that indeed some estimators exist, like the maximum
of a distribution, for which one can prove
conversely that  $F_{H^*}$ does {\em not} converge
to $F_{H}$, even after some normalization. 
On the other hand, for the purpose of getting a not too bad
estimate of the error bar, for example, bootstrapping is a very convenient
and suitable approach which has received high acceptance during recent years.

Now one can use $F_{H*}$ to calculate any desired quantity.
Most important is the case of a
 confidence interval $[h_l,h_u]$ such that the total probability
 outside the interval is $\alpha$, for given significance level $\alpha$,
i.e.\ $F_{H^*}(h_u)-F_{H^*}(h_l)=1-\alpha$. In particular, one
can distribute the weight $\alpha$ equally below and above the interval,
which allows to determine $h_l,h_u$ 
\begin{equation} 
F_{H^*}(h_u)=F_{H^*}(h_l)=\alpha/2\,.
\label{eq:bsConfidence}
\end{equation}
 Similar to the confidence intervals
presented in Sec.\ \ref{sec:confidence}, $[h_l,h_u]$ also 
re\-pre\-sents a confidence interval
for the unknown parameter $\theta$ which is to be estimated from the
estimator (if it is unbiased). Note that $[h_l, h_u]$ can be
non-symmetric about the actual estimate $h(x_0,x_1, \ldots, x_{n-1})$.
This will happen if the distribution $F_{H^*}$ is skewed.

For simplicity, as we have seen in Sec.\ \ref{sec:confidence},
one can use the variance $\Var[H]$ to describe the statistical
uncertainty of the estimator. As mentioned on page \pageref{page:applyMSE},
this corresponds basically to a $\alpha=0.32$ uncertainty. 

\sourcesC{randomness}{bootstrap.c\\bootstrap\_test.c}{5}
The following C function 
calculates $\Var[H^*]$, as approximation of 
the unknown $\Var[H]$. One has to pass as arguments
the number $n$ of sample points,
an array containing the sample points, the number $K$
of bootstrap iterations, and a pointer to the function
\verb!f! which represents the estimator. \verb!f! has to take two arguments:
the number of sample points and an array containing a sample.
For an explanation of function pointers, see Sec.\ \ref{sec:pointers}.
The function \verb!bootstrap_variance()!  returns $\Var[H^*]$.

{\small 
\linenumbers[1]
\begin{verbatim}
double bootstrap_variance(int n, double *x, int n_resample,
                         double (*f)(int, double *))
{
  double *xb;                                  /* bootstrap sample */
  double *h;                            /* results from resampling */
  int sample, i;                                  /* loop counters */
  int k;                                        /* sample point id */
  double var;                             /* result to be returned */
  h = (double *) malloc(n_resample * sizeof(double));
  xb = (double *) malloc(n * sizeof(double));
  for(sample=0; sample<n_resample; sample++)
  {
    for(i=0; i<n; i++)                                 /* resample */
    {
      k = (int) floor(drand48()*n);         /* select random point */
      xb[i] = x[k];
    }
    h[sample] = f(n, xb);                   /* calculate estimator */
  }
\end{verbatim}
\newpage
\begin{verbatim}
  var = variance(n_resample, h);      /* obtain bootstrap variance */
  free(h);
  free(xb);
  return(var);
}
\end{verbatim}
\nolinenumbers} 

The bootstrap samples $\{\hat x_i\}$ are stored in the array \verb!xb!, while
the sampled estimator values $\{h_k^*\}$ are stored in the array \verb!h!.
These arrays are allocated in lines 10--11.
In the main loop (lines 12--20) the bootstrap samples
are calculated, each time the estimator is obtained and the 
result is stored in \verb!h!. Finally, the variance of the sample
$\{h_k^*\}$ is calculated (line 22). Here, the function
\verb!variance()! is used, which works similarly to the
function \verb!mean()!, see exercise (\ref{ex:variance}).
Your are asked to implement a bootstrap function for general
confidence interval in exercise (\ref{ex:bootstrap}).

The most obvious way is to call \verb!bootstrap_variance()! with the
estimator \verb!mean! as forth argument.
For a distribution which is ``well behaved''
(i.e., where a sum of few random variables resembles the
Gaussian distribution),
 you will get  a variance that is,
at least if \verb!n_resample! is reasonably large, very close
to the standard Gaussian ($\alpha=0.32$) error bar.

\begin{figure}[!ht]
\begin{center}
\includegraphics[width=0.6\textwidth]{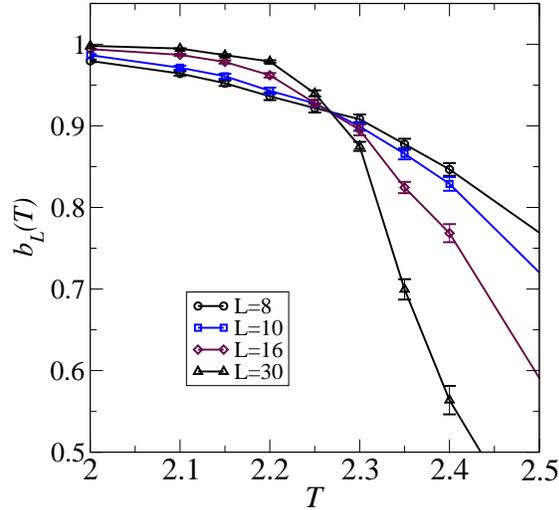}
\caption{Plot of Binder cumulant of two-dimensional model ferromagnet
as function of temperature $T$ (dimensionless units). 
Each system consists of $L\times L$ particles. The curves
for different system sizes  $L$ cross very close to the phase
transition temperature $T_c=2.269$ (known from analytical calculations
of this model). The error bars shown can be obtained
using a bootstrap approach.
\label{fig:binder}}
\end{center}
\end{figure}

\index{Binder cumulant|(}\index{cumulant!Binder|(}
For calculating properties of the sample mean, the bootstrap approach
works fine, but in this case one could also be satisfied with the standard
Gaussian confidence interval. The bootstrap approach is more
interesting for non-standard estimators. One prominent example
from the field of statistical physics is the so-called 
{\em Binder cumulant}\/
 \cite{binder1981}, which is given by:
\begin{equation}
b(x_0,x_1, \ldots, x_{n-1})=
0.5\left(3-\frac{\overline{x^4}}
{[\overline{x^2}]^2}\right)\,
\label{eq:binder}
\end{equation}
\sourcesC{randomness}{binder\_L8.dat\\binder\_L10.dat\\binder\_L16.dat\\
binder\_L30.dat}{7}
where $\overline{\rule{0pt}{5pt}\ldots}$ is again the sample mean,
for example $ \overline{x^2} = \sum_{i=0}^{n-1} x_i^2$.
The Binder cumulant is often used to determine phase transitions
via simulations, where only systems consisting 
of a finite number of particles  can be studied.
For example, consider a ferromagnetic\index{ferromagnet} system held at some
temperature\index{temperature} $T$. At low temperature, below the 
{\em Curie temperature}\index{Curie temperature}\index{temperature!Curie}
$T_c$, the system will exhibit a macroscopic magnetization $m$.
On the other hand, for temperatures above $T_c$, $m$ will
on average converge to zero when increasing the system size.
This transition is fuzzy, if the system sizes are small. Nevertheless,
when evaluating the Binder cumulant for different sets of sample 
points $\{m(T,L)_i\}$ which are obtained at several temperatures $T$ and 
for different system sizes $L$, the $b_L(T)$ curves for different $L$
will all cross \cite{landau2000} (almost) 
at $T_c$, which allows for a very precise
determination of $T_c$. A sample result for a two-dimensional
(i.e.\ layered) model ferromagnet exhibiting $L\times L$ particles
is shown in Fig.\ \ref{fig:binder}.
The Binder cumulant has been useful for the simulation of many other
systems like disordered materials, gases, optimization problems, 
liquids, and graphs describing social systems.

A confidence interval for the Binder cumulant 
is very difficult (or even impossible) to obtain using standard
error analysis. Using bootstrapping, it is straightforward. You can
use simply the function \verb!bootstrap_variance()! shown above
while providing as argument a function which evaluates the Binder
cumulant for a given set of data points.
\index{Binder cumulant|)}\index{cumulant!Binder|)}

\index{parametric bootstrap approach|(}\index{bootstrap approach!parametric|(}
So far, it was assumed that the empirical distribution function $F_{\hat X}$
was used to determine an approximation of $F_H$. Alternatively,
one can use some additional knwoledge which might be available.
Or one can make additional assumptions, via using a distribution
function $F_{\myvec{\lambda}}$ which is parametrized by
a vector of parameters $\myvec{\lambda}$. For an exponential
distribution, the vector would just consist of one parameter,
the expectation value,  while for a Gaussian distribution, 
$\myvec{\lambda}$ would consist of the expectation value and 
the variance. In principle,
arbitrary complex distributions with many parameters are possible.
To make $F_{\myvec{\lambda}}$ a ``good'' approximation of $F_{X}$, one has
to adjust the parameters such that the distribution function
 represents the sample
$\{x_i\}$ ``best'', resulting in a vector $\myvec{\hat \lambda}$ of parameters.
 Methods and tools to perform this {\em fitting}\index{fit}
of parameters are presented in Sec.\ \ref{sec:fitting}.
Using $F_{\myvec{\hat \lambda}}$ one can proceed as above: Either
one calculates $F_{\hat H}$ exactly based on $F_{\myvec{\hat \lambda}}$, 
which is most of the time too cumbersome. 
Instead, usually one performs simulations where one takes repeatedly 
samples $\{\hat x_0, \hat x_1, \ldots, \hat x_{n-1}\}$ 
from simulating $F_{\myvec{\hat \lambda}}$ and calculates each time
the estimator $h^*=h(\hat x_0, \hat x_1, \ldots, \hat x_{n-1})$. This
results, as in the case of the empirical bootstrap discussed above,
 in a sample distribution function $F_{H^*}$ which is further
analyzed. This approach, where  $F_{\myvec{\lambda}}$ is used
instead of  $F_{\hat X}$, is called 
{\em parametric bootstrap}\index{parametric bootstrap approach|)}%
\index{bootstrap approach!parametric|)}.

\begin{sloppypar}
Note that the bootstrap approach does not require that the 
sample points are statistically independent of each other. 
For instance,   the sample could be generated using
a Markov chain Monte Carlo
 simulation \cite{newman1999,landau2000,robert2004,liu2008}, 
where each data point $x_{i+1}$
is calculated using some random process, but also depends on the previous 
data point $x_{i}$.
More details on how to quantify correlations are given in 
Sec.\ \ref{sec:hypothesis}.
Nevertheless, if the sample 
follows the distribution $F_X$, everything is fine when using
bootstrapping and for example
a confidence interval will not depend on the fraction of ``independent''
data points. One can see this easily by assuming that you replace
each data point in the original sample $\{x_{i}\}$ by ten copies,
hence making the sample ten times larger without adding any information.
 This will not affect any
of the following bootstrap calculations, since the size of the
sample does not enter explicitly. The fact that bootstrapping is
not susceptible to correlations between data points is in contrast
to the classical calculation of confidence intervals explained
in Sec.\ \ref{sec:confidence}, where independence of data
is assumed and the number of independent data points enters
formulas like Eq.\ (\ref{eq:confidence}). Hence, when calculating
the error bar according  to Eq.\ (\ref{eq:confidence}) using the ten-copy 
sample, it will be {\em incorrectly} smaller by a factor $\sqrt{10}$,
since no additional information is available compared to the
original sample.
\end{sloppypar}

\index{bootstrap approach|)}

\index{jackknife technique|(}
It should be mentioned that bootstrapping is only one
of several resampling techniques. Another well known approach
is the {\em jackknife approach}, 
where one does not sample randomly
using $F_{\hat X}$ or a fitted  $F_{\myvec{\lambda}}$. Instead 
the sample $\{x_i\}$ is divided into $B$ blocks of  equal
size $n_b=n/B$ (assuming that $n$ is a multiple of $B$). 
Note that choosing $B=n$ is possible and not uncommon.
Next, a number $B$ of so-called {\em jackknife samples} $b=1,\ldots,B$ 
are formed from the original sample  $\{x_i\}$
by omitting exactly the sample points from the $b$'th block and
including all other points of the original sample. 
Therefore, each of these jackknife samples consists of $n-n_b$
sample points. 
For each jackknife sample, again the estimator is calculated,
resulting in a sample $\{h_k^{(j)}\}$ of size $B$.
Note that the sample distribution function $F^{(j)}$ of this sample
is {\em not} an approximation of the estimator distribution
function $F_H$! Nevertheless, it is useful.
For instance, the variance $\Var[H]$ can be estimated from
 $(B-1)S_h^2$, where $S_h^2$  is the sample variance  of
$\{h_k^{(j)}\}$. No proof 
of this is presented here. It is just noted that when increasing the number $B$
of blocks, i.e., making the different jackknife samples more alike, because
fewer points are excluded,
the sample of estimators values  $\{h_k^{(j)}\}$ will fluctuate less.
Consequently, this dependence on the number of blocks is exactly
compensated via the factor $(B-1)$. Note that for the jackknife method, 
in contrast to the boostrap approach, the statistical independence
of the original sample is required. If there are correlations
between the data points, the jackknife approach can be combined
with the so-called blocking method \cite{flyvbjerg1998}.
More details on the
jackknife approach can be found in \cite{efron1994}.
\index{jackknife technique|)}

Finally, you should be aware that there are cases where 
resampling approaches clearly fail. The most obvious example is the
calculation of confidence intervals for histograms, 
see Sec.\ \ref{sec:histogram}. A bin which exhibits no
sample points, for example, where the probability is very small, will
never get a sample point during resampling either. Hence, the error bar
will be of zero width. This is in contrast to the confidence interval
shown in Eq.\ \ref{eq:confHistoB}, where also bins with zero
entries exhibit a finite-size confidence interval. Consequently, you
have to think carefully before deciding which approach you will use
to determine the reliability of your results.

\index{resampling|)}

\index{analyzing data|)}
\index{data!analysis|)}

\section{Data plotting\label{sec:plotting}} 

\index{plotting data|(ii}
\index{data!plotting|(ii}
So far, you have learned many methods for analyzing data. Since you 
do not just want to look at tables filled with numbers, you should
visualize the data in viewgraphs. Those viewgraphs which contain the
essential results of your work can be used in presentations or
publications.
To analyze and plot data, several commercial and non-commercial programs are
available. Here, two free programs are discussed, 
{\it gnuplot}\/, and {\em xmgrace}\/.  {\it Gnuplot}\/
is small, fast, allows two- and three-dimensional curves to be generated
and transformed, as well as 
arbitrary functions to be fitted to the data (see Sec.\ \ref{sec:fitting}). 
On the other hand {\it xmgrace}\/ is more flexible and produces better output. 
It is recommended to use
{\it gnuplot}\/   for viewing and fitting data online, 
while {\it xmgrace}\/
is to be preferred for producing figures to be shown in presentations
 or publications.

\subsection{{\it gnuplot}\label{sec:gnuplot}}

\index{gnuplot@{\em{}gnuplot}|(ii}

The program {\it gnuplot}\/ is invoked by entering {\tt gnuplot} in a
shell, or from a menu of the graphical user interface of your operating 
system.
For a complete manual see \cite{PRA-texinfo}.

As always, our examples refer to a UNIX window system like X11, 
but the program is available for
almost all operating systems. After startup, in the window of your shell
or the window which pops up for {\tt gnuplot} the prompt (e.g.\ {\tt
  gnuplot$>$}) appears
and the user can enter commands in textual form, results are shown in
additional windows or are written into files. 
For a general introduction you can type just {\tt help}.

Before giving an example, it should be
pointed out that gnuplot {\em scripts}\/ \index{gnuplot@{\em{}gnuplot}!script}
 can be generated by simply writing the commands
into a file, e.g.\ {\tt command.gp}, and calling 
\verb!gnuplot command.gp!.

\sources{randomness}{sg\_e0\_L.dat}
The typical case is that you have available a data file of $x-y$ data or
with $x-y-dy$ data (where $dy$ is the error bar of the $y$ data points).
 Your \label{page:SGe0}
file might look like this, where the ``energy'' $e_0$
of a system\footnote{ 
It is the
ground-state energy \index{ground state!energy}
of a three-dimensional $\pm J$ spin glass \index{spin glass},
a protypical system in statistical physics. These spin glasses
model the magnetic behavior of alloys like iron-gold.} 
is stored as a function of the 
``system size'' $L$. The filename is {\tt sg\_e0\_L.dat}. The first column
contains the $L$ values, the second the energy values and the third the
standard error of the energy. Please note that lines starting with
``{\tt{}\#}'' are comment lines which are ignored on reading:

{\small \begin{verbatim}
# ground state energy of +-J spin glasses
# L    e_0   error
  3 -1.6710 0.0037
  4 -1.7341 0.0019
  5 -1.7603 0.0008
  6 -1.7726 0.0009
  8 -1.7809 0.0008
 10 -1.7823 0.0015
 12 -1.7852 0.0004
 14 -1.7866 0.0007
\end{verbatim}}

\begin{figure}[!ht]
\begin{center}
\includegraphics[width=0.8\textwidth]{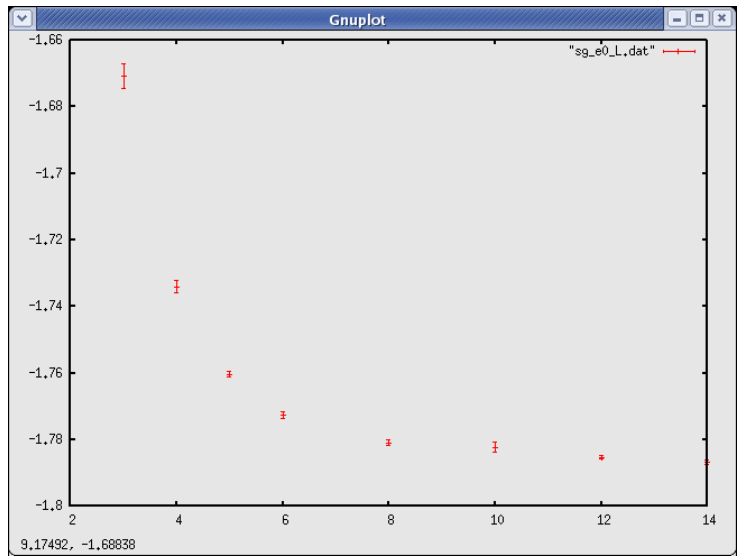}
\end{center}
\caption{{\it Gnuplot} window showing the result of a plot command.
\label{fig:gnuplotA}}
\end{figure}

To plot the data enter
{\small \begin{verbatim}
gnuplot> plot "sg_e0_L.dat" with yerrorbars
\end{verbatim}}
\noindent
which can be abbreviated as {\tt p "sg\_e0\_L.dat" w e}. Please do not
forget the quotation marks around the file name. Next, a window
pops up, showing the result, see Fig.\ \ref{fig:gnuplotA}.

For the {\tt plot} command many options and styles are available, e.g.\ {\tt
  with lines} produces lines instead of symbols. It is not explained here
how to set line styles or symbol sizes and colors, because this
is usually not necessary for a quick look at the data. For ``nice'' plots
used for presentations, we recommend {\em xmgrace}, see next section. 
Anyway, \verb!help plot! will tell you all you have to know
about the \verb!plot! command.

Among the important options of the {\tt plot} command is that one
can specify ranges. This can be done by specifying the range directly
after the command, e.g.\ 

{\small \begin{verbatim}
 gnuplot> plot [7:20]  "sg_e0_L.dat" with yerrorbars
\end{verbatim}}
\noindent
will only show the data for $x\in[7,20]$. Also an additional $x$ range
can be specified like in

{\small \begin{verbatim}
plot [7:20][-1.79:-1.77]  "sg_e0_L.dat" with yerrorbars
\end{verbatim}}
\noindent
If you just want to set the $y$ range, you have to specify \verb![ ]!
for the $x$-range. You can also fix the ranges via the \verb!set xrange!  
and the \verb!set yrange! commands, such that you do not have to give them each
time with the plot command, see \verb!help set xrange! or
\verb!help unset xrange! for unsetting a range.

{\tt Gnuplot} knows a lot of built-in functions like $\sin(x)$,
$\log(x)$, powers, roots, Bessel functions, error function,\footnote{The
error function is  erf$(x)=(2/\sqrt{\pi})\int_0^x dx'\exp(-x'^2)$.} 
 and many more. For a complete list type \verb!help functions!.
These function can be also plotted. Furthermore,
using these functions and standard arithmetic expressions, you
can also define your own functions, e.g.\ you can define a function
\verb!ft(x)! for the Fischer-Tippett pdf (see Eq. (\ref{eq:FiherTippet}))
for parameter $\lambda$ (called \verb!lambda! here) and show the function
via

{\small \begin{verbatim}
gnuplot> ft(x)=lambda*exp(-lambda*x)*exp(-exp(-lambda*x))
gnuplot> lambda=1.0
gnuplot> plot ft(x)
\end{verbatim}}
\noindent
You can also include arithmetic expressions in the plot command.
To plot a shifted and scaled Fischer-Tippett pdf you can type:

{\small \begin{verbatim}
gnuplot> plot [0:20] 0.5*ft(0.5*(x-5))
\end{verbatim}}

The Fischer-Tippett pdf has a tail which drops off exponentially. This
can be better seen by a logarithmic scaling of the $y$ axis.

\begin{figure}[!ht]
\begin{center}
\includegraphics[width=0.8\textwidth]{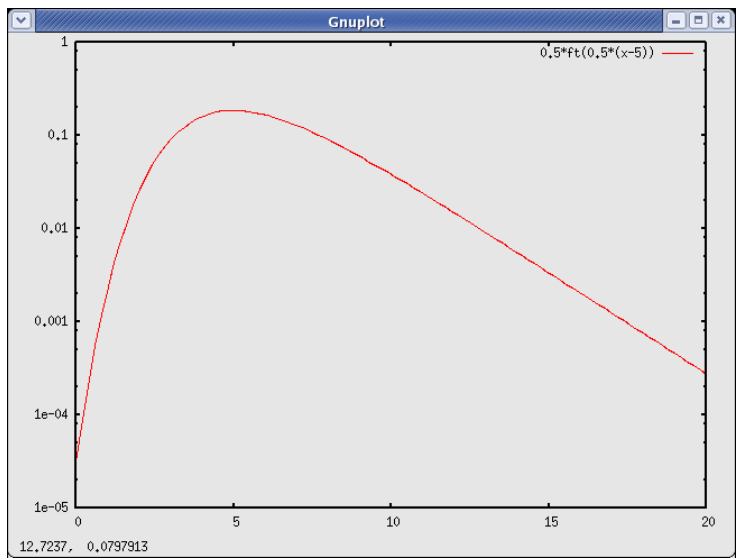}
\end{center}
\caption{{\it Gnuplot} window showing the result of plotting
a shifted and rescaled Fischer-Tippett pdf with logarithmically
scaled $y$-axis.
\label{fig:gnuplotLogscale}}
\end{figure}

{\small \begin{verbatim}
gnuplot> set logscale y
gnuplot> plot [0:20] 0.5*ft(0.5*(x-5))
\end{verbatim}}
 
\noindent 
will produce the plot shown in Fig.\ \ref{fig:gnuplotLogscale}.

  Furthermore, it is also possible to plot several functions in 
one plot, via separating them via commas, e.g.\ to compare a
Fischer-Tippett pdf to the standard Gaussian pdf, here the
predefined constant \verb!pi! is used:

{\small \begin{verbatim}
gnuplot> plot ft(x), exp(-x*x/2)/sqrt(2*pi)
\end{verbatim}}

It is possible to
read files with multi columns via the {\tt using} {\em data modifier}, e.g.

{\small \begin{verbatim}
gnuplot> plot "test.dat" using 1:4:5 w e
\end{verbatim}}
\noindent
displays the fourth column as a function of the first, with error bars
given by the 5th column. The elements behind the {\tt using} are called
entries.
Within the \verb!using! data modifier
you can also perform transformations and calculations. 
Each entry, where some calculations
should be performed have to be embraced in \verb!(  )! brackets.
Inside the brackets you can refer to the different columns of the input
file via \verb!$1! for the first column, \verb!$2! for the second, etc.
You can generate arbitrary expressions inside the brackets, i.e.\
use data from different columns (also combine several columns in one entry),
operators, variables, 
predefined and self-defined functions and so on. For example, in 
Sec.\ \ref{sec:fitting}, you will see that the data from
the \verb!sg_e0_L.dat! file follows
approximately a power law behavior $e_0(L)=e_{\infty}+aL^b$ with 
$e_{\infty}\approx-1.788$, $a\approx 2.54$ and $b\approx -2.8$. To visualize
this, we want to show $e_0(L)-e_{\infty}$ as a function of $L^b$.
This is accomplished via:

{\small \begin{verbatim}
gnuplot> einf=-1.788
gnuplot> b=-2.8
gnuplot> plot "sg_e0_L.dat" u ($1**b):($2-einf)
\end{verbatim}}
\noindent
Now the {\it gnuplot} window will show the data as a straight line (not shown,
but see Fig.\ \ref{fig:xmgrace:final}). 

 So far, all output has appeared on the screen. It is possible to
redirect the output, for example, to an encapsulated
 postscript file \index{postscript@{\em postscript} file format}
(by setting {\tt
  set terminal postscript} and redirecting the output {\tt set output
   "test.eps"}).  When you now enter a plot command, the
corresponding postscript file will be generated.

 Note that not only several functions but also several data 
files or a mixture of both can be combined into
one figure. 
To remember what a plot  exported to files means, 
you can set axis labels of the figure by
typing e.g.\ \verb!set xlabel "L"!, which becomes active when the next
\verb!plot! command is executed. Also you can use \verb!set title!
or place arbitrary labels via \verb!set label!. Use the \verb!help!
command to find out more.

\begin{figure}[!ht]
\begin{center}
\includegraphics[width=0.8\textwidth]{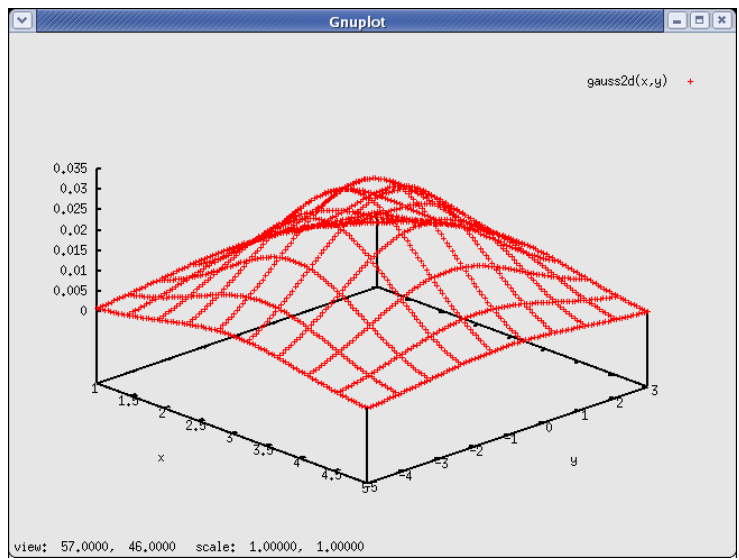}
\end{center}
\caption{ {\it Gnuplot} window showing the result of plotting
a two-dimensional function using {\tt splot}. 
\label{fig:gnuplot3d}
}
\end{figure}

  Also three-dimensional plotting (in fact a projection into two
dimensions) is possible
using the {\tt splot} command (enter {\tt help splot} to obtain more
information).  Here, as example, we plot a two-dimensional Gaussian
distribution:

{\small \begin{verbatim}
gnuplot> x0=3.0
gnuplot> y0=-1.0
gnuplot> sx=1.0
gnuplot> sy=5.0
gnuplot> gauss2d(x,y)=exp(-(x-x0)**2/(2*sx)-(y-y0)**2/(2*sy))\
> /sqrt(4*pi**2*sx**2*sy**2)
gnuplot> set xlabel "x"
gnuplot> set ylabel "y"
gnuplot> splot [x0-2:x0+2][y0-4:y0+4]  gauss2d(x,y) with points
gnuplot>
\end{verbatim}}
\noindent
Note that the long line containing the definition of the (two-argument)
function \verb!gauss2d()! is split up into two lines using a backslash
at the end of the first line. Furthermore, some of the variables are
used inside the interval specifications at the beginning of the
\verb!splot! command.  Clearly, you also can plot
data files with three-dimensional data. 
The resulting plot appearing in the output window
is shown in Fig.\  \ref{fig:gnuplot3d}.
You can drag the mouse inside the window showing the plot,
which will alter the view.

Finally, to stop the execution of {\it gnuplot}, enter the 
command \verb!exit!. These examples should give you already a good
impression of what can be done with {\it gnuplot}. More can be found
in the documentation or the online help. 
How to fit functions to data using {\it gnuplot}
is explained in Sec.\ \ref{sec:fitting}.
It is also possible to make,
with some effort, publication-suitable plots, but it is simpler
to achieve this with {\it xmgrace}, which is presented in the following
section.

\index{gnuplot@{\em{}gnuplot}|)}

 \subsection{\em{}xmgrace}
\index{xmgrace@{\em{}xmgrace}|(}

The {\it xmgrace}\/ (X Motiv GRaphing, Advanced Computation and 
Exploration of data) program is much more powerful than
{\it gnuplot}\/ and produces nicer output, commands are issued by clicking on
menus and buttons and it offers WYSIWYG.
The {\it xmgrace}\/ program offers almost every feature you can imagine for
two-dimensional data plots, including multiple plots (insets), 
fits, fast Fourier transform, interpolation. The look of the plots may be 
altered in any kind of way you can imagine like choosing fonts,
sizes, colors, symbols, styles for lines, bar charts etc. Also, you can create
manifold types of labels / legends and it is possible to add elements
like texts, labels, lines or other geometrical objects in the plot.
The plots can be exported to various format, in particular encapsulated
postscript (\verb!.eps!)
Advanced users also can program it or use it for real-time visualization
of simulations.  
On the other hand, its handling is a little bit
slower compared to {\it gnuplot} 
and the program has the tendency to fill your screen with windows. 
 For full
information, please consult the online help, the manual or the
program's web page \cite{xmgrace}.

 \begin{figure}[!ht]
\begin{center}
\includegraphics[width=0.6\textwidth]{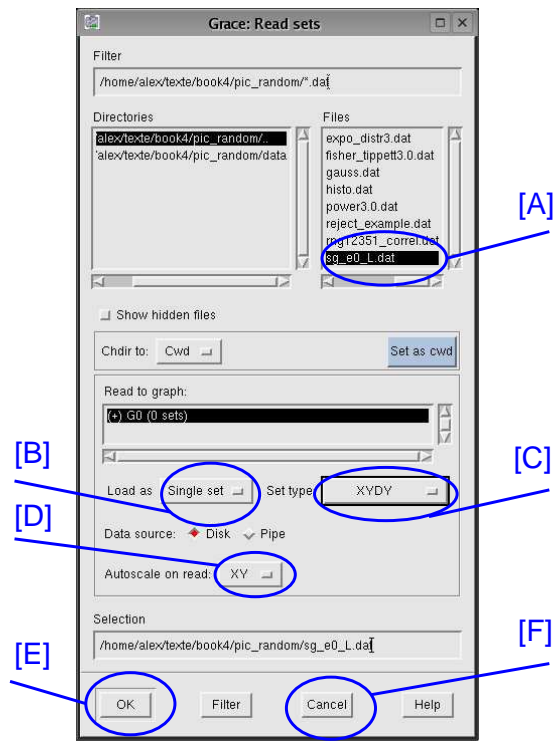}
\end{center}
\caption{The {\sf Grace:Read Set} window of the {\it xmgrace}\/ program.
Among others, you can select a file {\sf [A]}, choose the type of the input
file {\sf [B]}, choose the format of the data {\sf [C]}, what axes
should be rescaled automatically on input {\sf [D]}. You can
actually load the data by hitting on the {\sf OK} button 
{\sf [E]} and closing the window by hitting on the {\sf Cancel} button
{\sf [F]}.
\label{fig:xmgrace:filer}
}
\end{figure}

\begin{figure}[!ht]
\begin{center}
\includegraphics[width=0.9\textwidth]{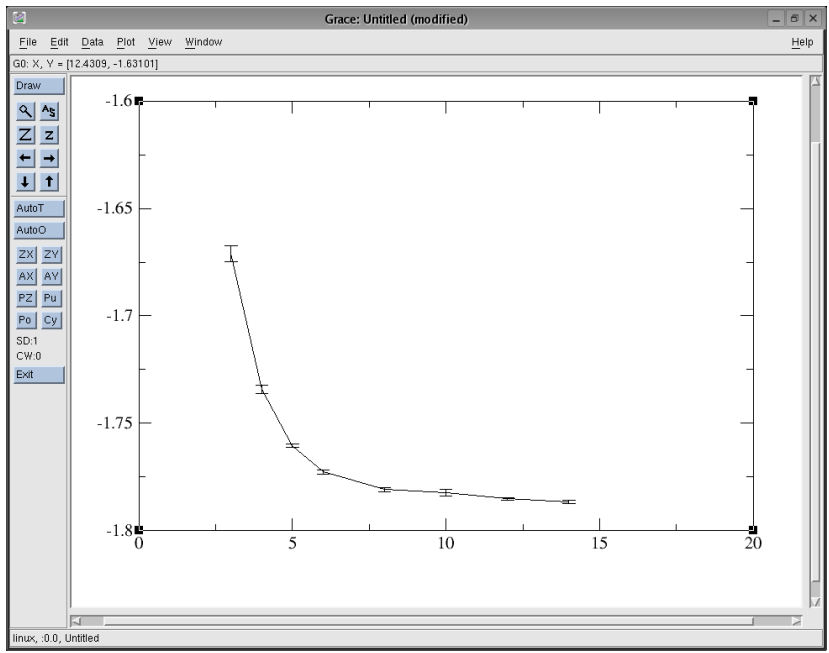}
\end{center}
\caption{The main  {\it xmgrace}\/ window after the data set has been
loaded (with auto scale).
\label{fig:xmgrace:loaded}
}
\end{figure}

 Here, just the main steps to produce a simple but nice plot are shown
and some further directions are mentioned.
 You will be given here the most important steps
to create a similar plot to the first example, shown for
the {\it gnuplot} program,  but ready for publication.
First you have to 
start the program 
by typing {\tt xmgrace} into a shell (or to start it from some 
window/operating system menu). Then you choose the 
{\sf \underline{D}ata} menu\footnote{The underlined character appears
also in the menu name and refers to the key one has to hit together
with {\sf Alt} button, if one wants
to open the menu via key strokes.}, next 
the {\sf \underline{I}mport} sub menu 
and finally the {\sf \underline{A}SCII..}  sub sub menu. 
Then a ``{\sf Grace:Read Set}''
window will pop up (see Fig.\ \ref{fig:xmgrace:filer})
and you can choose the data file to be loaded 
(here \verb!sg_e0_L.dat!) {\sf [A]}, the type of the input file 
({\sf Single Set}) {\sf [B]},
the format of the data ({\sf XYDY}) {\sf [C]}. This means you have
three columns, and the third one is an error bar for the second. 
Then you can hit on the
{\sf OK} button {\sf [E]}. The data will be loaded and shown in the
main window (see Fig.\ \ref{fig:xmgrace:loaded}). 
The axis ranges have been adjusted to the data, because
the ``{\sf Autoscale on read}'' is set by default to ``{\sf XY}'' {\sf [D]}.
You can quickly change the part of the data shown by the buttons
(magnifier, {\sf AS}, {\sf Z}, {\sf z}, $\leftarrow$, $\rightarrow$,
$\downarrow$, $\uparrow$) on the left of the main window just below
the {\sf Draw} button.

 Note that another important input file type is ``{\sf Block data}''
where the files consist of many columns of which you only
want to show some. When you hit th {\sf OK} button {\sf [E]},
another window ({\sf Grace:Edit block data}) 
will pop up, where you have to select the columns which you
actually want to display. For the data format (also when loading block data), 
some other important
choices are {\sf XY} (no error bars) and {\sf XYDYDY} 
(full confidence interval, maybe non-symmetric). 
Finally, you can
close the file selection window,  by hitting on the {\sf Cancel button}
{\sf [F]}. The {\sf OK} and {\sf Cancel} buttons are common to all
{\it xmgrace} windows and will not be mentioned explicitly in the
following.

 In the form the loaded data is shown by default, it is not suitable 
for publication purposes, because the symbols, lines and fonts are usually
too small/ too thin. To adjust many details of your graph, you should
go to the {\sf \underline{P}lot} menu. First, you choose the 
{\sf Plot a\underline{p}pearance...} 
sub menu. A corresponding window will pop up.
Here, you should just unselect the ``{\bf Fill}'' toggle box (upper right
corner), because otherwise the bounding box included in the \verb!.eps!
file will not match the plot and your figure will overwrite other parts
of e.g.\ your manuscript. The fact that your plot has no background now
becomes visible through the appearance of some small dots in the main
{\it xmgrace} window, but this does not disrupt the output when exporting
to \verb!.eps!.

 \begin{figure}[!ht]
\begin{center}
\includegraphics[width=0.6\textwidth]{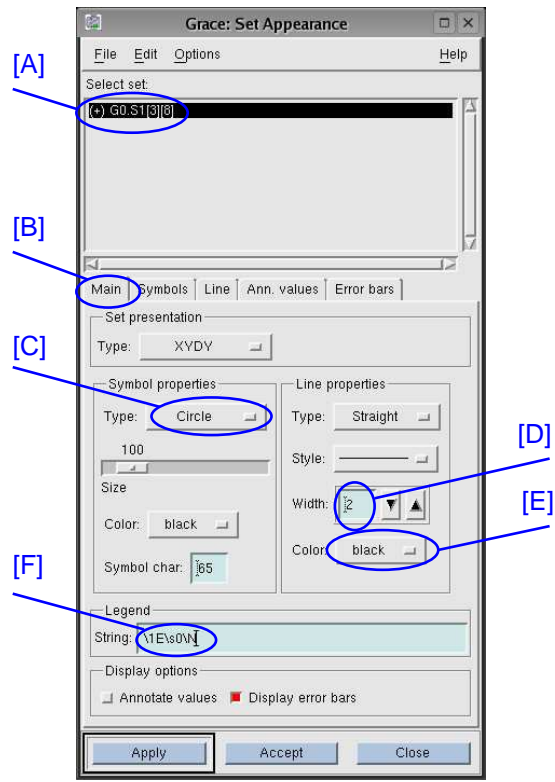}
\end{center}
\caption{The {\sf Grace:Set Appearance} window of the {\it xmgrace}\/ program.
First you have to select the set or sets
 which should be addressed by the changes
{\sf [A]}. Due to the large amount of adjustable parameters, the window
is organized into different tabs. The most import one is ``{\sf{}Main}''
{\sf [B]}, which is shown here.
Among others, you can select a symbol type {\sf [C]} (below: symbol size,
symbol color), choose the width of the lines {\sf [D]} (also: line type,
style) and the color {\sf [E]}. Furthermore, the label for this data
appearing in the legends can be states {\sf [F]}.
\label{fig:xmgrace:set}
}
\end{figure}

 Next, you choose the  {\sf \underline{S}et appearance...} sub menu from the 
{\sf \underline{P}lot} menu. The corresponding window will pop up,
see Fig.\ \ref{fig:xmgrace:set}. You can pop this window also by
double-clicking inside the graph.
  This window allows to change the actual display
style of the data. You have to select the data set or sets  {\sf [A]}
to which the changes  will be
be applied  to when hitting the {\sf Apply} button at the lower left of the
window. Note that the list of sets in this box 
will contain several sets
if you have imported more than one data set. Each of them can 
have (and usually should) its own
style. The box where the list of sets appears is also used to 
administrate the sets. If you hit the right mouse button, while the mouse
pointer is inside this box, a menu will pop up, where you can 
for instance copy or delete sets, hide or unhide them, or rearrange them. 

 The options in this window are arranged within different tabs,
the most important is the ``{\sf Main}'' tab {\sf [B]}. Here 
you can choose whether you want to show symbols for your data points
and which type {\sf [C]}, also the symbol sizes and colors.
If you want to show lines as well ({\sf Line properties} area at the
right), you can choose the {\sf style} like ``{\sf straight}'' and others, but
also ``{\sf none}'' is no lines should be displayed. The style
can be full, dotted, dashed, and different dotted-dashed styles.
 For presentations and publications it is important that lines are well
visible, in this example a line width of 2 is chosen {\sf [D]} and
a black color {\sf [E]}. For presentations you can distinguish
different data sets also by different colors, but for publications
in scientific journals you should keep in mind that the figures
are usually printed in black and white, hence light colors are not
visible.\footnote{Acting as referee reading scientific papers submitted
to journals, I experienced many times
that I could not recognize or distinguish some data because they were obviously
printed in a light color, or with a thin line width, or with tiny symbols 
\ldots.}

 Each data set can have a legend (see below how to activate it). Here, the
legend string can be stated. You can enter it directly, with the
help of some formatting commands which are characters preceded by
a  backslash \verb:\ :. The most important ones are
\begin{itemize}
\item \verb!\\! prints a backslash. \label{page:xmgrace:strings}
\item \verb!\0! selects the {\rm Roman} font, which is also
the default font. A font
is active until a new one is chosen.
\item \verb!\1! selects the {\it italic} font, used in equations. 
\item \verb!\x! selects a symbol font, which contains e.g.\ Greek
characters. For example \verb!\xabchqL! will
generate $\alpha\beta\chi\eta\theta\Lambda$, just to mention some
important symbols.
\item \verb!\s! generates a subscript, while \verb!\N!
switches back to normal. For example
\verb!\xb\s2\N\1(x)! generates $\beta_2(x)$.
\item \verb!\S! generates a superscript, for instance \verb!\1A\S3x\N-5!
 generates $A^{3x}-5$. 

\item The font size can be changed with \verb!\+! and \verb!\-!.

\item With \verb!\o! and \verb!\O! one can start and stop overlining,
respectively, for instance \verb!\1A\oBC\OD! generates $A\overline{BC}D$.
Underlining can be controlled via \verb!\u! and \verb!\U!.

\end{itemize}

  By default, error bars are shown (toggle box lower right corner).
At least you should increase the line width for the symbols
({\sf Symbols} tab) and increase the base and rise line widths for error bars
({\sf Error bars} tab).

 You should know that, when you are creating another plot,
 you do not have to redo all these and other 
adjustments of styles. 
Once you have found your standard, you can
save it using the {\sf Save Parameters...} sub menu from the 
{\sf \underline{P}lot} menu. You can conversely load a parameter set via
the {\sf \underline{L}oad Parameters...} sub menu of the same menu.

 \begin{figure}[!ht]
\begin{center}
\includegraphics[width=0.7\textwidth]{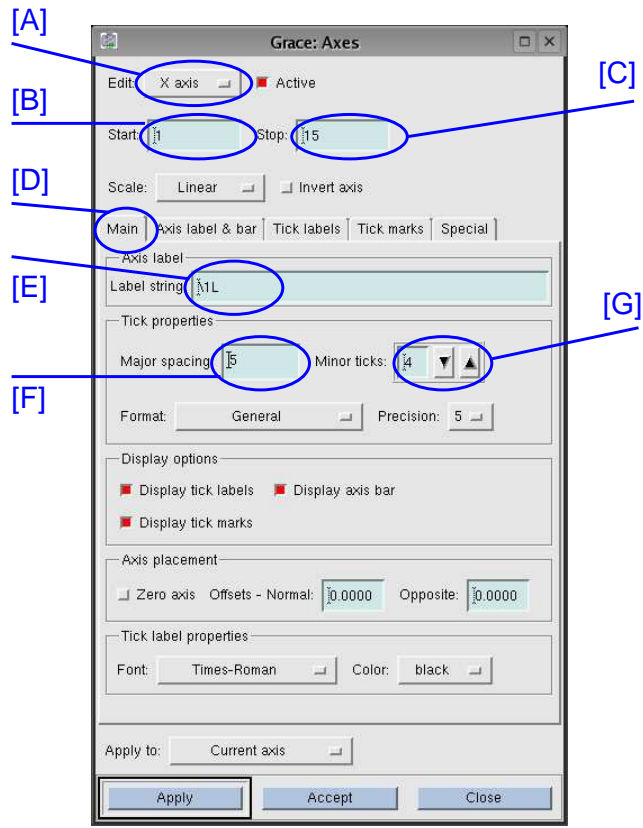}
\end{center}
\caption{The {\sf Grace: Axes} window of the {\it xmgrace}\/ program.
First you have to select the axis 
 which should be addressed by the changes
{\sf [A]}. Among others, you can change the range in the
{\bf Start} {\sf [B]} and {\bf Stop} {\sf [C]} fields.
Here the {\sf Main} tab {\sf [D]} is shown. You can enter an axis label
in the {\sf Label string} field {\sf [E]} and select the spacing of the
major and minor ticks {\sf [F,G]}
\label{fig:xmgrace:axes}
}
\end{figure}

Next, you can adjust the properties of the axes, by choosing 
the  {\sf \underline{S}et appearance...} sub menu from the 
{\sf \underline{P}lot} menu
or by  double-clicking on an axis. The corresponding window will pop up,
see Fig.\ \ref{fig:xmgrace:axes}.
 You have to select the axis where the current changes apply to {\sf [A]}. 
For the $x$ axis you should set the range in the fields 
{\bf Start} {\sf [B]} and {\bf Stop} {\sf [C]}, here to the values
1 and 15. Below these two fields you find the important
{\sf Scale} field, where you can choose linear scaling (default),
logarithmic or reciprocal, to mention the important ones.

The most important adjustments you can perform within the
{\sf Main} tab {\sf [D]}. Here you enter the label shown below the axis
in the  {\sf Label string} field {\sf [E]}. The format of the
string is the same as for the data set legends. Here you enter
just \verb!\1L!, which will show as $L$. The major spacing of the major 
(with labels) and minor ticks can be chosen in the corresponding fields
{\sf [F,G]}. Below there is a {\sf Format field}, where you can
choose how the tick labels are printed. Among the many formats,
the most common are {\sf General} ($1,5,10,\ldots$),
{\sf Exponential} (\verb!1.0e+00!, \verb!5.0e+00!, \verb!1.0e+01!,\ldots),
and {\sf Power}, which is useful for logarithmic scaled axes
($10^1,10^2,10^3,\ldots$). For the tick labels, you can also choose
a {\sf Precision}.
This and other fields of this tab you can leave at their
standard values here.
Nevertheless, you should also adjust the {\sf Char size} of the axis labels 
(tab {\sf Axis label \& bar}) and of the tick labels (tab
{\sf Tick labels}). For publications, character sizes above 150\%
are usually well readable. 
Note that in the  {\sf Axis label \& bar} tab, there is a field 
{\sf Axis transform} where you can enter formulas to transform the
axis more or less arbitrarily, see the manual for details.
All tabs have many other fields, which
are useful as well, but here we stay with the standard
choices. Note that sometimes the {\sf Special} tab is useful, where you can
enter all major and minor ticks individually.

To finish the design of the axes, you can perform similar changes
to the $y$ axis, with {\sf Start} field  $-1.8$, {\sf Stop} field $-1.6$,
{\sf Label string} field \verb!\1E\s0\N(L)! and the same character sizes
as for the $x$ axis for axis labels and tick labels in the corresponding tabs.
Note that the axis label will be printed vertically. If you do not like
this, you can choose the {\sf Perpendicular to axis} orientation
in the {\sf Layout} field of the {\sf Axis label \& bar} tab.

Now you have already a nice graph. To show you some more
of the capabilities of {\it xmgrace}, we refine it a bit.
Next, you generate an inset, i.e.\ a small subgraph inside the main
graph. This is quite common in scientific publications. For this
purpose, you select the {\sf underline{E}dit} menu and there
the {\sf A\underline{r}range graph...} sub menu. The corresponding
window appears. We want to have just one inset, i.e.\ in total 2 graphs.
For this purpose, you select in the {\sf Matrix} region of the window
the {\sf Cols:} field to 1 and the {\sf Rows:} field to 2. Then you hit
on the {\sf Accept button} which applies the changes and closes the window.
You now have two graphs, one containing the already loaded data, the other
one being empty. These two graphs are currently shown next to each other, one
at the top and one at the bottom. 

 \begin{figure}[!ht]
\begin{center}
\includegraphics[width=0.5\textwidth]{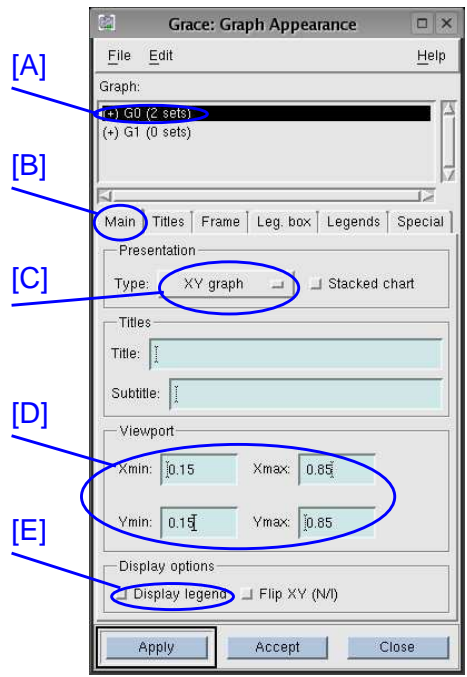}
\end{center}
\caption{The {\sf Grace: Graph Appearance} window of 
the {\it xmgrace}\/ program.
At the top one can select to which graph changes should apply {\sf [A]}.
The window is divided into different tabs {\sf [B]}, here the {\sf Main} tab
is shown. The {\sf Type} of the graph can be selected {\sf [C]}, also
{\sf Title} and {\sf Subtitle} (empty here). The extensions of
the graph can be selected in the {\sf Viewport} area {\sf [D]}.
This allows to make one graph an inset of another.
Using the {\sf Display legend} toggle {\sf [E]} the legend can be switched
on and off.
\label{fig:xmgrace:graph}
}
\end{figure}

To make the second graph an inset of the
first, you choose  the {\sf \underline{G}raph appearance...} sub menu from the 
{\sf \underline{P}lot} menu. At the top a list of the available graphs
is shown {\sf [A]}. Here you select the first graph \verb!G0!.
You need only the {\sf Main} tab {\sf [B]}, other tabs are for changing
styles of titles, frames and legends. We recommend to choose
{\sf Width} 2 in the {\sf Frame} tab. 
In the {\sf Main} tab, you can choose the {\sf Type} of graph {\sf [C]}, 
e.g.\ {\sf XY graph}, which we use here (default),
{\sf Polar graph} or {\sf Pie chart}. You only have to change the 
{\sf Viewport} coordinates {\sf [D]} here. These coordinates
are relative coordinates, i.e.\ the standard full viewport including
axes, labels and titles  is $[0,1]\times[0,1]$.
For the main graph \verb!G0!,
you choose {\sf Xmin} and {\sf Ymin} 0.15 and {\sf Xmax} and {\sf Ymax} 0.85.
Note that below there is a toggle box {\sf Display legend} {\sf [E]},
where you can control whether a legend is displayed. If you want to have
a legend, you can control its position in the {\sf Leg. box} tab.
Now the different graphs overlap. This does not bother you, because
next you select graph \verb!G1! in the list at the top of the window.
We want to have the inset in the free area of the plot, in the upper right
region. Thus, you enter the viewport coordinates  {\sf Xmin} 0.38,
{\sf Ymin} 0.5,  {\sf Xmax} 0.8 and  {\sf Ymax} 0.8.

 \begin{figure}[!ht]
\begin{center}
\includegraphics[width=0.5\textwidth]{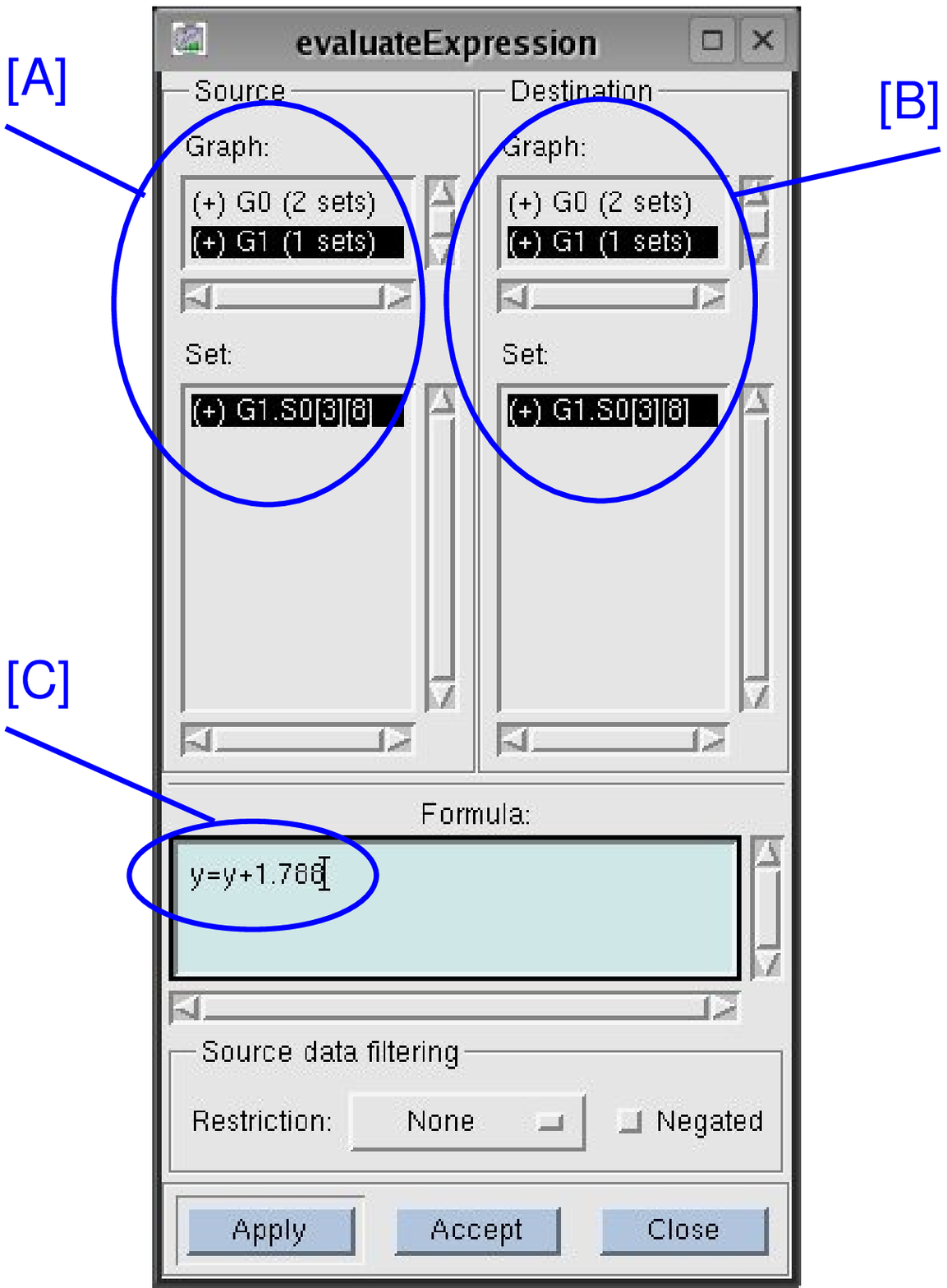}
\end{center}
\caption{The {\sf evaluateExpression} window of 
the {\it xmgrace}\/ program.
At the top you can select {\sf Source} {\sf [A]} and {\sf Destination}
 {\sf [B]}  sets of the transformation. The actual transformation is entered at
the bottom {\sf [C]}.
\label{fig:xmgrace:evaluate}
}
\end{figure}

Now the second graph is well placed, but empty. 
We want to show a scaled version
of the data in the inset. Hence, you import the data again in the same way
as explained above, while
choosing {\sf Read to graph} \verb!G1! in the {\sf Grace: Read sets} window. 
In Sec.\ \ref{sec:fitting}, you will see that the data follows
approximately a power law behavior $e_0(L)=e_{\infty}+aL^b$ with 
$e_{\infty}\approx-1.788$, $a\approx 2.54$ and $b\approx -2.8$. To visualize
this, we want to show $e_0(L)-e_{\infty}$ as a function of $L^b$.
Hence, we want to {\em transform} the data. You choose from
the {\sf \underline{D}ata} menu the {\sf \underline{T}ransformations} sub menu
and there the {\sf \underline{E}valuate expression} sub sub menu. Note that here
 you can
also find many other transformations, e.g.\ Fourier transform, interpolation
and curve fitting. Please consult the manual for details.
In this case, the {\sf evaluateExpression} window pops up, see
Fig.\ \ref{fig:xmgrace:evaluate} (if you did not close
the windows you have used before, your screen will be already pretty
populated). A transformation always takes the data points
from one {\em source} set, applies a formula to all data points (or to a subset
o points) and stores the result in a {\em destination} set. These
sets can be selected at the top of the window in the {\sf Source} {\sf [A]}
and {\sf Destination} {\sf [B]} fields for graph and set separately. 
Note that the
data in the destination set is overwritten. If you want to write the
transformed data to a new set, you can first copy an existing set
(click on the right mouse button in the {\sf Destination Set} window
and choose {\sf Duplicate}). In our case, we want to replace the data,
hence you select for source and destination the data set from graph
\verb!G1!. The transformation is entered below {\sf [C]}, here you
first enter \verb!y=y+1.788! to shift the data. The you hit the
{\sf Apply} button at the bottom. Next you change the transformation to
\verb!x=x^(-2.8)! and hit the
{\sf Apply} button again. When you now select the second graph by clicking
into it, and hit the {\sf AS} (auto scale) button on the left
of the main window, you will see that the data points follow a 
nice straight line in the inset,
which confirms the behavior of the data.

\begin{figure}[!ht]
\begin{center}
\includegraphics[width=0.9\textwidth]{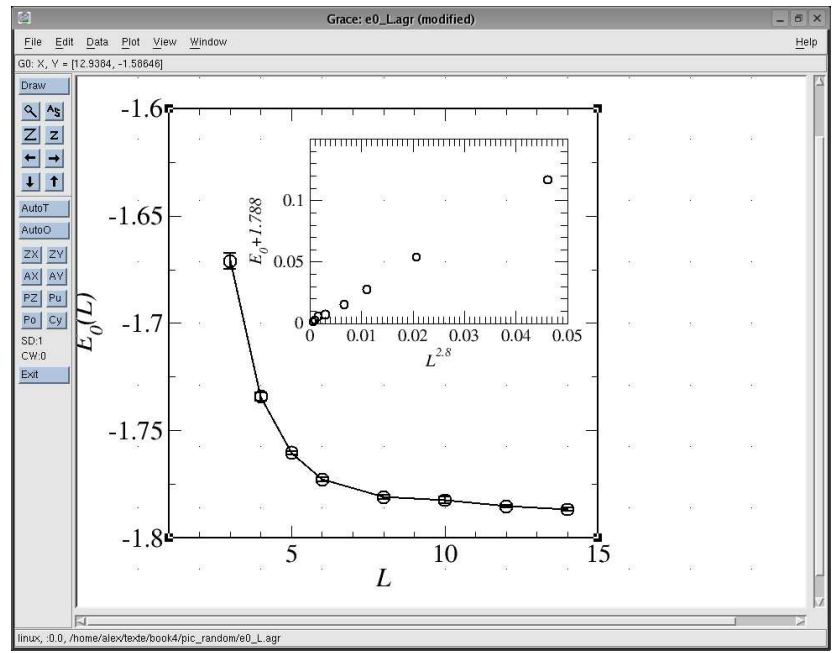}
\end{center}
\caption{The main  {\it xmgrace}\/ window after all adjustments have
been made.
\label{fig:xmgrace:final}
}
\end{figure}

Again you should select symbols, line stiles, and axis labels
for the inset. Usually smaller font sizes are used here. Note that all
operations always apply to the {\em current} graph, which can 
be selected for example 
by clicking near the corners of the boundary boxes of the graph (which does
not always work, depending on which other windows are open) or by double 
clicking on the corresponding graph in the graph list in the
{\sf Grace: Graph Appearance} window. The final main window
is shown in Fig.\ \ref{fig:xmgrace:final}. Note that the left axis label
is not fully visible. This is no problem when exporting the file
as encapsulated postscript; everything will be shown. 
But if you do not like it,
you can adjust the {\sf Xmin} value of graph \verb!G0!.

 Finally, if you choose the menu {\sf \underline{W}indow} and the sub menu
{\sf Drawing \underline{o}bjects} a window will pop up which enable 
 many graphical elements like texts, lines, boxes and ellipses
(again with a variety of choices for colors, styles, sizes etc.)
tobe added/changed and deleted
in your plot. We do not go into details here.

\noindent
\begin{minipage}[t]{0.6\textwidth}
Now you should save your plot using the {\sf File} menu and
the {\sf S\underline{a}ve as...} sub menu, e.g.\ with file
name \verb!sg_e0_L.agr!, where \verb!.agr! is the typical postfix
of {\it xmgrace} source files. When
\end{minipage}
\sourcesB{randomness}{sg\_e0\_L.dat}\\[2.5pt]
you want to create another plot
with similar layout later, it is convenient to start from this saved file
by copying it to a new file 
and subsequently using again {\it xmgrace} to modify the new file.

To export your file as encapsulated postscript, suitable for including
it into presentations or publications (see Sec.\ \ref{sec:publish}),
you have to choose the {\sf File} menu and
the {\sf Prin\underline{t} setup...} sub menu. In the window, which pops
up, you can select the {\sf Device} {\sf EPS}. The file name
will automatically switch to  \verb!sg_e0_L.eps! (this seems not to work
always,
in particular if you edit several files, one after the other, please
check the file names always). Having
hit on the {\sf Accept} button, you can select  the {\sf File} menu and
the {\sf \underline{P}rin\underline{t}} sub menu, which will generate
the desired output file.\footnote{Using the tool {\tt epstopdf}
you can convert the postcript file also to a {\em pdf} file.
With other tools like {\em convert} or {\em gimp} you can
convert to many other styles.}

Now you have a solid base for viewing and plotting, hence
we can continue with advanced analysis techniques. You can experiment
with plotting using {\it xmgrace} in exercise (\ref{ex:xmgrace}).

\index{xmgrace@{\em{}xmgrace}|)}

\index{plotting data|)}
\index{data!plotting|)}

\section{Hypothesis testing and (in-)dependence 
of data \label{sec:hypothesis}}
\index{analyzing data|(}
\index{data!analysis|(}
\index{hypothesis!testing|(}

In the previous section, you have learned how to visualized data,
mainly data resulting from the basic analysis methods
presented in Sec.\ \ref{sec:statistics}. In this section,
we proceed with more elaborate analysis methods.
One important way to analyze data of simulations is to test
hypotheses\index{hypothesis} concerning the results. The hypothesis
to be tested is usually called {\em null hypothesis} H$_0$.
Examples for null hypotheses  are:

\begin{itemize}
\item[(A)] In a traffic system, opening a new track 
will decrease the mean value of the  travel time 
$\overline{t}_{{\rm{}A}\to{\rm{}B}}$ for a 
connection  A$\to$B below a target threshold $t_{\rm target}$.

\item[(B)]  Within an acquaintance network, 
a change of the rules describing how people meet will
change the distribution of the number of people each
person knows.
\label{page:hypotheses}

\item[(C)] The distribution of ground-states 
energies in disordered magnets follows
a Fisher-Tippett distribution.

\item[(D)] Within a model of an ecological system,
the population size of foxes is dependent on the 
population size of beetles.

\item[(E)]  For a protein dissolved in water at
room temperature, adding a certain salt
to the water  changes the structure of the protein.
\end{itemize}

One now can model these situations and use simulations
to address the above questions. 
The aim is to find methods which tell us whetheror not, depending on
the results of the simulations, we should {\em accept}
a null hypothesis. There is no general approach. The way
we can test H$_0$ depends on the formulation of the null hypothesis. 
In any case, our result will again be based on a set of measurements,
such as a sample of independent data points 
$\{x_0,x_1, \ldots, x_{n-1}\}$, formally obtained
by sampling from random variables $\{X_0,X_1, \ldots, X_{n-1}\}$ (here
again, all described by the same distribution function $F_X$).
To get a solid
statistical interpretation, we use a 
{\em test statistics}\index{test statistics}, which is a function
of the sample $t=t(x_0,x_1, \ldots, x_{n-1})$.
Its distribution describes a corresponding random variable $T$. This means,
you can use any estimator (see page \pageref{page:estimator}),
which is also a function of the sample,
as test statistics. Nevertheless, there are many test statistics,
which usually are not used as estimators. 

To get an idea of what a test statistics $t$ may look like, 
 we discuss now test statistics for the above list of examples. 
For (A), one can use obviously the sample mean. This has 
to be compared to the threshold value. This will be performed
 within a statistical interpretation, enabling a null hypothesis
to be accepted or
rejected, see below. For (B) one needs
to compare the distributions of the number of acquaintances before
and after the change, respectively. Comparing two distributions can be done
in many ways. One can just compare some moments, or
define a distance between them based on the difference in area
between the distribution function, just to mention
two possibilities. For discrete random variables,
the mean-squared difference is particularly suitable,
leading to the so-called chi-squared test,
see Sec.\ \ref{sec:chi2}.  For the example (C), the task is similar to (B),
only that the empirical results are compared to a given distribution
and that the corresponding random variables are continuous.
Here, a method based on the maximum distance between two distribution functions
is used widely, called Kolmogorov-Smirnov (KS) test 
(see Sec.\ \ref{sec:KSTest}).
To test hypothesis (D), which means to check for statistical
independence, one can record a two-dimensional
histogram of the population size of foxes and beetles. This is
compared with the distribution where both
populations are assumed to be independent, i.e.\ with the
product of the two single-population distribution functions. Here,
a variant of the chi-squared test is applied, see Sec.\ 
\ref{sec:correlation}.
In the case (E), the sample is not a set of just one-dimensional
numbers, instead the simulation results are conformations of
proteins given by $3N-$dimensional
vectors of the positions $\myvec{r}_i$ $(i=1,\ldots,N)$
of $N$ particles. Here, one could introduce 
a method to compare two protein conformations
$\{\myvec{r}_i^A\}$, $\{\myvec{r}_i^B\}$ in the following way:
First, one  ``moves'' the second protein towards the first one
such that the positions of the center of 
masses agree. Second, one considers the axes through
 the center of masses and through the first
atoms, respectively. One rotates the second protein around its center
of mass such that these
axes become parallel. Third, the second protein is rotated around the
above axis
such that the distances between the last atoms of the two proteins are 
minimized. Finally, for these normalized positions $\{\myvec{r}^{B\star}_i\}$,
one calculates the squared difference of all
pairs of atom positions $d=\sum_i (\myvec{r}^A_i-\myvec{r}^{B\star}_i)^2$
which serves as test function. For a statistical analysis,
the distribution of $d$ for one thermally fluctuating protein
can be determined via a simulation and then compared to the average value
observed when changing the conditions. We do not go into further details here.

The general idea to test a null hypothesis using a test statistics
in a statistical meaningful way is as follows: 
\begin{enumerate}
\item You have to know,
at least to an approximate level, the probability distribution 
function $F_T$ of the test
statistics {\em under the assumption that the null
hypothesis is true}. This is the main step and will be covered in detail
below.
\item You select a certain significance level 
$\alpha$\label{significance level}. Then you calculate
an interval $[a_l, a_u]$ such that the cumulative probability of $T$ 
outside the interval equals to $\alpha$, for instance by distributing
the weight 
equally outside the interval via $F(a_l)=\alpha/2$, $F(a_u)=1-\alpha/2$.
Sometimes one-sided intervals are more suitable, e.g.\ $[\infty,a_u]$
with $F(a_u)=1-\alpha$, see below concerning example (A).

\item You calculate the actual value $t$ of the 
test statistics from your simulation. If
$t\in[a_l,a_u]$ then you {\em accept} the hypothesis, otherwise 
you reject it. Correspondingly, the interval $[a_l,a_u]$ is called 
{\em acceptance interval}\index{acceptance interval}.
\end{enumerate}
Since this is a probabilistic interpretation, there is a small
probability $\alpha$ that you do not accept the null hypothesis,
although it is true. This is called a 
{\em type I error}\index{type I error}\index{error!type I}
(also called {\em false negative}),\index{false negative}
but this error is under control, because $\alpha$ is known.

\begin{figure}[!ht]
\begin{center}
\includegraphics[width=0.6\textwidth]{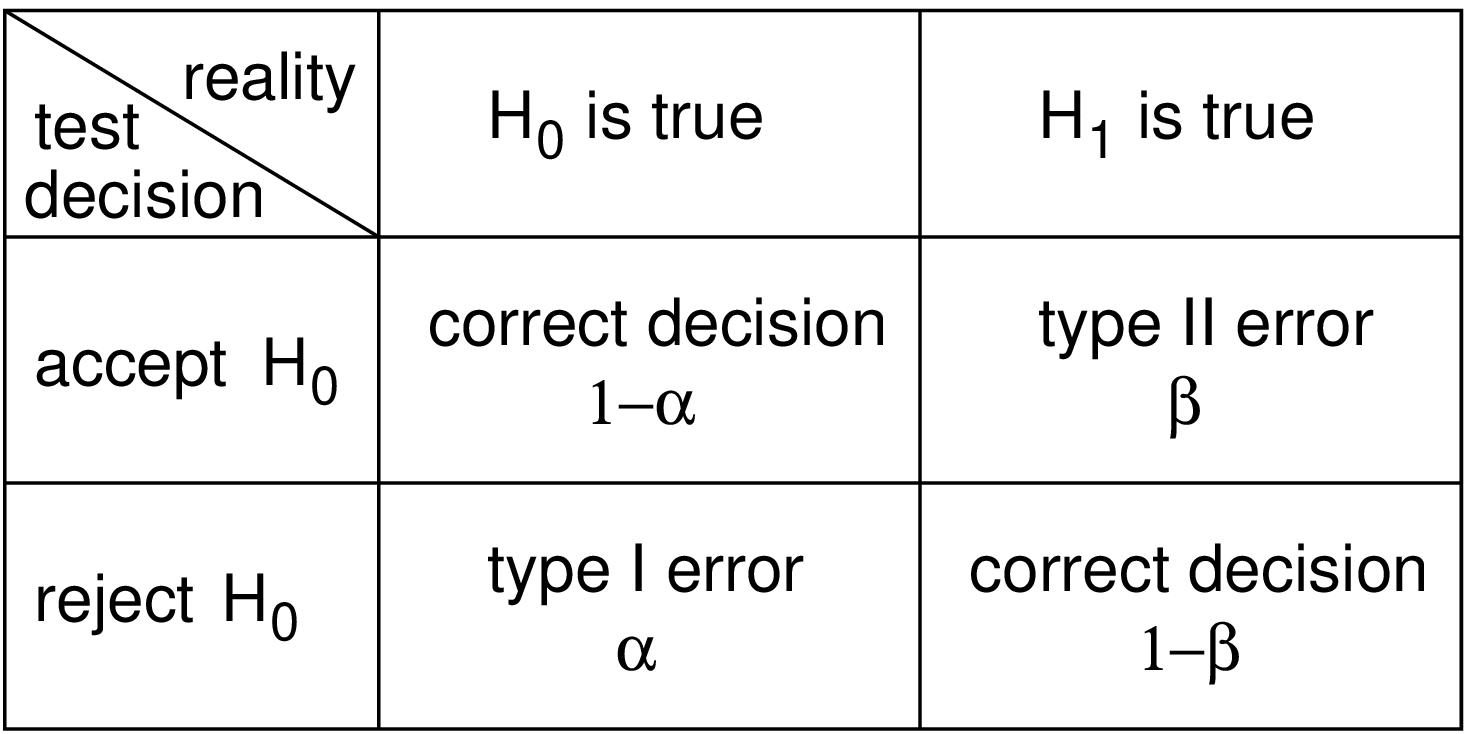}
\end{center}
\caption{The null hypothesis H$_0$ might be true, or the alternative
(usually unknown) hypothesis H$_1$. The test of the null hypothesis
might result in an acceptance or in a rejection. This leads to the
four possible scenarios which appear with the stated probabilities.
\label{fig:errors}}
\end{figure}

On the other hand,
it is important to realize that in general the fact that the value of the
test statistics falls inside the acceptance interval does {\em
not} prove that the null hypothesis is true!  
A different hypothesis H$_1$ could
indeed hold, just your test statistics
is not able to discriminate between the two hypotheses. Or,
with a small probability $\beta$, you might obtain some
value for the test statistics
which is unlikely for H$_1$, but likely for H$_0$. Accepting the
null hypothesis, although it is not true, is called a 
{\em type II error}\index{type II error}\index{error!type II}
(also called {\em false positive})\index{false positive}.
 Usually, H$_1$ is not known, hence $\beta$ cannot 
be calculated explicitly. 
The different cases and the corresponding 
possibilities are summarized in Fig.\ \ref{fig:errors}.
To conclude: If you want to prove a hypothesis H
(up to some confidence level $1-\alpha$), it is better to use
the opposite of H as null hypothesis, if this is possible.

Indeed, in general 
the null hypothesis must be suitably formulated, such that it can be
tested, i.e.\ such that the distribution function $F_T$ describing $T$
can be obtained, at least in principle.
 For example (A), since the test statistics $T$ is a sample mean,
it is safe to assume a Gaussian distribution for $T$: One
can perform enough simulations rather easily, such that
the central limit theorem applies. We use as null hypothesis the
opposite of the formulated hypothesis (A). Nevertheless,
it is impossible to
 calculate an acceptance interval for the Gaussian distribution
based on the assumption that the mean is {\em larger} than a given value.
Here, one can change the null hypothesis, such that instead an expectation
value {\em equal to} 
$t_{\rm target}$ is assumed. Hence, the null hypothesis assumes that the 
test statistics has a 
Gaussian distribution with expectation value $t_{\rm target}$.
The variance of $T$ is unknown, but one can use,
as for the calculation of error bars, the sample variance $s^2$ divided
by $n-1$. Now one calculates on this basis an interval $[a_l,\infty]$
with $F_T(a_l)=\alpha$. Therefore, one rejects the null hypothesis
if $t<a_l$, which happens with probability $\alpha$. On the other hand,
if the true expectation value is even larger than $t_{\rm target}$, then the
probability of finding a mean with $t<a_l$ becomes
even smaller than $\alpha$, i.e.\ less likely. 
Hence, the hypothesis (A) can be accepted or rejected 
 on the basis of a fixed expectation value.

For a general hypothesis test, to evaluate the
distribution of the test statistics $T$, one can perform a 
{\em Monte Carlo simulation}\index{Monte Carlo simulation|ii}. This means
one draws repeatedly samples of size $n$ according to a
distribution $F_X$ determined by the null hypothesis. 
Each time one calculates the test statistics $t$ and records
a histogram of these values (or a sample distribution function
$F_{\hat T}$) which is an approximation of $F_T$. In this way, 
the corresponding 
statistical information can be obtained. To save computing time,
in most cases no Monte Carlo simulations are performed, but some
knowledge is used to calculate or approximate $F_T$.

In the following sections, the cases corresponding to examples (B), (C), (D)
are discussed in detail.  This means, it is explained how one
can test for equality of discrete distributions 
via the chi-squared test and
for equality of continuous distributions via the KS test. Finally,
some methods for testing concerning (in-)dependence of data
and for quantifying the degree of dependence are stated.

\subsection{Chi-squared test \label{sec:chi2}}
\index{chisquaredtest@{chi-squared test}|(ii}
\index{test!chisquared@{chi-squared}|(ii}

The chi-squared test is a method to compare histograms and 
discrete probability distributions. 
The test works also for discretized (also called 
{\em binned})\index{binned probability distributions}%
\index{probability!distribution!binned}
continuous probability distributions, where the probabilities
are obtained by integrating the pdf over the different bins.
The test comes in two variants: 
\begin{itemize}
\item 
Either you want to compare the
histogram $\{h_k\}$ for bins $B_k$ (see Sec.\ \ref{sec:histogram})
describing the sample $\{x_0,x_1,$ \ldots, $x_{n-1}\}$
to a given  discrete or discretized probability mass function
with probabilities $\{p_k\}=\Prob(x\in B_k)$. The null 
hypothesis H$_0$ is: 
 ``{\em the sample follows a distribution given by $\{p_k\}$''.}

Note that the probabilities are fixed and independent of the data sample.
If the probabilities are parametrized and the parameter
is determined by the sample (e.g.\ by the mean of the data) such that
the probabilities fit the data best, related methods as described in Sec.\ 
\ref{sec:fitting} have to be applied.

\item Alternatively, you want to compare two histograms 
$\{h_k\},\,\{\hat h_k\}$
obtained from two different samples $\{x_0,x_1,$ \ldots, $x_{n-1}\}$
and $\{\hat x_0, \hat x_1,$ \ldots, $\hat x_{n-1}\}$
defined for the same bins $B_k$. The null hypothesis H$_0$ is:
 ``{\em the two samples follow the same  
distribution''}.\footnote{Note that here we 
assume that the two samples have the same
size, which is usually easy to achieve in simulations. A different
case occurs when also the number of sample points is a random
variable, hence a difference in the number of sample points makes
the acceptance of H$_0$ less likely, see \cite{PRA-numrec1995}.}
\end{itemize}

In case the test is used to compare intrinsically discrete data,
the intervals $B_k$ can  conveniently be 
chosen such that each possible outcome
corresponds to one interval. Note that due to the binning process,
the test can be applied to high-dimensional data as well, 
where the sample is a set of vectors. Also non-numerical
data can be binned.
In these cases each bin
represents either a subset of the high-dimensional space
or, in general, a subset of the possible outcomes. 
 For simplicity,
we restrict ourselves here to one-dimensional numerical samples.

\begin{figure}[!ht]
\begin{center}
\includegraphics[width=0.8\textwidth]{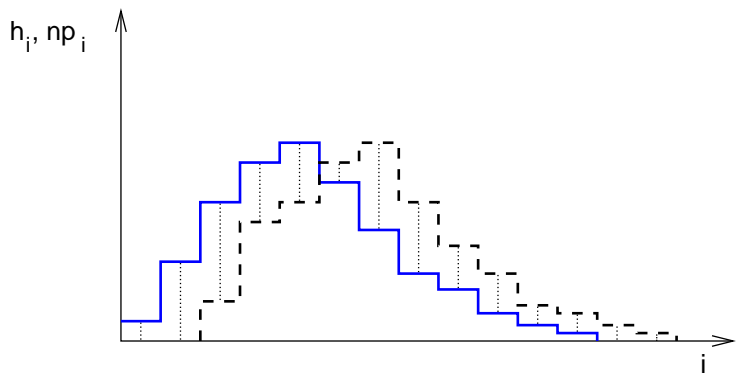}
\end{center}
\caption{Chi-squared statistics: A histogram (solid line) 
is compared to a discrete
probability distribution (dashed line). 
For each bin, the sum of the squared differences of the bin counter
$h_k$ to the expected number of counts $n p_k$ is calculated (dotted 
vertical lines), see Eq.\ (\ref{ex:chi2A}).
 In this case, the differences are quite notable,
thus the probability that the histogram was obtained via random
experiments from a random variable described by the probabilities $\{p_k\}$
(null hypothesis) will be quite small.
\label{fig:chi2Example}}
\end{figure}

We start with the first case, where a sample histogram is compared
to a probability distribution,  corresponding to example (C) on page 
\pageref{page:hypotheses}. The test statistics, called
$\chi^2$, is defined as:
\begin{equation}
\chi^2 = {\sum_k}^\prime\frac{(h_k-np_k)^2}{np_k}\,
\label{ex:chi2A}
\end{equation}
with $np_k$ being the expected number of sample points in bin $B_k$.
The prime at the sum symbol indicates that bins with $h_k=np_k=0$
are omitted. The number of contributing bins is denoted by $K^\prime$.
If the pmf $p_k$ is nonzero for an infinite
number of bins,  the sum is truncated for terms 
$n p_k\ll 1$. This means
that the number of contributing bins will be always finite.
Note that bins exhibiting $h_k>0$ but $p_k=0$ are not omitted. This
results in an infinite value of $\chi^2$, which is reasonable, because
for data with $h_k>0$ but $p_k=0$, the data cannot be described
by the probabilities $p_k$.

The chi-squared distribution with $\nu=K^\prime-1$ degrees
of freedom  (see Eq.\ (\ref{eq:chi2})) describes
the chi-squared test statistics, 
if the number of bins and the number of bin entries is large.
 The term $-1$ in the number of degrees of freedom
comes from the fact that the total number
of data points $n$ is equal to the total number of expected data points 
$\sum_k n_k p_k = n \sum_k p_k = n$, hence the $K'$ different summands
are not statistically independent.
The probability density of the chi-squared distribution 
is given in Eq.\ (\ref{eq:chi2}). To perform the actual test,
it is recommended to use the implementation in the {\em GNU scientific library}
(GSL) \index{GNU scientific library}
(see Sec.\ \ref{sec:gsl}).


\noindent
\begin{minipage}[t]{0.6\textwidth}
Next, a C function \verb!chi2_hd()! 
is shown which calculates the cumulative probability 
({\em p-value}) \index{p-value|ii} that
a value of $\chi^2$ or larger is obtained,
given the null hypothesis that the
\end{minipage}
\sourcesB{randomness}{chi2.c}\\[3pt]
 sample was 
generated
using the probabilities $p_k$.
Arguments of \verb!chi2_hd()! 
are the number of bins, and two arrays \verb!h[]! and \verb!p[]!
containing the histogram $h_k$ and the probabilities $p_k$, respectively:


{\small 
\linenumbers[1]
\begin{verbatim}
double chi2_hd(int n_bins, int *h, double *p)
{
  int n;                      /* total number of sample points */
  double chi2;                                  /* chi^2 value */
  int K_prime;                  /* number of contributing bins */
  int i;                                            /* counter */

  n = 0;
  for(i=0; i<n_bins; i++)
    n += h[i];      /* calculate total number of sample_points */
  
  chi2 = 0.0; K_prime = 0;
  for(i=0; i<n_bins; i++)                   /* calculate chi^2 */
  {
    if(p[i] > 0)
    {
      chi2 += (h[i]-n*p[i])*(h[i]/(n*p[i])-1.0);
      K_prime ++;
    } 
    else if(h[i] >0)       /* bin entry for zero probability ? */
    {
      chi2 = 1e60;
      K_prime ++;
    }
  }
  return(gsl_cdf_chisq_Q(chi2, K_prime-1));
}
\end{verbatim}
\nolinenumbers} 
\noindent
First, in lines 8--10, the total number of sample points
is obtained from summing up all histogram entries. In the main loop,
lines 12--25, the value of $\chi^2$ is calculated. In parallel,
the number of contributing bins is determined. Finally (line 26)
the p-value is obtained using the GSL function
\verb!gsl_cdf_chisq_Q()!. This p-value can be compared with
the significance level $\alpha$. If the the p-value is larger, the
null hypothesis is accepted, otherwise rejected.

Note that the result for the p-value 
clearly depends on the number of bins, and,
if applicable, on the actual choice of bins. Nevertheless,
all reasonable choices, although maybe leading to somehow different 
numerical results, will lead 
 to the same decisions concerning the null hypothesis in most cases.

Next, we consider the case, where we want to compare two
histograms $\{h_k\},\{\hat h_k\}$ corresponding to example (B) on page 
\pageref{page:hypotheses}. In this case the $\chi^2$ statistics reads 
\begin{equation}
\chi^2 = {\sum_k}^\prime\frac{(h_k-\hat h_k)^2}{h_k+\hat h_k}\,
\label{eq:chi2B}
\end{equation}
The sum runs over all bins where $h_k\neq 0$ or $\hat h_k \neq 0$,
and $K^\prime$ being the corresponding number of contributing bins.
Consequently, the bins which should be included are uniquely defined,
in contrast to the case where a histogram is compared to a distribution
defined for infinitely many outcomes.
Note that in the denominator the sum of the bin entries occurs, not
the average. The reason is that the chi-squared distribution is a
sum of standard Gaussian distributed numbers (variance 1) and here,
where the differences of two (approximately) Gaussian quantities are
taken, the resulting variance is the sum of the individual variances,
approximated roughly by the histogram entries.
To calculate the p-value, again the chi-squared distribution with 
$\nu=K^\prime-1$ degrees of freedom is to be applied.
Here, no C implementation is shown, rather we refer the reader
to exercise (\ref{ex:chi2}).

\index{chisquaredtest@{chi-squared test}|)}
\index{test!chisquared@{chi-squared}|)}

\subsection{Kolmogorov-Smirnov test \label{sec:KSTest}}
\index{Kolmogorov-Smirnov test|(}
\index{test!Kolmogorov-Smirnov|(}

Next, we consider the case where the statistical properties
of a sample  $\{x_0,x_1,$ \ldots, $x_{n-1}\}$, obtained from a
repeated experiment using a
 continuous random variable, is to be compared 
to a given distribution function $F_X$.
One could, 
in principle, compare  a histogram and a correspondingly binned
probability distribution using the chi-squared test
explained in the previous section. Unfortunately, the binning is artificial
and has an influence on the results (imagine few very large bins).
Consequently, the method presented in this section is usually preferred,
since it requires no binning. Note  that if the distribution
function  is parametrized and if the parameter
is determined by the sample (e.g.\ by the mean of the data) such that
the $F_{X}$ fits the data best, the methods from Sec.\ 
\ref{sec:fitting} have to be applied.

The basic idea of the {\em Kolmogorov-Smirnov} test
is to compare the distribution function to the
empirical sample distribution function 
$F_{\hat X}$\index{sample!distribution function}%
\index{distribution function!sample} defined
in Eq.\ (\ref{eq:FhatX}). Note that $F_{\hat X}(x)$ is piecewise constant
with jumps of size $1/n$ at the positions $x_i$ (assuming that
each data point is contained uniquely in the sample).

Here again, one has several choices for the test
statistics. For instance, one could calculate the area between
$F_X$ and $F_{\hat X}$. Instead, usually just the maximum difference between
the two functions is used:
\begin{equation}
d_{\max} \equiv \max_x \left|F_X(x)-F_{\hat X}(x)\right|
\label{eq:Dmax}
\end{equation}

\begin{figure}[!ht]
\begin{center}
\includegraphics[width=0.8\textwidth]{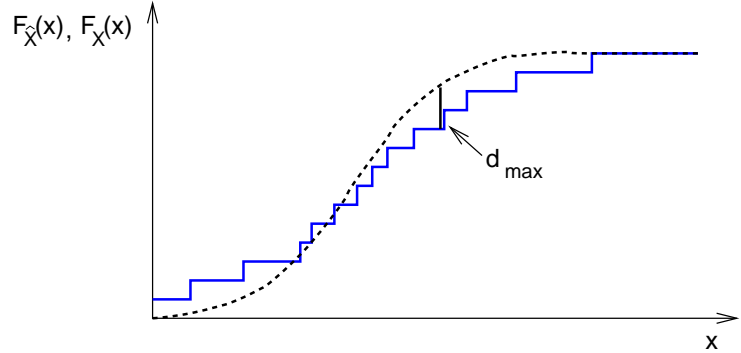}
\end{center}
\caption{Kolmogorov-Smirnov test: A sample
distribution function (solid line) 
is compared to a given 
probability distribution function (dashed line). 
The sample statistics $d_{\max}$ 
is the maximum difference between the two functions.
\label{fig:KSExample}}
\end{figure}

Since the sample distribution function changes only at the sample points,
one has to perform the comparison just before and just after
the jumps. Thus, Eq.\ (\ref{eq:Dmax}) is equivalent to

$$
d_{\max} \equiv \max_{x_i} \left\{\left|F_X(x_i)-1/n-F_{\hat X}(x_i)\right|,
                   \left|F_X(x_i)-F_{\hat X}(x_i)\right|\right\}\,
$$
This sample statistics is visualized in Fig.\ \ref{fig:KSExample}.
 
 The p-value, i.e.\ the probability of a value of $d_{\max}$
as measured ($d_{\max}^{\rm measured}$)
 or worse, given the null hypothesis\index{null hypothesis}
 that the sample
is drawn from $F_X$, is approximately given by (see
\cite{PRA-numrec1995} and references therein):
\begin{equation}
\label{eq:KS:probab}
\Prob(d_{\max} \ge d_{\max}^{\rm measured})
= Q_{\rm KS}\left( [\sqrt{n}+0.12+0.11/\sqrt{n}] d_{\max}^{\rm measured}
 \right)
\end{equation}
This approximation is already quite good for $n\ge 8$.
Here, the following auxiliary
probability function is used:
\begin{equation}
  \label{eq:QKS}
  Q_{\rm KS}(\lambda) = 2\sum_{i=1}^{\infty} (-1)^{i+1} e^{-2i^2\lambda^2}\,
\end{equation}
\sources{randomness}{ks.c}
with $Q_{\rm KS}(0)=1$ and $Q_{\rm KS}(\infty)=0$.
This function can be implemented most easily by a direct summation
\cite{PRA-numrec1995}. The function \verb!Q_ks()! 
receives the value of $\lambda$
as argument and returns $Q_{\rm KS}(\lambda)$:


{\small 
\linenumbers[1]
\begin{verbatim}
double Q_ks(double lambda)
{
  const double eps1 = 0.0001;  /* relative margin for stop */
  const double eps2 = 1e-10;   /* relative margin for stop */
  int i;                                   /* loop counter */
  double sum;                               /* final value */
  double factor;            /* constant factor in exponent */
  double sign;                               /* of summand */
  double term, last_term;        /* summands, last summand */

  sum = 0.0; last_term = 0.0; sign = 1.0;    /* initialize */
  factor = -2.0*lambda*lambda;
  for(i=1; i<100; i++)                           /* sum up */
  {
    term = sign*exp(factor*i*i);
    sum += term;
    if( (fabs(term) <= eps1*fabs(last_term)) || 
        (fabs(term) <= eps2*sum))
      return(2*sum);
    sign =- sign;
    last_term = term;
  }
  return(1.0);                /* in case of no convergence */
}
\end{verbatim}
\nolinenumbers} 
\noindent
The summation (lines 13--22) is performed for at most 100 iterations.
If the current term is small compared to the previous one or
very small compared to the sum obtained so far, the summation
is stopped (line 17--18). If this does not happen within 100 iterations, 
the sum has not converged (which means
$\lambda$ is very small) and $Q(0)=1$ is returned.

This leads to the following C implementation for the KS test. The
function \verb!ks()! expects as arguments the number of sample points
\verb!n!, the sample \verb!x[]! and a pointer \verb!F!
to the distribution function:


{\small 
\linenumbers[1]
\begin{verbatim}
double ks(int n, double *x, double (*F)(double))
{
  double d, d_max;                    /* (maximum) distance */
  int i;                                    /* loop counter */
  double F_X;            /* empirical distribution function */
  
  qsort(x, n, sizeof(double), compare_double);

  F_X = 0; d_max = 0.0;
  for(i=0; i<n; i++)                    /* scan through F_X */
  {
    d = fabs(F_X-F(x[i]));   /* distance before jump of F_X */
    if( d> d_max)
      d_max = d;
    F_X += 1.0/n;
    d = fabs(F_X-F(x[i]));    /* distance after jump of F_X */
    if( d> d_max)
      d_max = d;
  }
  return(Q_ks( d_max*(sqrt(n)+0.12+0.11/sqrt(n))));
}
\end{verbatim}
\nolinenumbers} 
\noindent
First the sample is sorted (line 7). This allows for a simple
implementation of the sample distribution function, because
at each sample data point, in the order of occurrence, the
value of $F_{\hat X}$ is increased by $1/n$. When obtaining the
maximum distance (lines 10--19), one has to compare $F_{\hat X}$
to the distribution function $F_X$ just before (lines 12--14)
and after (lines 15--18) the jumps.
Note that this implementation works also for samples, where some data points
occur multiple times.

For the actual test, one calculates the p-value for the given sample
using \verb!ks()!. If the p-value exceeds the indented significance
level $\alpha$, the null hypothesis is accepted, i.e.\
the data is compatible with the distribution with high probability.
Usually quite small significances are used, e.g.\ $\alpha=0.05$.
This means that even substantial values of $d_{\max}$ are accepted. Thus,
one rejects the null hypothesis only, as usual, in case the
probability for an error of type I is quite small.

It is also possible to compare two samples of sizes $n_1,n_2$
via the KS test. The test
statistics for the two sample distribution functions is again
the maximum distance. The probability to find a value of $d_{\max}$
as obtained or worse, given the null hypothesis\index{null hypothesis}
 that the samples
are drawn from the same distribution, is as above in Eq.\
(\ref{eq:KS:probab}), only one has to replace $n$ by the
``effective'' sample size $n_{\rm eff}= n_1n_2/(n_1+n_2)$,
for details see \cite{PRA-numrec1995} and references therein.
It is straightforward to implement this test when using
the C function \verb!ks()! shown above as template.

\index{Kolmogorov-Smirnov test|)}
\index{test!Kolmogorov-Smirnov|)}

\subsection{Statistical (in-)dependence \label{sec:correlation}}

\index{statistical dependence|(}

Here, we consider samples, which consist of pairs $(x_i,y_i)$
($i=0,1,\ldots,n-1$)
of data points. Generalizations to higher-dimensional data
is straightforward. The question is, whether the $y_i$ values
depend on the $x_i$ values (or vice versa). In this case, 
one also says that they are
{\em statistically related}. If yes, this means
that if we know one of the two values, we can predict the other
one with higher accuracy. The formal definition of statistical
(in-) dependence was given in Sec.\ \ref{sec:introProb}.
An example of statistical dependence occurs in weather
simulations: The amount of snowfall is statistically related 
to the temperature: If it is too warm or too cold, it will
not snow. This also shows, that the dependence of two variables
it not necessarily monotonous. In case one is interested in monotonous
and even linear dependence, one usually says that
the variables are {\em correlated}, see below.

\begin{figure}[!ht]
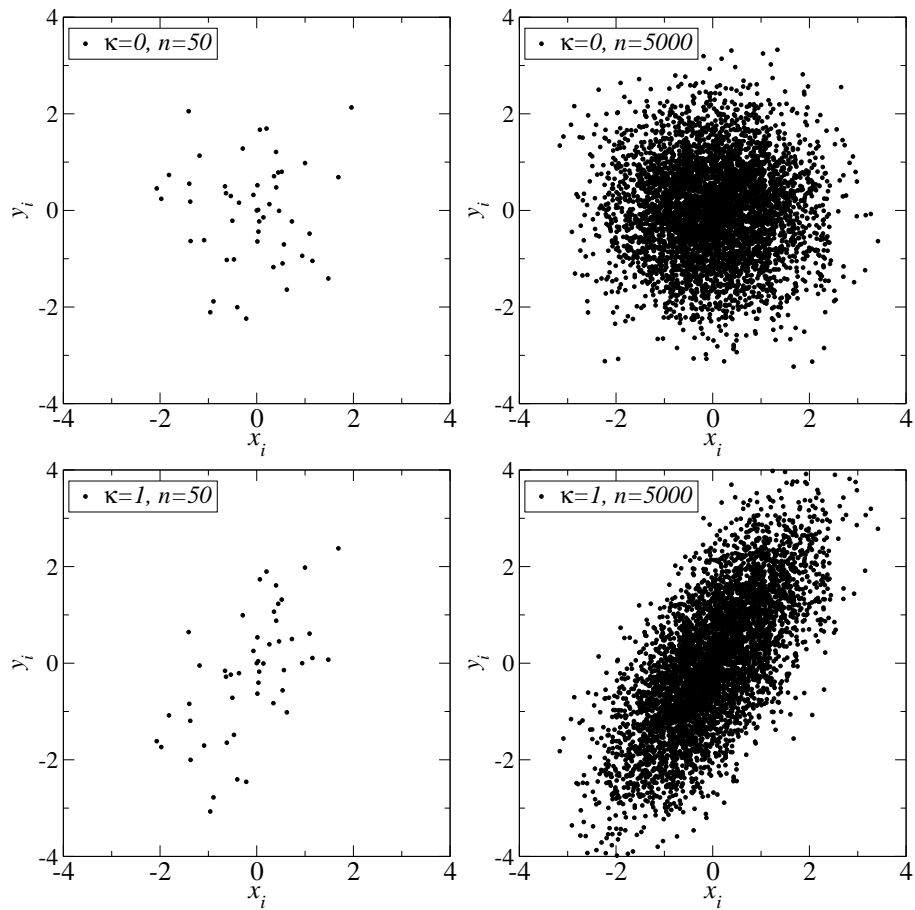

\begin{center}
\includegraphics[width=0.49\textwidth]{pic_random_small/points0A.eps}
\includegraphics[width=0.49\textwidth]{pic_random_small/points0B.eps}

\includegraphics[width=0.49\textwidth]{pic_random_small/points1A.eps}
\includegraphics[width=0.49\textwidth]{pic_random_small/points1B.eps}
\caption{Scatter plots for $n$ data points $(x_i,y_i)$ where
the $x_i$
numbers are generated from a standard Gaussian distribution
(expectation value 0, variance 1), while each $y_i$ number is drawn
from a Gaussian distribution with expectation value $\kappa x_i$ (variance 1).
\label{fig:correlation}}
\end{center}
\end{figure}

\sourcesTOPC{randomness}{points0A.dat\\points0B.dat\\points1A.dat\\points1B.dat}{7}
It is important to realize that we have to distinguish between
statistical {\em significance} of a statistical dependence
 and the {\em strength} of the dependence. 
Say that our test tells us that the $x$ values are statistically related
with high probability. This usually just means that
we have a large sample.
On the other hand, the strength of the statistical dependence can be still
small.
It could be, for eaxample, 
that a given value for $x$ will influence the probability
distribution for $y$ only slightly. One the other hand,
the strength can be large, which means, for example, knowing $x$
almost determines $y$. But if we have only few samples points,
we cannot be very sure whether the data points are related or not. 
Nevertheless, there is some
connection: the larger the strength,
 the easier it is to show that the dependence is
significant.
For illustration consider a sample where the $x_i$
numbers are generated from a standard Gaussian distribution
(expectation value 0, variance 1), while each $y_i$ number is drawn
from a Gaussian distribution with expectation value $\kappa x_i$ 
(variance 1).\footnote{This is an example, where the random variables
$Y_i$ which described the sample are not identical.}
Hence, if $\kappa=0$, the data points are independent.
Scatter plots, where each sample point ($x_i,y_i$) is shown as
dot in the $x-y$ plane are exposed in Fig.\ \ref{fig:correlation}.
Four possibilities are presented, 
$k=0/1$ combined with $n=50/5000$. 
Below, we will also present what the methods we use here will tell us about
these data sets.

In this section, first a variant of the chi-squared 
test is presented, which enables us to check
whether data is independent. Next, the {\em linear correlation
coefficient} is given, which states the strength of linear
correlation. Finally, it is discussed how one can quantify the
dependence {\em within} a sample, for example between sample
points $x_i,x_i+\tau$.

To test statistical dependence for a sample $\{(x_0,y_0),$
$(x_1,y_1),$ \ldots, $(x_{n-1},y_{n-1})\}$, one considers
usually the null hypothesis: H$_0=$ ``The $x$ sample points and the $y$
sample points are independent.'' To test H$_0$ one puts
the pairs of sample points into two-dimensional histograms
$\{h_{kl}\}$. The counter $h_{kl}$ receives a count,
if for data point $(x_i,y_i)$ we have $x_i\in B^{(x)}_k$ and 
$y_i \in B^{(y)}_l$, for suitably determined bins
$\{ B^{(x)}_k\}$ and $\{B^{(y)}_l\}$. 
Let $k_x$ and $k_y$ the number of bins in $x$ and $y$ direction,
respectively. Next,
one calculates single-value (or one-dimensional) 
histograms $\{\hat h^{(x)}_k\}$ and
$\{\hat h^{(y)}_l\}$ defined by
\begin{eqnarray}
\hat h^{(x)}_k & = & \sum_l h_{kl} \nonumber \\
\hat h^{(y)}_l & = & \sum_k h_{kl} \label{eq:rowColum}
\end{eqnarray}
These one-dimensional histograms describe how many counts in a certain
bin arise for one variable, regardless of the value of the
other variable. It is assumed that all entries of these histograms
are not empty. If not, the bins should be adjusted accordingly.
Note that $n=\sum_k \hat h^{(x)}_k = \sum_l \hat h^{(y)}_l
= \sum_{kl} h_{kl}$ holds. 

Relative frequencies, which
are estimates of probabilities, are obtained by normalizing with $n$,
i.e.\ $\hat h^{(x)}_k/n$ and $\hat h^{(y)}_l/n$. If the two variables
$x_i,y_i$
are independent, then the relative frequency to obtain a pair
of values $(x,y)$ in bins $\{ B^{(x)}_k\}$ and $\{B^{(y)}_l\}$
should be the product of the single-value relative frequencies.
Consequently, by multiplying with $n$ one obtains the corresponding
expected  number $n_{kl}$ of counts, under the assumption that H$_0$ holds:
\begin{equation}
n_{kl} = n \frac{\hat h^{(x)}_k}{n}\frac{\hat h^{(y)}_l}{n} =
\frac{\hat h^{(x)}_k\hat h^{(y)}_l}{n}
\end{equation}
\begin{sloppypar}
These expected numbers are compared to the actual numbers 
in the two-dimensional histogram $\{h_{kl}\}$ via the
$\chi^2$ test statistics, comparable to Eq.\ (\ref{ex:chi2A}):
\end{sloppypar}

\begin{equation}
\chi^2 = \sum_{kl} \frac{(h_{kl}-n_{kl})^2}{n_{kl}}
\end{equation}
The statistical interpretation of $\chi^2$ is again provided by
the chi-squared distribution. The number of degrees of freedom is
determined by the number of bins ($k_xk_y$) in the two-dimensional histogram
minus the number of constraints. The constraints are given
by Eq.\ (\ref{eq:rowColum}), except that the total number of counts being
$n$ is contained twice, resulting in $k_x+k_y-1$. Consequently,
the number of degrees of freedom is 
\begin{equation}
\nu=k_x k_y-k_x-k_y+1\,.
\end{equation}
Therefore, under the 
assumption that the $x$ and $y$ sample points are independent,
$p=1-F(\chi^2, \nu)$  gives the probability (p-value) of observing a 
test statistics of $\chi^2$ or larger. 
$F$ is here the distribution function of the chi-square distribution,
see  Eq.\ (\ref{eq:chi2}).
This p-value\index{p-value} has to be compared to the
significance level $\alpha$. If $p<\alpha$, the null hypothesis is rejected.

\sources{randomness}{chi2indep.c}
The following C function implements the chi-squared independence test 
\verb!chi2_indep()!. It receives the number of bins
in $x$ and $y$ direction as arguments, as well as a two-dimensional array,
which carries the histogram:

{\small 
\linenumbers[1]
\begin{verbatim}
double chi2_indep(int n_x, int n_y, int **h)
{
  int n;                      /* total number of sample points */
  double chi2;                                  /* chi^2 value */
  int k_x, k_y;                 /* number of contributing bins */
  int k, l;                                        /* counters */
  int *hx, *hy;                  /* one-dimensional histograms */

  hx = (int *) malloc(n_x*sizeof(int));            /* allocate */
  hy = (int *) malloc(n_y*sizeof(int));

  n = 0;            /* calculate total number of sample_points */
  for(k=0; k<n_x; k++)
    for(l=0; l<n_y; l++)
      n += h[k][l]; 
  
  k_x = 0;                  /* calculate 1-dim histogram for x */
  for(k=0; k<n_x; k++)
  {
    hx[k] = 0;
    for(l=0; l<n_y; l++)
      hx[k] += h[k][l];
    if(hx[k] > 0)                   /* does x bin contribute ? */
      k_x++;
  }

\end{verbatim}
\newpage
\begin{verbatim}
  k_y = 0;                  /* calculate 1-dim histogram for y */
  for(l=0; l<n_y; l++)
  {
    hy[l] = 0;
    for(k=0; k<n_x; k++)
      hy[l] += h[k][l];
    if(hy[l] > 0)                   /* does y bin contribute ? */
      k_y++;
  }

  chi2 = 0.0; 
  for(k=0; k<n_x; k++)                      /* calculate chi^2 */
    for(l=0; l<n_y; l++)
      if( (hx[k] != 0)&&(hy[l] != 0) )
        chi2 += pow(h[k][l]-(double) hx[k]*hy[l]/n, 2.0)/
          ((double) hx[k]*hy[l]/n);

  free(hx);
  free(hy);
  return(gsl_cdf_chisq_Q(chi2, k_x*k_y - k_x -k_y + 1));
}
\end{verbatim}
\nolinenumbers} 
\noindent
First, the one-dimensional histograms are allocated (lines 9--10).
Then the total number of counts, i.e.\ the sample size, is calculated
(lines 12--15). In lines 17--26, the one-dimensional histogram for
the $x$ direction is obtained. Also the effective number of bins
in that direction is calculated. In lines 27--35, the same happens
for the $y$ direction. The actual value of the $\chi^2$ test
statistics is determined in lines 37--42. Finally,
the allocated memory is freed (lines 44-45) and the p-value
calculated (line 46), again  the GSL function
\verb!gsl_cdf_chisq_Q()! is used.

The p-values\index{p-value}
 for the sample sets shown in Fig.\ \ref{fig:correlation}
are as follows: $p(\kappa=0,n=50)=0.077$, $p(\kappa=0,n=5000)=0.457$,
$p(\kappa=1,n=50)=0.140$, $p(\kappa=1,n=5000)<10^{-100}$. Hence,
the null hypothesis of independence would not be rejected
(say $\alpha=0.05$) for the case $\kappa=1,n=50$, which is actually
correlated. On the other hand, if the number of samples is large
enough, there is no doubt.

Once it is established that a sample contains dependent data,
one can try to measure the strength of dependence.
A standard way is to use
the {\em linear correlation coefficient}%
\index{linear correlation coefficient|ii} 
(also called {\em Pearson's $r$})\index{pearsonsr@{Pearson's $r$}}
given by  
\begin{equation}
\label{eq:linearCorrelation}
r \equiv \frac{\sum_i (x_i-\overline{x})(y_i-\overline{y})}
             {\sqrt{\sum_i (x_i-\overline{x})^2} 
              \sqrt{\sum_i (y_i-\overline{y})^2}}\,.
\end{equation}
This coefficient assumes, as indicated by the name, that
a linear correlation exists within the data. The implementation
using a C function is straight forward, see exercise 
(\ref{ex:lcc}). For the data shown in Fig.\ \ref{fig:correlation},
the following correlation coefficients are obtained:
 $r(\kappa=0,n=50)=0.009$, $r(\kappa=0,n=5000)=0.009$,
$r(\kappa=1,n=50)=0.653$, $r(\kappa=1,n=5000)=0.701$.
Here, also in the two cases, where the statistics is low, the
value of $r$ reflects whether or not the data is correlated. Nevertheless,
this is only the case because we compare strongly correlated data
to uncorrelated data. If we compare weakly but significantly
correlated data, we will still get a small value of $r$. Hence,
to test for significance, it is better to use the hypothesis test based
on the $\chi^2$ test statistics.

\begin{sloppypar}
Finally, note that a different type of correlation may arise: So far it was
always assumed that the different sample points $x_i,x_j$ 
(or sample vectors) are statistically 
independent of each other. Nevertheless,
it could be the case, for instance, that the sample is generated using
a Markov chain Monte Carlo
 simulation \cite{newman1999,landau2000,robert2004,liu2008}, 
where each data point $x_{i+1}$
is calculated using some random process, but also depends on the previous 
data point $x_{i}$, hence $i$ is a kind of artificial sample time of
the simulation. This dependence decreases with growing time distance between
sample points.
One way to see how quickly this dependence decreases is
to use a variation of the correlation coefficient 
Eq.\ (\ref{eq:linearCorrelation}), i.e.\ a {\em correlation function}:
\end{sloppypar}

\begin{eqnarray}
\tilde C(\tau) & = & 
\frac{1}{n-\tau}\sum_{i=0}^{n-1-\tau} x_i x_{i+\tau} \nonumber\\
& & - \left(\frac{1}{n-\tau}\sum_{i=0}^{n-1-\tau}x_i\right) \times
 \left( \frac{1}{n-\tau}\sum_{i=0}^{n-1-\tau}x_{i+\tau}\right)
\end{eqnarray}
\noindent
The term $ \frac{1}{n-\tau}\sum_{i=0}^{n-1-\tau}x_i \times
  \frac{1}{n-\tau}\sum_{i=0}^{n-1-\tau}x_{i+\tau}$
will converge to $\overline{x}^2$ for $n\to\infty$ if it
can be assumed that the distribution of the sample points
is stationary, i.e.\ does not depend on the sample time. 
Therefore, $\tilde C(\tau)$ is approximately  
$ \frac{1}{n-\tau}\sum_{i=0}^{n-1-\tau}(x_i-\overline{x})
(x_{i+\tau}-\overline{x})$, comparable to the nominator of the 
linear correlation coefficient
Eq.\ (\ref{eq:linearCorrelation}).
Usually one normalizes the correlation function by $\tilde C(0)$,
which is just the sample variance in the stationary case,
see Eq.\ (\ref{eq:sampleVariance}):
\begin{equation}
C(\tau)=\tilde C(\tau)/C(0)\,.
\end{equation}

\begin{figure}[!ht]
\begin{center}
\includegraphics[width=0.6\textwidth]{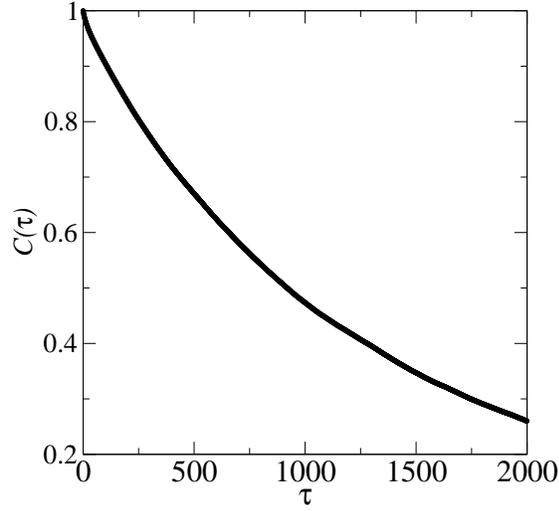}
\end{center}
\caption{Correlation function $C(\tau)$ for a simulation of
a ferromagnetic system, $x_i$ being the magnetization
at time step $i$. (For experts: 
Ising system of size $16\times 16$ spins simulated
with single-spin flip Metropolis Monte Carlo at a (reduced) 
temperature $T=2.269$ close
to the phase transition temperature, where correlation times $\tau_c$ 
are large). 
\label{fig:correl}}
\end{figure}

Consequently, for any data, for example obtained from a 
Markov chain Monte Carlo simulation, $C(0)=1$ will always hold,
 Then $C(\tau)$ decreases
with increasing difference $\tau$, see for example Fig.\ \ref{fig:correl}. 
Very often the functional form
is similar to an exponential $\sim \exp(-\tau/\tau_c)$.
In theory, $C(\tau)$ should converge to zero for $\tau\to\infty$, but due
to the finite size of the sample, usually strong fluctuations appear
for $\tau$ approaching $n$. A typical time $\tau_c$ 
which measures how fast the dependence of
the sample points  decreases is given by $C(\tau_c)=1/e$,
which is consistent with the above expression, if
the correlation function decreases exponentially.
At twice this distance, the correlation is already substantially
decreases (to $1/e^2$).
Consequently, if you want to obtain error bars for samples
obtained from dependent data, you could include for instance  only points
$x_0,x_2{\tau_c}, x_{4\tau_c}, x_{6\tau_c},\ldots$ in a sample, or just
use $n/(2\tau_c)$ instead of $n$ in any calculation of error bars.
Although these error bars are different from those
if the sample was really independent, it gives
a fairly good impression of the statistical error.

Alternatively, to obtain a typical time $\tau_c$ without calculating 
a correlation function, you can also
use the {\em blocking method} \cite{flyvbjerg1998}. Within this
approach, you
iteratively merge neighboring data points via 
$x^{(z+1)}_{i}=(x^{(z)}_{2i}+x^{(z)}_{2i+1})/2$
and $n^{(z+1)}=n^{(z)}/2$ (iteration level $z=0$ corresponds
to the original sample). You calculate the standard error
bar $\sigma^{(z)}/\sqrt{n^{(z)}-1}$
for each iteration level. Once it reaches a plateau
at level $z_c$, the data is (almost) independent and the
true error bar is given by the level value. Then 
$\tau_c=2^{z_c}$ is a typical time of independence of the data points.

If you are really just interested in error bars,
i.e.\ you do not need to know the value of $\tau_c$,
 you could also use the bootstrap approach\index{bootstrap approach}
which is not susceptible to dependence of data, see Sec.\
\ref{sec:bootstrap}.

\index{statistical dependence|)}
\index{hypothesis!testing|)}

\section{General estimators\label{sec:generalEstimate}}

In Sec.\ \ref{sec:statistics}, different methods are presented
of how to estimate  parameters which can be obtained
directly and simply from the given sample $\{x_0,x_1,$ \ldots, $x_{n-1}\}$.
In this section, a general method is considered which enables
estimators to be obtained
 for arbitrary parameters of probability distributions.
The method is based on the 
{\em maximum-likelihood principle}, which is exposed in Sec.\
\ref{sec:maxlikelihood}. This principle can be extended
to the {\em modeling of data}, where often a sample of triplets  
$\{(x_0,y_0,\sigma_0),\,(x_1,y_1,\sigma_1),\,$ \ldots, 
$(x_{n-1},y_{n-1},\sigma_{n-1})\}$ is given.
Typically the $x_i$ data points represent some control parameter,
which can be chosen in the simulation, such as the temperature
of a gas. It is assumed that all $x_i$ values
are different. Consequently, the simulation has been carried 
out at $n$ different values of the control parameter.
The $y_i$ data points are averages of measurements (e.g.\ the density
of the gas)
obtained in the simulations for the fixed value $x_i$ of the control parameter.
The $\sigma_i$ values are the corresponding error 
bars.\footnote{Sometimes also the $x_i$ data points are measured
quantities which are also characterized by error bars.
The generalization of the methods to this case
is straightforward.}
Modeling the data means that one wants to determine a relationship
$y=y(x)$. Usually some assumptions or knowledge about the
relationship are available, which means
one has available one  parametrized test function 
$y_{\myvec{\theta}}(x)$.
Consequently, the set of parameters has to be adjusted
$\myvec{\theta}$ such that the function $y_\myvec{\theta}(x)$
{\em fits} the sample ``best''. This is called {\em data fitting}
and will be explained in Sec.\ \ref{sec:fitting}.
This approach can also be used to compare several fitted test functions
to determined which represents the most suitable model.

\subsection{Maximum likelihood\label{sec:maxlikelihood}}
\index{maximum likelihood|(}

Here, we consider the following task: For a given sample 
$\{x_0,x_1,$ \ldots, $x_{n-1}\}$ and a probability distribution
represented by a pmf $p_{\myvec{\theta}}(x)$ or a pdf  
$f_{\myvec{\theta}}(x)$,  we want to determine
the parameters $\myvec{\theta}=(\theta_1,\ldots,\theta_{n_{\rm p}})$ 
such that the pmf or pdf represents the
data ``best''. This is written in parentheses, because there is
not unique definition what ``best'' means, or even a mathematical
way to derive a suitable criterion. If one assumes no
prior knowledge about the parameters, one can use the
following principle:

\begin{definition}
The {\em maximum-likelihood principle}\index{maximum likelihood!principle}
 states that the parameters
$\myvec{\theta}$ should be chosen such that the likelihood
of the data set, given the parameters, is maximal.
\end{definition}

In case of a discrete random variable, if it can be assumed that
the different data points are independent,
the likelihood of the data  
is just given by the product of the single data point
probabilities. This defines the 
{\em likelihood function}\index{likelihood function}
\begin{equation}
L(\myvec{\theta})\equiv p_{\myvec{\theta}}(x_1) p_{\myvec{\theta}}(x_2)
 \ldots p_{\myvec{\theta}}(x_{n-1})
= \prod_{i=0}^{n-1}p_{\myvec{\theta}}(x_i)
\end{equation}

 For the continuous case, the probability to obtain during
a random experiment  exactly
a certain sample is zero. Nevertheless, for a small
uncertainty parameter $\epsilon$, the probability
to obtain  
 a value in the interval $[\tilde x-\epsilon,\tilde x+\epsilon]$
is $\Prob(\tilde x-\epsilon \le X < \tilde x+\epsilon)=$
$\int_{\tilde x-\epsilon}^{\tilde x+\epsilon} f_{\myvec{\theta}}(x)\,dx$
$\approx f_{\myvec{\theta}}(\tilde x)2\epsilon$. Since $2\epsilon$
enters just as a factor, it is not relevant to determining the
maximum. Consequently, for the continuous
case, one considers the following 
likelihood function\index{likelihood function}
   
\begin{equation}
L(\myvec{\theta})\equiv f_{\myvec{\theta}}(x_1) f_{\myvec{\theta}}(x_2)
 \ldots f_{\myvec{\theta}}(x_{n-1})
= \prod_{i=0}^{n-1}f_{\myvec{\theta}}(x_i)
\end{equation}

To find the maximum of a likelihood function $L(\myvec{\theta})$
analytically, one has to calculate the first derivatives 
with respect to all parameters, respectively, and
requires them to be zero. Since calculating the derivative of
a product involves the application
of the product rule, it is usually more convenient to 
consider the {\em log-likelihood
function}\index{log-likelihood function}
\begin{equation}
l(\myvec{\theta}) \equiv \log L(\myvec{\theta})\,.
\end{equation}
This turns the product of single-data-points pmfs or pdfs into
a sum, where the derivatives are easier to obtain. Furthermore,
since the logarithm is a monotonous function, the maximum
of the likelihood function is the same as the maximum
of the log-likelihood function.
Hence, the parameters which suit ``best'' are determined within
the maximum-likelihood approach by the set of equations
\begin{equation}
\pd{l(\myvec{\theta})}{\theta_k} \stackrel{!}{=} 0 \quad
(k=1,\ldots,n_{\rm p})
\label{eq:pd}
\end{equation}
Note that the fact that the first derivatives are zero only
assures that an extremal point is obtained. 
Furthermore, these equations often have several solutions.
Therefore, one has to check explicitly which solutions are indeed
maxima, and which is the largest one.
Note that maximum-likelihood estimators, since they are
functions of the samples, are also
random variables ${\rm{}ML}_{\theta_k,n}(X_0,\ldots,X_{n-1})$.

As a toy example, we consider the exponential distribution with
the pdf given by Eq.\ (\ref{eq:exponentialDistr}). It has
one parameter $\mu$. The log-likelihood function for a sample
$\{x_0,x_1,$ \ldots, $x_{n-1}\}$ is in this case
\begin{eqnarray*}
l(\mu) & = & \log \prod_{i=0}^{n-1}f_{\mu}(x_i) \\
& = & \sum_{i=0}^{n-1} \log 
\left\{ \frac{1}{\mu} \exp\left( -\frac{x_i}{\mu} \right) \right\} \\
& = & \sum_{i=0}^{n-1}\left( \log\left\{\frac{1}{\mu}\right\} 
 -\frac{x_i}{\mu}   \right)\\
& = & n \log\left\{\frac{1}{\mu}\right\} -\frac{n}{\mu} \overline{x}
\end{eqnarray*}
Taking the derivative with respect to $\mu$ we obtain:
\begin{eqnarray*}
0 \stackrel{!}{=}\pd{L(\myvec{\theta})}{\mu}
= n \frac{-1}{\mu^2} \mu - \frac{-n}{\mu^2} \overline{x}
= \frac{-n}{\mu^2} (\mu-\overline{x})
\end{eqnarray*}
This implies $\mu=\overline{x}$. It is easy to verify
that this corresponds to a maximum.
Since the expectation value
for the exponential distribution is just $\E[X]=\mu$,
this is compatible with the result from Sec.\ \ref{sec:statistics},
where it was shown
that the sample mean is an unbiased estimator of the expectation value.

If one applies the maximum-likelihood principle to a Gaussian
distribution with parameters $\mu$ and $\sigma^2$, one obtains
(not shown here, see for example \cite{dekking2005}) 
as maximum-likelihood  estimators the sample mean $\overline{x}$ (for $\mu$) 
and the sample variance $s^2$ (for $\sigma^2$), respectively. 
This means (see Eq.\ (\ref{eq:varianceBiased}))
that the maximum-likelihood estimator for $\sigma^2$ is biased. Fortunately,
we know that the bias disappears asymptotically for $n\to\infty$. Indeed, it
can be shown, under rather mild conditions on the underlying
distributions, that all maximum-likelihood estimators 
${\rm{}ML}_{\theta_k,n}(X_0,\ldots,X_{n-1})$
for a parameter $\theta_k$ are asymptotically unbiased, i.e.
\begin{equation}
\lim_{n\to \infty} \E[{\rm{}ML}_{\theta_k,n}] = \theta_k 
\end{equation}

In contrast to the exponential and Gaussian cases, 
for many applications the maximum-likelihood parameter is
not directly related to a standard sample estimator. Furthermore,
${\rm{}ML}_{\theta_k,n}$ can often
even not be determined analytically. In this case, one
has to optimize the log-likelihood function numerically,
for example, using the corresponding methods from the 
{\em GNU scientific library} (GSL) \index{GNU scientific library}
(see Sec.\ \ref{sec:gsl}).

\sources{randomness}{max\_likely.c}
As example, we consider Fisher-Tippett distribution,
see Eq.\ (\ref{eq:FiherTippet}), shifted to exhibit the maximum
at $x_0$ instead of at 0. Hence, we have two parameters $\lambda$ and
$x_0$ to adjust. The function to be optimized (the {\em target function}),
i.e.\ the
log-likelihood function here, must be of a special format when using 
the minimization functions of the 
GSL.  This first argument of the target function 
contains the pdf parameters to be adjusted, i.e.\
the main argument vector of the target function.
This argument must be  of the
type \verb!gsl_vector!, which is a GSL type for vectors.
One needs to include \verb!<gsl/gsl_vector.h>! to use this data type.
These vectors are created using \verb!gsl_vector_alloc()!,
set elements via \verb!gsl_vector_set()!, access elements via
\verb!gsl_vector_get()! and delete the vectors via \verb!gsl_vector_free()!.
The usage of these functions should be self-explanatory from the
examples below, but you may also have a look at 
the GSL documentation \cite{gsl}.

The second argument of the target function contains {\em one}
pointer to all additional data needed to calculate the target
function, i.e.\ the sample in this case. Thus, the sample must be stored
in {\em one}
 chunk of memory. For this purpose, we use the following structure type:

{\small 
\begin{verbatim}
typedef struct
{
  int       n;       /* number of sample points; */
  double   *x;                         /* sample */
}
sample_t;
\end{verbatim}
} 

Since the GSL package contains actually
 minimization functions, while
we are interested in a maximum,
the actual log-likelihood function returns minus
the log-likelihood.  The log-likelihood function  reads as follows:
\newpage
{\small 
\linenumbers[1]
\begin{verbatim}
double ll_ft(const gsl_vector *par, void *param)
{
  double lambda, x0;                     /* parameters of pdf */
  sample_t *sample;                                 /* sample */
  double sum;          /* sum of log-likelihood contributions */
  int i;                                      /* loop counter */

  lambda = gsl_vector_get(par, 0);                /* get data */ 
  x0 = gsl_vector_get(par, 1); 
  sample = (sample_t *) param;

  sum = sample->n*log(lambda);    /* calculate log likelihood */
  for(i=0; i<sample->n; i++)
    sum -= lambda*(sample->x[i]-x0) + 
           exp(-lambda*(sample->x[i]-x0));

  return(-sum);                    /* return - log likelihood */
}
\end{verbatim}
\nolinenumbers} 

First, we convert the pointers passed as arguments to the data format
that we find useful (lines 8--10). Next, the actual log likelihood
$$
l(\lambda,x_o)= n\log \lambda - \lambda \sum_{i=0}^{n-1}(x_i-x_0)
-\sum_{i=0}^{n-1} \exp(-\lambda(x_i-x_0))
$$
is calculated in lines 12--15 and finally returned with inverted
sign (line 17).

The GSL has built in several minimization algorithms. They are
all put under one of two frameworks. One framework is for algorithms
which require the target function and its first derivatives. 
The other framwork contains
algorithms where just the target function is sufficient.
Here we use  the {\em simplex algorithm}\index{simplex algorithm},
which belongs to the latter form. It works by spanning a 
simplex,\footnote{A simplex is a convex set in an $n$-dimensional space
generated by $n+1$ corner points.}
evaluating the target functions at the corners of the simplex,
and iteratively changing
the simplex until it is very small and contains the solution. Note that
the algorithm is only able to find local minima, and only one of them.
If several minima exist, the choice of the initial parameters
strongly influence the final results; Here, one maybe has to try several
parameters. For details see \cite{gsl}.
Here we only show how to use the minimizer. The minimizer itself
is stored in a special data structure of type
\verb!gsl_multimin_fminimizer!. The target function 
has to be put into a 
``surrounding'' variable of type \verb!gsl_multimin_function!.
Furthermore, one needs two \verb!gsl_vector! variables to store
the current estimate for the optimum (specifying the position of the simplex)
and  to store the size of the simplex. Also, \verb!par! is used here to
state the dimension of the target function argument (2) and
\verb!sample! to store the sample.

These variables are declared
as follows:
\newpage

{\small \begin{verbatim}
  int num_par;                        /* number of parameters */
  sample_t sample;                                  /* sample */

  gsl_multimin_fminimizer *s;           /* the full mimimizer */
  gsl_vector *simplex_size;        /* (relative) simplex size */ 
  gsl_vector *par; /* params to be optimized = args of target */
  gsl_multimin_function f;  /* holds function to be optimized */
\end{verbatim}}

The actual allocation and initialization of these variables
may look as follows:

{\small \begin{verbatim}
  sample.n = 10000;                          /* initilization */
  sample.x = (double *) malloc(sample.n*sizeof(double));
  num_par = 2;

  f.f = &ll_ft;                    /* initialize minimization */
  f.n = num_par;
  f.params = &sample;
  simplex_size = gsl_vector_alloc(num_par);  /* alloc simplex */
  gsl_vector_set_all(simplex_size, 1.0);      /* init simplex */
  par = gsl_vector_alloc(num_par);  /* alloc + init arguments */
  gsl_vector_set(par, 0, 1.0);
  gsl_vector_set(par, 1, 1.0);
  s =
    gsl_multimin_fminimizer_alloc(gsl_multimin_fminimizer_nmsimplex, 
                                  num_par);
  gsl_multimin_fminimizer_set(s, &f, par, simplex_size);
\end{verbatim}}

\begin{sloppypar}
The set-up of the minimizer object comes in two steps,
first allocation using \verb!gsl_multimin_fminimizer_alloc()!,
then initialization  via \verb!gsl_multimin_fminimizer_set()!
while passing  the target function, the starting point \verb!par!
and the (initial) simplex size.\footnote{The simplex is spanned
by {\tt par} and the $n$ vectors given by {\tt par} plus ($0,\ldots,0,$
{\tt simplex\_size[i]}, $0,\ldots,0$) for $i=1,\ldots,n$.}
The \verb!sample.x[]! array has to be filled with the actual 
sample (not shown here).
\end{sloppypar}

The minimization loop looks as follows:

{\small \begin{verbatim}
  do                                  /* perform minimization */
  {
    iter++;
    status = gsl_multimin_fminimizer_iterate(s);  /* one step */
    if(status)    /* error ? */ 
      break;
    size = gsl_multimin_fminimizer_size(s);    /* converged ? */
    status = gsl_multimin_test_size(size, 1e-4);
  }
  while( (status == GSL_CONTINUE) && (iter<100) );
        \end{verbatim}}
The main work is done in \verb!gsl_multimin_fminimizer_iterate()!.
Then it is checked whether an error has occurred. Next,
the size of the simplex is calculated and finally  tested
whether the size falls below some limit, $10^{-4}$ here.

\begin{sloppypar}
The actual estimate of the parameters can be obtained
via \verb!gsl_vector_get(s->x, 0)! and \verb!gsl_vector_get(s->x, 1)!.
Note that finally all allocated memory should be freed:
\end{sloppypar}

\index{maximum likelihood|)}
{\small \begin{verbatim}
  gsl_vector_free(par);                    /* free everything */
  gsl_vector_free(simplex_size);
  gsl_multimin_fminimizer_free(s);
  free(sample.x);
        \end{verbatim}}

As an example, $n=10000$ data points were generated according to
a Fisher-Tippett distribution with parameters $\lambda=3.0$, $x_0=2.0$.
With the above starting parameters, the minimization converged
 to the values $\hat\lambda=2.995$ and
$\hat x_0=2.003$ after 39 iterations.

\subsection{Data fitting\label{sec:fitting}}
\index{data fitting|(} \index{fitting|(}

In the previous section, the parameters of a probability distribution
are chosen such that the distribution describes the data best. Here,
we consider a more general case, called 
{\em modeling of data}\index{modeling of data}\index{data modeling}.
As explained above, here a sample of triplets  
$\{(x_0,y_0,\sigma_0),\,(x_1,y_1,\sigma_1),\,$ \ldots, 
$(x_{n-1},y_{n-1},\sigma_{n-1})\}$ is given.
Typically, the $y_i$ are measured values obtained from a simulation
with some control parameter (e.g.\ the temperature) 
fixed at different values $x_i$; $\sigma_i$ is the corresponding 
error bar of $y_i$.
Here, one wants to determine parameters 
$\myvec{\theta}=(\theta_1,\ldots,\theta_{n_{\rm p}})$
such that the given parametrized function
$y_{\myvec{\theta}}(x)$ fits the data ``best'', one says 
one wants to {\em fit} the function to the data. Similar to the
case of fitting a pmf or a pdf, there is no general principle
of what ``best'' means.

Let us assume that the $y_i$ are random variables,
i.e.\ comparing different simulations. Thus, the measured
values  are scattered around
their ``true'' values $y_{\myvec{\theta}}(x_i)$. 
This scattering can be described approximately by a Gaussian distribution
with mean $y_{\myvec{\theta}}(x_i)$ and variance $\sigma_i^2$:
\begin{eqnarray}
\label{eq:distrMeasurements}
q_{\myvec{\theta}}(y_i)\sim \exp\left(-\frac{ 
    (y_i -y_{\myvec{\theta}}(x_i))^2}{2\sigma_i^2}   \right) \,.
\end{eqnarray}
This assumption is often valid, e.g.\ when each sample point $y_i$
is itself a sample mean obtained from a simulation performed at
control parameter value $x_i$, and $\sigma_i$ is the
corresponding error bar.
The log-likelihood function for the full data sample is 
\begin{eqnarray*}
l(\myvec{\theta}) & = & \log \prod_{i=0}^{n-1} q_{\myvec{\theta}}(y_i) \\
& \sim & - \sum_{i=0}^{n-1}\frac{1}{2}  
\left( \frac{y_i -y_{\myvec{\theta}}(x_i)}{\sigma_i}\right)^2
\end{eqnarray*}
Maximizing $l(\myvec{\theta})$ is equivalent to minimizing 
$-2l(\myvec{\theta})$,
hence one minimizes the 
{\em mean-squared difference}\index{mean-squared difference}
\begin{equation}
\chi^2_{\myvec{\theta}} = \sum_{i=0}^{n-1}
\left( \frac{y_i -y_{\myvec{\theta}}(x_i)}{\sigma_i}\right)^2
\end{equation}
\index{least-squares fitting|(ii}\index{fitting!least-squares|(ii}
This means the parameters 
$\myvec{\theta}$ are determined such that function 
$y_{\myvec{\theta}}(x)$ follows the data points
$\{(x_0,y_0),\ldots(x_{n-1},y_{n-1})\}$ as close as possible,
where the deviations are measured in terms of the error bars
$\sigma_i$. Hence, data points with smaller error bar
enter with more {\em weight}. The full procedure is called 
{\em least-squares fitting}.

The minimized mean-squared difference is a random variable. Note that
the different terms are not statistically independent, since
they are related by the $n_{\rm p}$ parameters 
$\myvec{\hat \theta}$ which
are determined via minimizing $\chi^2_{\myvec{\theta}}$.
As a consequence, the distribution of $\chi^2_{\myvec{\hat\theta}}$ 
is approximately given
by chi-squared distribution (see Eq.\ (\ref{eq:chi2}) for the pdf)
with $n-n_{\rm p}$ degrees 
of freedom. This distribution 
can be used to evaluate the statistical significance
of a least-squares fit, see below.

In case, one wants to model the underlying distribution function
for a sample as in Sec.\ \ref{sec:maxlikelihood}, 
say for a continuous distribution, it is possible
in principle to use the least-squares approach as well. In
this case one would fit 
the parametrized pdf to a histogram pdf, which has also the above
mentioned sample format $\{(x_i,y_i,\sigma_i)\}$.
Nevertheless,
 although the least-squares principle is derived
using the maximum-likelihood principle, usually
different parameters are obtained if one fits a pdf to
a histogram pdf compared to obtaining these parameters from
a direct maximum-likelihood approach. Often \cite{bauke2007},
the maximum-likelihood method gives more accurate results. Therefore,
one should use a least-squares fit mainly for a 
fit of a non-pmf/non-pdf function to a data set.

Fortunately, to actually perform
 least-squares fitting, you do not have to write
your own fitting functions, because there are very good
fitting implementations readily available.
Both programs presented in
Sec.\ \ref{sec:plotting}, {\it gnuplot}\/ and {\it xmgrace}\/, offer
fitting to arbitrary functions. It is advisable to use {\em
  gnuplot}\/, since it offers  higher flexibility for that purpose
and gives you more information useful to estimate the quality of a fit. 

As an example, let us suppose that you want to fit an algebraic function
of the form $f(L)=e_\infty +aL^b$ to the data set  of the file {\tt
  sg\_e0\_L.dat} shown on page \pageref{page:SGe0}. 
First, you have to define the function and
supply some rough (non-zero) estimations for the unknown parameters.
Note that the exponential operator is denoted by {\tt **} and the
standard argument for a function definition is {\tt x}, but this
depends only on your choice:

{\small \begin{verbatim}
gnuplot> f(x)=e+a*x**b
gnuplot> e=-1.8
gnuplot> a=1
gnuplot> b=-1
\end{verbatim}}

The actual fit is performed via the {\tt fit} command. The program
uses the nonlinear least-squares
 Levenberg-Marquardt algorithm \cite{PRA-numrec1995}, which allows
 a fit data to almost all arbitrary functions. To issue the command,
 you have to state the fit function, the data set and the parameters
 which are to be adjusted. 
First, we consider the case where just two columns of the data
are used or available (in this case, {\it gnuplot} assumes $\sigma_i=1$).
For our example you enter:

{\small \begin{verbatim}
gnuplot> fit f(x) "sg_e0_L.dat" via e,a,b
\end{verbatim}}

 Then {\it gnuplot}\/ writes log information to the output describing the
fitting process. After the fit has converged it prints for the given example:

 {\small \begin{verbatim}
After 17 iterations the fit converged.
final sum of squares of residuals : 7.55104e-06
rel. change during last iteration : -2.42172e-09

degrees of freedom (ndf) : 5
rms of residuals      (stdfit) = sqrt(WSSR/ndf)      : 0.00122891
variance of residuals (reduced chisquare) = WSSR/ndf : 1.51021e-06

Final set of parameters            Asymptotic Standard Error
=======================            ==========================

e               = -1.78786         +/- 0.0008548    (0.04781%)
a               = 2.5425           +/- 0.2282       (8.976%)
b               = -2.80103         +/- 0.08265      (2.951%)


correlation matrix of the fit parameters:

               e      a      b      
e               1.000 
a               0.708  1.000 
b              -0.766 -0.991  1.000 
 
\end{verbatim}}

  The most interesting lines are those where the results 
$\myvec{\hat\theta}$ for your
parameters along with the standard error bar\index{error bar}
are printed.\footnote{These ``error bars'' are calculated in a way
which is in fact correct only when fitting linear functions; hence,
they have to be taken with care.} Additionally, the quality
of the fit can be estimated by the information provided in the three
lines beginning with ``{\tt degree of freedom}''. \index{degree!of freedom}
The first of these lines
states the number of degrees of freedom, which is just $n-n_{\rm p}$. 
The mean-squared difference $\chi^2_{\myvec{\hat\theta}}$ 
is denoted as \verb!WSSR! in the {\it gnuplot}\/
output. A measure of quality of the fit \index{quality of fit}
\index{fitting!quality of} is the probability $Q$
that the value of the mean-squared difference
 is equal or larger compared to the value from the current fit, given the
assumption that the data points are distributed as in 
Eq.\ (\ref{eq:distrMeasurements})
 \cite{PRA-numrec1995}. 
The larger the value of
$Q$, the better is the quality of the fit.
As mentioned above, $Q$ can be evaluated from a chi-squared
distribution with $n-n_{\rm p}$ degrees of freedom.
To calculate $Q$ using the {\it gnuplot output}
you can use the little program {\tt Q.c}

{\small \linenumbers[1]
\begin{verbatim}
#include <stdio.h>
#include <stdlib.h>
#include <math.h>
#include <gsl/gsl_cdf.h>

int main(int argc, char *argv[])
{
  double WSSRndf;
  int ndf;

  if(argc != 3)
  {
    printf("USAGE %s <ndf> <WSSR/ndf>\n", argv[0]);
    exit(1);
  }
  ndf = atoi(argv[1]);
  sscanf(argv[2], "%lf", &WSSRndf);
  printf("# Q=%e\n", gsl_cdf_chisq_Q(ndf*WSSRndf, ndf));

  return(0);
}
\end{verbatim}
\nolinenumbers
}

\noindent
which uses the {\tt gsl\_cdf\_chisq\_Q()} function from
the GSL (see Sec.\ \ref{sec:gsl}). 
The program is called in the form {\tt Q <ndf>
  <WSSR/ndf>}, which can be taken from the {\it gnuplot}\/ output.
Note that in this case we obtain $Q=1$, which is so large, because
$\sigma_i=1$ was used, see below.

To watch the result of the fit along with the original data, just enter

{\small \begin{verbatim}
gnuplot> plot "sg_e0_L.dat" w e, f(x)
\end{verbatim}}

\begin{figure}[th]
\centerline{\psfig{file=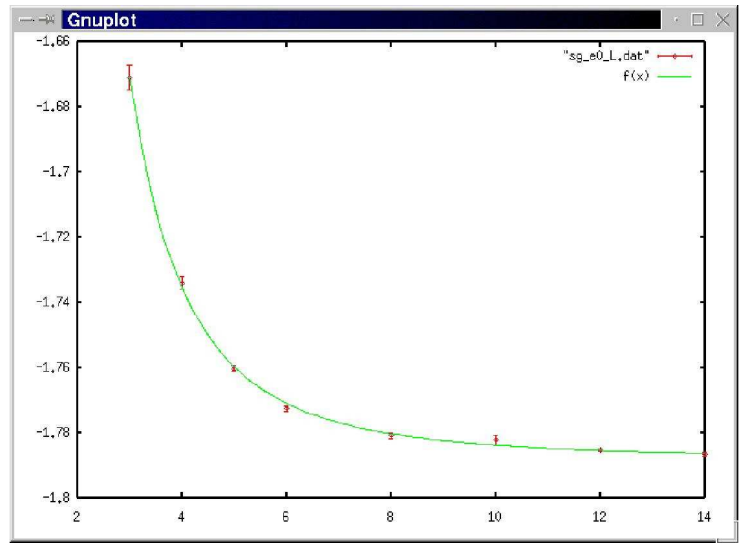,width=0.8\textwidth}}
\caption{{\it Gnuplot}\/ window showing the result of a fit command along with
  the input data.
\label{fig:gnuplotB}}
\end{figure}

\noindent
The result is displayed in Fig.\ \ref{fig:gnuplotB}.
Please note that the
 convergence depends on the initial choice of the parameters. The
 algorithm may be trapped into a local minimum in case the parameters
 are too far away from the best values. Try the initial values
 {\tt e=1}, {\tt a=-3} and {\tt b=1}! 
Furthermore, not all function parameters have to be subjected to
the fitting. Alternatively, you can set some parameters to fixed
values and omit them from the \verb!via! list at the end of the \verb!fit!
command. Remember that
 in the above example all data points enter into the
result with the same weight, i.e.\ $\sigma_i=1\,\forall i$ is assumed.
 You can tell the algorithm to consider
the error bars, for example supplied in the third column,
 by typing 

{\small \begin{verbatim}
gnuplot> fit f(x) "sg_e0_L.dat" using 1:2:3 via e,a,b
\end{verbatim}}
\noindent
 Then, data points with larger error bars have less influence
on the results. In this case a different result whith smaller
value of $Q$ will arise (try it !).

Finally, you can also restrict the data points which
are considered for the fit, which is applicable if only a subset
of the sample follows the function law you are considering. This can be done
in the same way as restricting the range of plotted values, for
instance using

{\small 
\begin{verbatim}
gnuplot> fit [5:12] f(x) "sg_e0_L.dat" using 1:2:3 via e,a,b
\end{verbatim}
}
\noindent
More information on how to use the {\tt fit} command,
such as fitting higher-dimensional data, can be
obtained when using the {\em gnuplot} online help
via entering {\tt help fit}.
\index{least-squares fitting|)}\index{fitting!least-squares|)}
\index{data fitting|)} \index{fitting|)}

\index{data!analysis|)}
\index{analyzing data|)}

 \clearpage

\begin{exercises}
\item {\bf Sampling from discrete distribution}
\label{ex:samplingDiscrete}
\index{random variable!discrete}
\index{discrete random variable}

\begin{minipage}[t]{0.55\textwidth}
Design, implement and test a function, which 
returns a random number which is distributed according to
some discrete distribution function stored in an array \verb!F!,
as describe in Sec.\ \ref{sec:drawDiscrete}.
\end{minipage}
\sourcesSOL{randomness}{poisson.c}

The function prototype reads as follows:

{\small
\begin{verbatim}
/******************** rand_discrete() *****************/
/** Returns natural random number distributed        **/
/** according a discrete distribution given by the   **/
/** distribution function in array 'F'               **/
/** Uses search in array to generate number          **/
/** PARAMETERS: (*)= return-paramter                 **/
/**       n: number of entries in array              **/
/**       F: array with distribution function        **/ 
/** RETURNS:                                         **/
/**     random number                                **/
/******************************************************/
int rand_discrete(int n, double *F)
\end{verbatim}}

For simplicity, you can use the \verb!drand48()! function from the
standard C library to generate random numbers 
distributed according to $U(0,1)$.

Furthermore,  design, implement and test a function, which allocates and
initializes the array \verb!F! for a Poisson distribution with 
parameter $\mu$, see Eq.\ (\ref{eq:poissonA}) for 
the\index{Poisson distribution}\index{distribution!Poisson}
probability mass function. The function should determine automatically
how many entries of \verb!F! are needed, depending on the paramater $\mu$.
The function prototype reads as follows:

{\small
\begin{verbatim}
/********************* init_poisson() *****************/
/** Generates  array with distribution function      **/
/** for Poisson distribution with mean mu:           **/
/** p(k)=mu^k*exp(-mu)/x!                            **/
/** The size of the array is automatically adjusted. **/
/** PARAMETERS: (*)= return-paramter                 **/
/**   (*) n_p: p. to number of entries in table      **/
/**        mu: parameter of distribution             **/
/** RETURNS:                                         **/
/**     pointer to array with distribution function  **/
/******************************************************/
double *init_poisson(int *n_p, double mu)
\end{verbatim}}

Hints: To determine the array sizes, 
you can first loop over the probabilities and take the
first  value \verb!k_0! where $p($\verb!k_0!$)=0$ within the precision
of the numerics. This value of \verb!k_0! serves as array size.
Alternatively, you start with some size and extend the array
if needed by doubling its size. 
For testing purposes, you can generate many numbers, calculate the mean and
compare it with $\mu$. Alternatively, you could record a histogram
(see Chap.\ \ref{chap:oop}) and compare with Eq.\ (\ref{eq:poissonA}).

\item {\bf Inversion Method for Fisher-Tippett distribution}
\label{ex:fisherTippett}
\index{Fisher-Tippett distribution} \index{distribution!Fisher-Tippett}

\begin{minipage}[t]{0.55\textwidth}
Design, implement and test a function, which 
returns a random number which is distributed according to the
Fisher-Tippett distribution Eq.\ (\ref{eq:FiherTippet})
with parameter $\lambda$.
Use the inversion method.
\end{minipage}
\sourcesSOL{randomness}{fischer\_tippett.c}

The function prototype reads as follows:

{\small
\begin{verbatim}
/******************** rand_fisher_tippett() ***********/
/** Returns random number which is distributed       **/
/** according the Fisher-Tippett distribution        **/
/** PARAMETERS: (*)= return-paramter                 **/
/**       lambda: parameter of distribution          **/
/** RETURNS:                                         **/
/**     random number                                **/
/******************************************************/
double rand_fisher_tippett(double lambda)
\end{verbatim}}

Remarks: For simplicity, you can use the \verb!drand48()! function from the
standard C library to generate random numbers 
distributed according to $U(0,1)$. To test your function, you can 
calculate the mean of the generated numbers, for instance,
and compare it with the expectation value $\sim 0.57721/\lambda$.

\item {\bf Variance of data sample}\label{ex:variance}
\index{sample!variance}

\begin{minipage}[t]{0.55\textwidth}
Design, implement and test a function, which 
calculates the variance $s^2$ of a sample of data points.
Use directly Eq.\ (\ref{eq:sampleVariance}), i.e.\
do {\em not} use an equivalent form of Eq.\ (\ref{eq:Var:Prop}), since this
form is more susceptible to rounding errors.
  \end{minipage}
\sourcesSOL{randomness}{variance.c}

The function prototype reads as follows:

{\small
\begin{verbatim}
/********************** variance() ********************/
/** Calculates the variance of n data points         **/
/** PARAMETERS: (*)= return-paramter                 **/
/**        n: number of data points                  **/
/**        x: array with data                        **/
/** RETURNS:                                         **/
/**     variance                                     **/
/******************************************************/
double variance(int n, double *x)
\end{verbatim}}

Remark: The so-called corrected double-pass algorithm \cite{chan1983}
aims at further reducing the rounding error. It is based
on the equation
$$
s^2 = \frac{1}{n} \left[ \sum_{i=0}^{n-1} (x-\overline{x})^2
- \frac{1}{n} \left( \sum_{i=0}^{n-1} (x_i-\overline{x}) \right)^2 \right]\,.
$$
The second would be zero for exact arithmetic and accounts for rounding
erros occurring in the second term. It becomes important
in particular if the expectation value is large. Perform
experiments for generating
Gaussian distributed number with $\sigma^2=1$ and $\mu=10^{14}$, without and
with the correction.

\item {\bf Bootstrap}\label{ex:bootstrap}
\index{bootstrap approach}

\begin{minipage}[t]{0.55\textwidth}
Design, implement and test a function, which 
uses bootstrapping to calculate  
the confidence interval at significance level $\alpha$ given in
Eq.\ (\ref{eq:bsConfidence}).
  \end{minipage}
\sourcesSOL{randomness}{bootstrap\_ci.c}

The function prototype reads as follows:

{\small
\begin{verbatim}
/***************** bootstrap_ci() *********************/
/** Calculates a confidence interval by 'n_resample' **/
/** times resampling the given sample points         **/
/** and each time evaluation the estimator 'f'       **/
/** PARAMETERS: (*)= return-paramter                 **/
/**           n: number of data points               **/
/**           x: array with data                     **/
/**  n_resample: number of bootstrap iterations      **/
/**       alpha: confidence level                    **/
/**           f: function (pointer) = estimator      **/
/**     (*) low: (p. to) lower boundary of conf. int.**/
/**    (*) high: (p. to) upper boundary of conf. int.**/
/** RETURNS:                                         **/
/**     (nothing)                                    **/
/******************************************************/
void bootstrap_ci(int n, double *x, int n_resample,
                  double alpha, double (*f)(int, double *),
                  double *low, double *high)
\end{verbatim}}

Hints: Use the function \verb!bootstrap_variance()!
as example. To get the entries at the positions defined via 
Eq.\ (\ref{eq:bsConfidence}), you can sort the bootstrap sample
first using \verb!qsort()!, see Sec.\ \ref{sec:c:library}.

\begin{sloppypar}
You can test your function by using the provided
main file \verb!bootstrap_test.c!, the auxiliary files
\verb!mean.c! and \verb!variance.c! and by compiling with
\verb!cc -o bt bootstrap_test.c bootstrap_ci.c mean.c -lm -DSOLUTION!.
Note that the macro definition \verb!-DSOLUTION! makes the \verb!main()!
function
to call \verb!bootstrap_ci()! instead of \verb!bootstrap_variance()!.
\end{sloppypar}

\item {\bf Plotting data}\label{ex:xmgrace}
\index{xmgrace@{\em{}xmgrace}|(ii}
\index{plotting data}
\index{data!plotting}

\begin{minipage}[t]{0.55\textwidth}
Plot the data file \verb!FTpdf.dat! using {\em xmgrace}.
The file contains a histogram pdf generated for the Fisher-Tippett distribution.
The file format is 1st column: bin number, 2nd: bin midpoint,
3rd: pdf value, 4th: error bar.
Use \end{minipage}
\sourcesSOL{randomness}{FTplot.agr}

the ``block data'' format to read the files (columns 2,3,4). 
Create a plot with inset.
The main plot should show the histogram pdf with error bars and
logarithmically scaled $y$ axis, the inset should show the data
with linear axes. 
Describe the plot using a text label placed in the plot.
Choose label sizes, line width and other styles suitably.
Store the result as \verb!.agr! file and export it to a postscript (eps)
file. 

The result should look similar to:

\begin{center}
\includegraphics[width=0.6\textwidth]{pic_random_small/FTplot.eps}
\end{center}


\item {\bf Chi-squared test}\label{ex:chi2}
\index{chisquaredtest@{chi-squared test}}
\index{test!chisquared@{chi-squared}}

\begin{minipage}[t]{0.55\textwidth}
Design, implement and test a function, which 
calculates the $\chi^2$ test statistics for two histograms 
$\{h_i\},\{\hat h_i\}$ according Eq.\ (\ref{eq:chi2B}). 
 The function should return the p-value,
i.e.\ the 

\end{minipage}
\sourcesSOL{randomness}{chi2hh.c}

cumulative probability 
(``p-value'') \index{p-value} that
a value of $\chi^2$ or larger is obtained under the assumption that
the two histograms were obtained by sampling from the same
(discrete) random variable.

The function prototype reads as follows:

{\small
\begin{verbatim}
/********************* chi2_hh() ***********************/
/** For chi^2 test: comparison of two histograms      **/
/** to probabilities: Probability to                  **/
/** obtain the corresponding  chi2 value or worse.    **/
/** It is assumed that the total number of data points**/
/*+ in the two histograms is equal !                  **/
/**                                                   **/
/** Parameters: (*) = return parameter                **/
/**  n_bins: number of bins                           **/
/**       h: array of histogram values                **/
/**      h2: 2nd array of histogram values            **/
/**                                                   **/
/** Returns:                                          **/
/**       p-value                                     **/
/*******************************************************/
double chi2_hh(int n_bins, int *h, int *h2)
\end{verbatim}}

Hints: Use the functio \verb!chi2_hd()! as example.
Include a test, which verifies that the total number of counts
in the two histograms agree.

To test the function: Generate two histograms according to a binomial
distribution with parameters $n=$ \verb!par_n!$=10$ and $p=$ 0.5 or
$p=$  \verb!par_p!. Perform a loop for different values of \verb!par_p!
and calculate the p-value each time using 
the \verb!gsl_cdf_chisq_Q()! function of  the {\em GNU scientific library}
(GSL) \index{GNU scientific library}
(see Sec.\ \ref{sec:gsl}).

\item {\bf Linear correlation coefficient}\label{ex:lcc}
\index{linear correlation coefficient}

\begin{minipage}[t]{0.55\textwidth}
Design, implement and test a function, which 
calculates the linear correlation coefficient $r$ 
to measure the strength of a correlation for a sample
$\{(x_0,y_0),$
$(x_1,y_1),$ \ldots, $(x_{n-1},y_{n-1})\}$.

\end{minipage}
\sourcesSOL{randomness}{lcc.c}

The function prototype reads as follows:


{\small
\begin{verbatim}
/**************************** lcc() ********************/
/** Calculates the linear correlation coefficient     **/
/**                                                   **/
/** Parameters: (*) = return parameter                **/
/**     n: number of data points                      **/
/**     x: first element of sample set                **/
/**     y: second element of sample set               **/
/**                                                   **/
/** Returns:                                          **/
/**       r                                           **/
/*******************************************************/
double lcc(int n, double *x, double *y)
\end{verbatim}}

Remark: Write a \verb!main()! function which generates
a sample in the following way:  the $x_i$
numbers are generated from a standard Gaussian distribution
$N(0,1)$  while each $y_i$ number is drawn
from a Gaussian distribution with expectation value $\kappa x_i$ (variance 1).
Study the result for different values of $\kappa$ and $n$.

\item {\bf Least-squares fitting}\label{ex:leastSquares}
\index{least-squares fitting}\index{fitting!least-squares}

\begin{minipage}[t]{0.55\textwidth}
Copy the program from exercise (\ref{ex:fisherTippett})
to a new program and change it such that numbers for a shifted
 Fisher-Tippett with parameters $\lambda$ and peak position $x_0$
are generated.
The numbers should be stored in a histogram and a
histogram pdf should be written to the standard output.

\end{minipage}
\sourcesSOL{randomness}{fitFT.gp\\fisher\_tippett2.c}

\begin{itemize}
\item Choose the histogram parameters (range, bin range)
such that the histograms match the generated data well.
\item Run the program to generate $n=10^5$ numbers
for parameters $x_0=2.0$ and $\lambda=3.0$. 
Pipe the histogram pdf to a file (e.g.\ using \verb!> ft.dat! at the
end of the call).
\item Plot the result using {\em gnuplot}. 
\item Define the pdf for the Fisher-Tippett distribution in {\em gnuplot} 
and fit the function to the data with $x_0$ and $\lambda$
as adjustable parameters. Choose a suitable range for the fit.
\item Plot the data together with the fitted function.
\item How does the result compare to the maximum-likelihood fit
presented in Sec.\ \ref{sec:maxlikelihood}?
\item Does the fit (in particular for $\lambda$) get better if
you increase the number of sample points to $10^6$?
\end{itemize}

Hints: The shift is implemented by just adding $x_0$
to the generated random number.
Use either the histograms from Chap.\ \ref{chap:oop},
or implement a ``poor-mans histogram'' via an array \verb!hist! (
see also in the \verb!main()! function
of the \verb!reject.c! program partly presented in Sec.\ \ref{sec:reject}).

\end{exercises}



\end{document}